# THE HYDROGEN ATOM ACCORDING TO WAVE MECHANICS – I. SPHERICAL POLAR COORDINATES

*J. F. Ogilvie**

Centre for Experimental and Constructive Mathematics, Department of Mathematics, Simon Fraser University, Burnaby, British Columbia V5A 1S6 Canada

Escuela de Química, Universidad de Costa Rica, Ciudad Universitaria Rodrigo Facio, San Pedro de Montes de Oca, San José, 11501-2050 Costa Rica

**Abstract**

In the first of five parts in a series, the Schroedinger equation is solved in spherical polar coordinates to yield wave functions that enable an accurate calculation of the frequencies and intensities of lines in the absorption spectrum of the Lyman series in the vacuum-ultraviolet region. Accurate plots of surfaces of amplitude functions illustrate the variation of shapes and sizes varying with quantum numbers *k*, *l*, *m* for comparison with the corresponding plots of amplitude functions in other systems of coordinates.

**Resumen**

En este primer artículo de cinco, la ecuación de Shroedinger se resuelve usando coordinadas polares esféricas para obtener funciones de onda que facilitan un cálculo preciso de las frecuencias e intensidades del especto de absorción de una serie de Lyman en la región de ultravioleta al vacío. Los gráficos precisos de las funciones de amplitud muestran la variación de las formas y tamaños con los números cuánticos *k*, *l* y *m* para comparar con los gráficos correspondientes a las funciones de amplitud en otros sistemas de coordenadas.

**Palabras clave:** átomo de hidrógeno, mecánica cuántica, coordenadas polares esféricas, orbitales, espectro atómico.

## I. INTRODUCTION

Several months after Heisenberg initiated quantum mechanics in 1925 [1], Schroedinger introduced wave mechanics with four articles, of translated title *Quantisation as a Problem of Proper Values* [2,3,4,5], that have since served as a general basis of calculations on microscopic systems for physical and chemical purposes. Heisenberg attacked first the problem of an anharmonic oscillator [1], which eventually involved operations with explicit matrices [6]. Pauli then applied a symbolic method, involving no explicit matrix, to generate the energies of states of the hydrogen atom, but "the calculation of transition probabilities (intensities) [was] omitted from consideration" [7]. In his first article on wave mechanics, Schroedinger solved his differential equation independent of time for the hydrogen atom in spherical polar coordinates on assuming an amplitude function of appropriate properties [2], and achieved an account of the energies of the discrete states that was intrinsically no great advance on Bohr's grossly flawed derivation and even on Pauli's quantum-mechanical achievement [7]. In contrast, in an astonishing achievement in his third article, Schroedinger not only analogously solved the differential equation for the hydrogen atom in paraboloidal coordinates [4] but also developed a perturbation theory and calculated the intensities of spectral lines; in a fourth part [5], on incorporating time as a variable, he eliminated the energy parameter from the partial-differential equation, producing a temporal dependence in the resulting wave function.

---

* Corresponding author: ogilvie@cecm.sfu.ca

Among the eleven systems of coordinates [8] that allow a separation of three spatial variables in the Helmholtz equation, or hence also the Laplace equation because that Helmholtz equation contains the laplacian operator, only four systems enable a complete separation of variables of Schroedinger's partial-differential equation for the hydrogen atom to yield ordinary-differential equations, specifically the two specified above plus ellipsoidal coordinates [9], for which only indirect solutions in series had been achieved [10] before the present work, and spheroconical coordinates for which no explicit algebraic solution has ever been reported. The objective of the few articles in this series is to present in turn the solution, derived directly with advanced mathematical software (Maple) for symbolic computation, of Schroedinger's temporally dependent or independent equation in each of the four systems of coordinates, accompanied with numerous accurate illustrations of surfaces of amplitude functions in the various systems, and then to discuss the ramifications of these multiple solutions in a chemical context. Although the governing equations are, of necessity, defined in other systems of coordinates with three spatial dimensions, we view these surfaces invariably in rectangular cartesian coordinates: a computer program translates effectively from another or original system of coordinates, in which the algebra and calculus are performed, to the system to which a human eye is accustomed. The scope of treatments in articles in this series is limited to that appropriate to Schroedinger's equations in a context of pioneer quantum mechanics in its wave-mechanical variety, so neglecting *relativistic* effects, effects of electronic and nuclear intrinsic angular momenta, radiative effects and other aspects that are typically omitted from general undergraduate courses in chemistry and physics.

The most fundamental application of quantum mechanics is in atomic physics, which has also chemical implications. The simplest chemical species is the hydrogen atom, $^1$H, which consists of a simple atomic nucleus – a proton in the most common instance – and one electron, bound through an electrostatic attraction that acts between these particles. Heisenberg recognized that the observable properties of an atom or molecule are the frequencies and intensities of its spectral lines, and that the fundamental properties of an atomic or molecular system that are involved in a calculation of a spectrum are the relative positions and momenta of the particles comprising that system [1]. Within pioneer quantum mechanics, nobody has yet succeeded in predicting the intensities of lines in the discrete spectrum of a hydrogen atom without involving explicitly Schroedinger's amplitude or wave functions, or equivalent. For this reason, even though the latter are incontestably artefacts of both a particular method of calculation and a selected system of coordinates, they seem at present to be unavoidable for a calculation of important observable properties of an atomic or molecular system. In this part I, we review the solution of Schroedinger's equation for the hydrogen atom in spherical polar coordinates, presenting merely the most pertinent equations and formulae as a basis to explain the quintessential mathematical and physical implications. Details of the derivation of the formulae appear elsewhere [11].

## II.    SCHROEDINGER'S EQUATION IN SPHERICAL POLAR COORDINATES

The magnitude of a central force on an object depends on only the distance of that object from the origin; the direction of the force is along the line joining the origin and the object. The coulombic attraction is a central force, which implies a conservative field and which signifies that it is expressible as the gradient of a potential energy. Schroedinger's equation for an electron moving in a central force field is invariably separable in spherical polar coordinates, which in Schroedinger's paper is called simply polar coordinates [2]. We assume the electron and the proton, or other atomic nucleus, to constitute point masses that interact according to Coulomb's law; a deviation from that law might imply a non-zero rest mass of a photon, for which no evidence exists, apart from the effects of the finite volume and shape of a massive atomic nucleus, and their isotopic variation, for which experimental evidence exists. We first relate these coordinates, i.e. radial coordinate $r$, polar angular coordinate $\theta$ and equatorial angular coordinate $\phi$, to cartesian coordinates $x, y, z$ as algebraic formulae, according to ISO standard 80000-2:2009,

$$x = r \sin(\theta) \sin(\phi), \quad y = r \sin(\theta) \cos(\phi), \quad z = r \cos(\theta)$$

with domains $0 \leq r < \infty$, $0 \leq \theta < \pi$, $0 \leq \phi < 2\pi$, so that axis $z$ in cartesian coordinates becomes the polar axis in spherical polar coordinates. For the motion of the electron relative to the atomic nucleus, the use of a reduced mass converts the problem of treating two interacting particles into a treatment of effectively a single particle subject to a force field; the motion of the atom as a whole through space is of little interest – only the internal motion produces observable properties readily observable in atomic spectra in absorption or emission. Coordinate $r$ signifies the distance between reduced mass $\mu$ and the origin; coordinate $\theta$ signifies the angle of inclination between a line joining that reduced mass to the origin and polar axis $z$ in cartesian coordinates; coordinate $\phi$ signifies the equatorial angle between a half-plane containing that line, between the reduced mass and the origin, and half-plane $x=0$; a half-plane extends from the polar axis to $\infty$ in any direction. The limiting cases are thus for $r$ a point at the origin as $r \to 0$, and for $\theta$ a line along positive axis $z$ as $\theta \to 0$ and along negative axis $z$ as $\theta \to \pi$. Surfaces of coordinates $r$, $\theta$ and $\phi$ as constant quantities are exhibited, with definitions, in figure 1. For use within an integrating element in subsequent integrals, the jacobian of the transformation of coordinates between cartesian and spherical polar, as defined above, is $r^2 \sin(\theta)$.

After the separation of the coordinates of the centre of mass of the H atom, Schroedinger's temporally dependent equation in explicit SI units,

$$\left( -\frac{\frac{\partial}{\partial r} \psi(r, \theta, \phi, t)}{4\pi^2 r} - \frac{\cos(\theta) \left( \frac{\partial}{\partial \theta} \psi(r, \theta, \phi, t) \right)}{8\pi^2 r^2 \sin(\theta)} - \frac{\frac{\partial^2}{\partial \theta^2} \psi(r, \theta, \phi, t)}{8\pi^2 r^2} - \frac{\frac{\partial^2}{\partial \phi^2} \psi(r, \theta, \phi, t)}{8\pi^2 r^2 \sin(\theta)^2} - \frac{\frac{\partial^2}{\partial r^2} \psi(r, \theta, \phi, t)}{8\pi^2} \right) h^2/\mu - \frac{Z e^2 \psi(r, \theta, \phi, t)}{4\pi \varepsilon_0 r} = \frac{i h \left( \frac{\partial}{\partial t} \psi(r, \theta, \phi, t) \right)}{2\pi}$$

contains within terms on the left side an electrostatic potential energy proportional to $r^{-1}$ and first and second partial derivatives of an assumed wave function $\Psi(r, \theta, \phi, t)$ with respect to spatial coordinates $r, \theta, \phi$, and on the right side a first partial derivative with respect to time $t$. Apart from fundamental physical constants, specifically electric permittivity of free space $\varepsilon_0$, Planck constant $h$ and protonic charge *e*, there appear constant parameters $Z$ for atomic number – $Z = 1$ for H – and $\mu = m_e M /(M + m_e)$ for the reduced mass of the atomic system with nuclear mass $M$; this reduced mass is practically equal to the electronic rest mass $m_e$. In the limit of infinite nuclear mass, the position of that nuclear mass coincides with the origin of the system of coordinates. After the separation of the variables and the solution of the four consequent ordinary-differential equations including definition of the integration constants and separation parameters, the full solution of the above equation has this form [*11*].

$$\Psi(r, \theta, \phi, t) = c \sqrt{\frac{Z \pi \mu e^2 k!}{\varepsilon_0 h^2 (k + 2l + 1)!}} \left(\frac{2 \mu \pi e^2 Z}{(k + l + 1) h^2 \varepsilon_0}\right)^{(l+1)} r^l$$

$$\text{LaguerreL}\left(k, 2l + 1, \frac{2 \pi \mu e^2 Z r}{h^2 \varepsilon_0 (k + l + 1)}\right) e^{\left(-\frac{\pi \mu e^2 Z r}{h^2 \varepsilon_0 (k + l + 1)}\right)} e^{(i m \phi)}$$

$$\sqrt{\frac{(2l+1)(l-|m|)!}{(l+|m|)!}} \text{LegendreP}(l, |m|, \cos(\theta)) e^{\left(-\frac{i \mu Z^2 e^4 \pi t}{4 h^3 \varepsilon_0^2 (k + l + 1)^2}\right)} \Big/ \left(2(k + l + 1)\sqrt{\pi}\right)$$

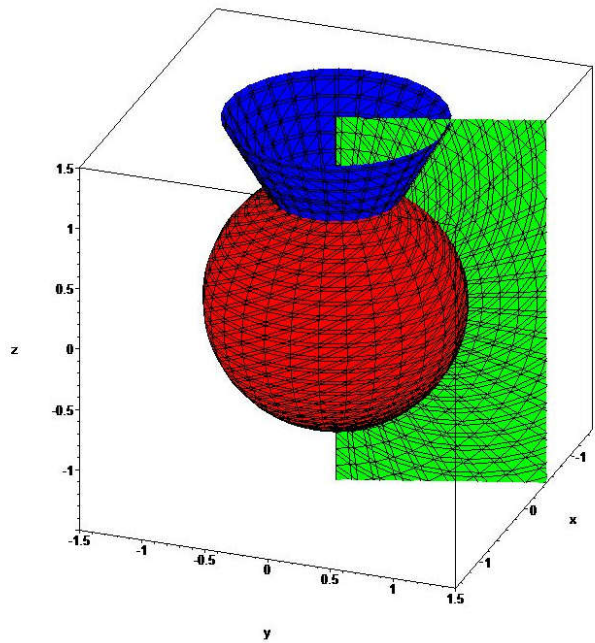

**FIGURE 1.** Definition of spherical polar coordinates $r$, $\theta$, $\phi$: a sphere (red) has radius $r = 1$ unit; a cone (blue) has polar angle $\theta = \pi/6$ rad with respect to polar axis $+z$; a half-plane (green) has equatorial angle $\pi/10$ rad with respect to plane $x = 0$.

The presence of i = $\sqrt{-1}$ in two exponential factors as product with $\phi$ or $t$ signifies that this formula is complex, thus containing real and imaginary parts. Coefficient *c*, which here does not denote the speed of light and which equals any complex number of magnitude unity such as a fourth root of unity – i. e. $c = \pm 1, \pm\sqrt{-1}$, appears because Schroedinger's equation is linear and homogeneous, or equally because Schroedinger's temporally independent equation is of form an eigenvalue relation,

$$H(r, \theta, \phi)\, \psi(r, \theta, \phi) = E\, \psi(r, \theta, \phi)$$

in which $H(r,\theta,\phi)$ denotes a hamiltonian operator with contributions from kinetic and potential energies. A conventional choice $c = 1$, which is arbitrary and lacks physical justification, signifies that some amplitude function $\psi(r,\theta,\phi)$, as solution from the temporally independent Schroedinger equation with $m=0$, appears in a real form, whereas most must be complex; with a mathematically valid alternative choice $c = i$, some amplitude functions would be entirely imaginary but most would still be complex. Choosing $c = -1$ or $-i$ merely reverses the phase of an amplitude function or its constituent parts. Parameters that appear in the solution but not in the partial-differential equation take discrete values, imposed by boundary conditions, as follows: $m$ is called the equatorial, or magnetic, quantum number that assumes only integer values and that arises in the solution of the angular equation to define $\Phi(\phi)$, as indicated below; $l$ is called the azimuthal quantum number that assumes values of only non-negative integers and that arises in the solution of the angular equation to define $\Theta(\theta)$, which also involves $m$ as its absolute value; product $Y(\theta,\phi) = \Theta(\theta)\,\Phi(\phi)$ constitutes a special function known as spherical harmonic to represent functions defined on the surface of a sphere; $k$ is a radial quantum number that assumes values of only non-negative integers, and arises in the solution of the radial equation to define $R(r)$, which also involves quantum number $l$. The names of quantum numbers $k$ and $m$ hence pertain to the coordinates from which they arise. There is no relative limitation of the values of $k$ and $l$, but, for a given value of $l$, $m$ assumes $2l + 1$ integer values from $-l$ to $+l$. The total wave function is thus a product

$$\Psi(r, \theta, \phi, t) = c\, R(r)\, \Theta(\theta)\, \Phi(\phi)\, \tau(t) = c\, \psi(r, \theta, \phi)\, \tau(t)$$

from the normalised solutions of the four separate ordinary-differential equations. The multiplicative terms in the total product that contain no variable serve as normalizing factors, to ensure that, for the amplitude function,

$$\int \psi(r,\theta,\phi)^* \, \psi(r,\theta,\phi)\, \mathrm{d}vol = 1 \,,$$

in which the integration, with $\psi(r,\theta,\phi)^*$ formed from $\psi(r,\theta,\phi)$ on replacing i by $-i$ and with volume element $\mathrm{d}vol$ incorporating the jacobian specified above, is performed over all space according to the domains of the spatial variables as specified above. Henceforth we take $c = 1$.

The coefficient of $t$ with i in one exponential term above has the physical dimensions and significance of a radial frequency, as Schroedinger noted [5], but we interpret that quantity in its particular context as energy $E$ of a particular state divided by Planck constant $h$, as Schroedinger also applied. We associate sum $k + l + 1$, which must assume a value of a positive integer, in the same exponent with experimental quantum number $n$ for energy,

$$n = k + l + 1 \,;$$

according to the formula for the discrete spectral lines of H derived by Balmer and elaborated by Rydberg, the energy of a discrete state of H is proportional to $-1/n^2$. That sum of integers occurs also in associated Laguerre function LaguerreL that represents in R(*r*) the radial dependence of wave function Ψ(*r*, θ, φ, *t*), but not in associated Legendre function LegendreP of the first kind, Θ(θ), that contains the angular dependence on θ.

The physical significance of equatorial quantum number *m* is that, for a given value of *l*, 2 *l* + 1 specifies the number of states, distinguished by their values of *m* from −*l* to +*l*, that have distinct energies for a H atom in the presence of an externally applied magnetic field; that field hence removes a degeneracy whereby multiple states have the same energy. Multiple amplitude functions, corresponding to particular values of *k*, *l*, *m* and explicitly numbering $(k + l + 1)^2$, yield the same energy of a H atom, according to the eigenvalue relation above, in the absence of an externally applied magnetic or electric field; for the solution of Schroedinger's temporally dependent or independent equation in spherical polar coordinates, the energy in the absence of a magnetic field is thus independent of equatorial quantum number *m*. The mathematical significance of azimuthal quantum number *l* is that it specifies the number of angular nodes of an amplitude function – i.e. the number of times that a particular amplitude function changes sign on θ varying between 0 and π rad. The mathematical significance of radial quantum number *k* is that it specifies the number of radial nodes of an amplitude function – i.e. the number of times that a particular amplitude function changes sign on *r* varying from 0 to ∞, or the number of zero points of product ψ(*r*,θ,φ)* ψ(*r*,θ,φ) = |ψ(*r*,θ,φ)|² between *r* = 0 and ∞ in any direction from the origin, Whereas equatorial quantum number *m* has hence a direct physical significance, the azimuthal, *l*, and radial, *k*, quantum numbers have directly a mathematical significance within only a restricted geometric context, and are therefore artifacts of this derivation in spherical polar coordinates; in contrast, product *l* (*l* + 1) has a physical significance as discussed below.

Among many properties of a hydrogen atom that one might explore after having derived explicit formulae for the amplitude or wave functions, we mention only the principal properties that concerned Heisenberg, namely the frequencies and intensities of spectral lines. The frequency ν of a spectral line is that of a photon emitted or absorbed by an atom, and bears no direct relation to any purported internal frequency of an atom that produces that spectrum. That optical frequency is specifically the difference of energies, $E_j$ and $E_{j'} > E_j$, of two states of an atom between which an optical transition occurs, divided by Planck's constant, according to Bohr's relation:

$$\nu = (E_{j'} - E_j ) / h$$

The energies of states of H defined with energy quantum number $n = k + l + 1$ are depicted in figure 2, in which energies are expressed in rydberg unit; in terms of fundamental physical constants,

$$1 \text{ rydberg} = m_e e^4 / 8\, e_0^2\, c\, h^3 = 2.179872325 \times 10^{-18} \text{ J}$$

For energies less than a limiting energy as $n \to \infty$ that corresponds to ionization of the atom, the energies are discrete, although of formally uncountable number, whereas above that threshold the energies are continuous.

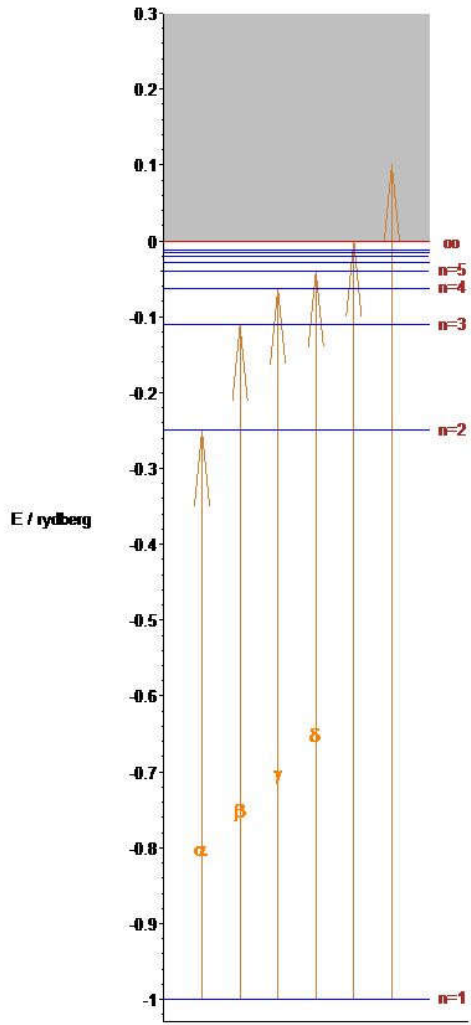

**FIGURE 2.** Energies of continuous states with positive energies and of discrete states with negative energies, relative to $E = 0$ that pertains to a proton and an electron infinitely separate and at rest; energies at five discrete values bear labels of energy quantum number $n$ from 1 to 5. In order of increasing length, four vertical arrows with greek letters indicate transitions between discrete states, observed in absorption in the Lyman series of lines in the vacuum-ultraviolet spectral region; two further arrows indicate a minimum ionization and a transition into the continuum of energies above the threshold of ionization.

The electric-dipolar moment appropriate to an optical transition in absorption from a state with quantum numbers $k$, $l$, $m$ to another state with quantum numbers $k'$, $l'$, $m'$ involving component $z$, with $z = r \cos(\theta)$, of the electric component of the electromagnetic field is calculated, in the length representation and with spherical polar coordinates, as quantity $< e\,z > = < e\,r \cos(\theta) >$ involving integration over the corresponding wave functions that hence include the temporal dependence.

$$< e\,z > = \int_0^\infty \int_0^\pi \int_0^{2\pi} \Psi_{k', l', m'}(r, \theta, \phi, t) * e\,r \cos(\theta)\, \Psi_{k, l, m}(r, \theta, \phi, t)\, r^2 \sin(\theta)\, d\phi\, d\theta\, dr$$

In this triple integral over the spatial coordinates, $\Psi^*$ denotes a complex-conjugate wave function in terms of spatial $r$, $\theta$, $\phi$ and temporal $t$ variables and of quantum numbers $k'$, $l'$, $m'$; a transformation from $\Psi$ to $\Psi*$ involves the replacement everywhere of i by –i. For this observable property, intensity, as for any other observable property, the chosen value of coefficient $c$ in the definition of the wave function above is hence immaterial. The intensity of an optical transition is proportional to the square of the above matrix element for the electric-dipolar moment of that transition, which is thus independent of the sense of $z$ and of time $t$. In Dirac's notation with *bra* and *ket*, for a H atom ($Z = 1$), an optical transition corresponding to vertical arrow α in figure 2 is indicated from the state of least accessible energy specified with quantum numbers $k = 0$, $l = 0$, $m = 0$, so $n = k + l + 1 = 1$, to the first excited state, specified with quantum numbers $k' = 0$, $l' = 1$, $m' = 0$, so $n = 2$. This matrix element of dipolar moment is calculated symbolically to have this value:

$$<0, 1, 0 \mid e\,z \mid 0, 0, 0> \ = \ \frac{128\,\sqrt{2}\,h^2\,\varepsilon_0\,\mathrm{e}^{\left(\left(-\frac{3\,i\,\mu\,Z^2\,e^4\,\pi}{16\,h^3\,\varepsilon_0^2}\right)t\right)}}{243\,Z\,e\,\pi\,\mu}$$

The states, denoted $^2$S for the electronic ground state and $^2$P for an excited state accessible therefrom in absorption, are distinct from the amplitude or wave functions expressed in a particular coordinate system with which calculations might be made, but for the particular transition indicated we associate the amplitude function specified with quantum numbers $k=0$, $l=0$, $m=0$ uniquely with state $^2$S and amplitude function with quantum numbers $k=0$, $l=1$, $m=0$ with state $^2$P. In the latter formula we interpret the coefficient, with i, of $t$ in the exponent to specify the optical frequency, i.e. the angular frequency of the photon that is absorbed when a H atom undergoes the pertinent transition with a gain of one unit, in terms of $h/2\pi$, of angular momentum through conservation of that quantity; the circular frequency is hence

$$\nu \ = \ \frac{3\,\mu\,Z^2\,e^4}{32\,h^3\,\varepsilon_0^2}$$

A contrast with the above interpretation of the coefficient of i $t$ in the exponent of the wave function itself, as energy divided by Planck constant, is noteworthy. In figure 3, we show quantitatively the absorption spectrum of the H atom below the threshold of ionization: the scale of the abscissa variable has unit $10^{15}$ Hz = PHz for frequency ν; because the intensities of transitions decrease rapidly with increasing frequency, for illustrative purposes the ordinate scale is logarithmic in a quantity $10^3\,f$; oscillator strength $f$ is a dimensionless quantity that serves as an appropriate measure of intensity. The greek letters above the spectral lines in figure 3 pertain to the designations of features of the spectrum of the H atom in the vacuum-ultraviolet region, in which transitions in absorption occur from the ground state |0,0,0> to states denoted |$k$, 1, 0 > with values of $k$ increasing from zero, and correlate with the same greek letters in figure 2. The absorption spectrum in the continuous region above the minimum energy of ionization is calculated quantitatively elsewhere [*12*]; like the discrete spectrum, the intensity per unit energy in the continuous spectrum diminishes rapidly with energy increasing above the threshold of ionization.

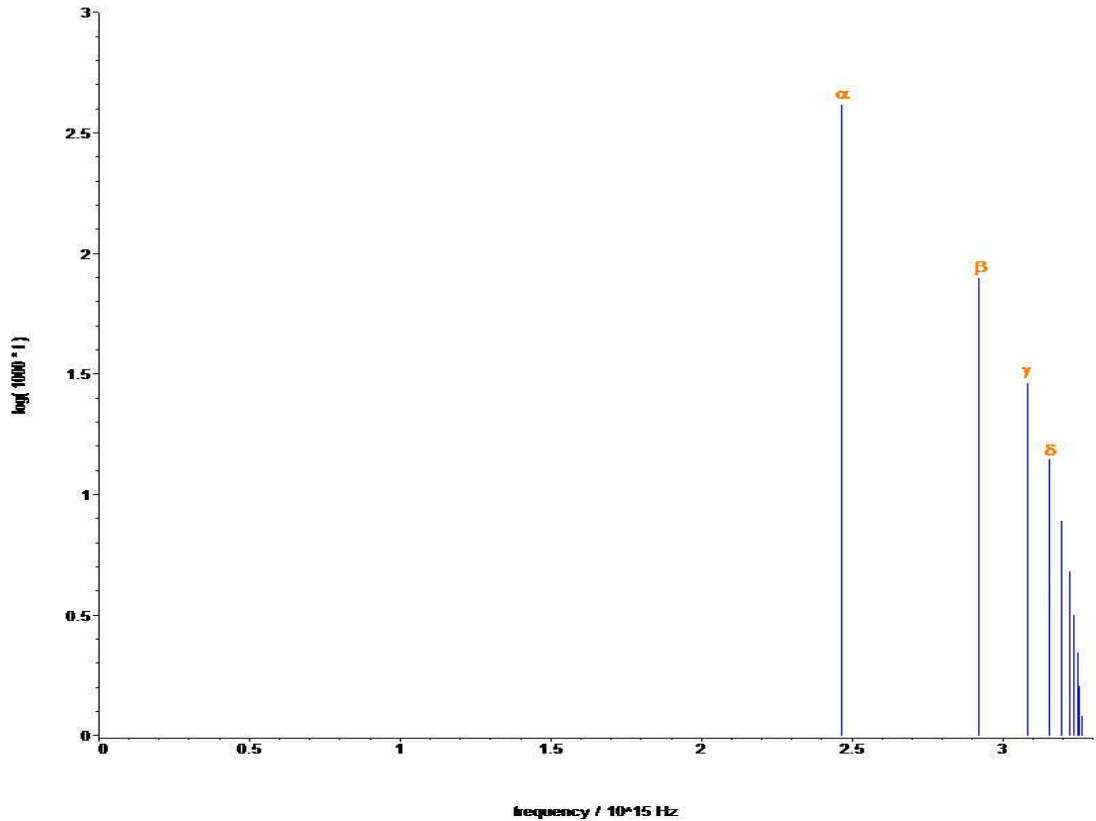

**FIGURE 3.** Quantitative representation of the absorption spectrum of the H atom for transitions from the ground state to ten discrete excited states as lines at particular frequencies/$10^{15}$ Hz with intensities according to $\log_{10}(10^3 f)$; $f$ denotes the oscillator strength.

### III.    GRAPHICAL REPRESENTATIONS OF AMPLITUDE FUNCTION $\psi(r,\theta,\phi)$

Not only for comparison with graphical representations of amplitude functions calculated in coordinates of other systems but also to present quantitatively accurate shapes and sizes of these functions, here follow plots of the surfaces of selected amplitude functions.  As a plot involving three independent variables – spatial coordinates $r$, $\theta$, $\phi$ – and one dependent variable $\psi(r,\theta,\phi)$ would require four spatial dimensions, the best way to proceed with two dimensions, or three pseudo-dimensions, is to exhibit a surface of constant $\psi$ at a value selected to display the overall spatial properties in a satisfactory manner.  Because in many published papers and textbooks these surfaces are portrayed inaccurately, we explain our procedure to produce an accurate plot, deploying first for this purpose amplitude function $\psi_{0,0,0}$.  Its formula,

$$\psi_{0,0,0} = \frac{e^3 \, \mu^{(3/2)} \, Z^{(3/2)} \, \pi \, e^{\left(-\frac{\pi \mu Z e^2 r}{h^2 \varepsilon_0}\right)}}{h^3 \, \varepsilon_0^{(3/2)}} \, ,$$

shows no dependence on angular variables, merely an exponential decay with distance *r* of the reduced mass from the origin of coordinates that effectively marks the location of the atomic nucleus.  By assumption, the electron interacts with the nuclear matter only through the electrostatic interaction; with point particles, the probability of finding the electron is maximum at the atomic nucleus, at which the amplitude function has a cusp: i.e. d$\psi$/d*r* is discontinuous at the origin.  We plot this dependence on *r* in figure 4 to demonstrate pertinent properties.

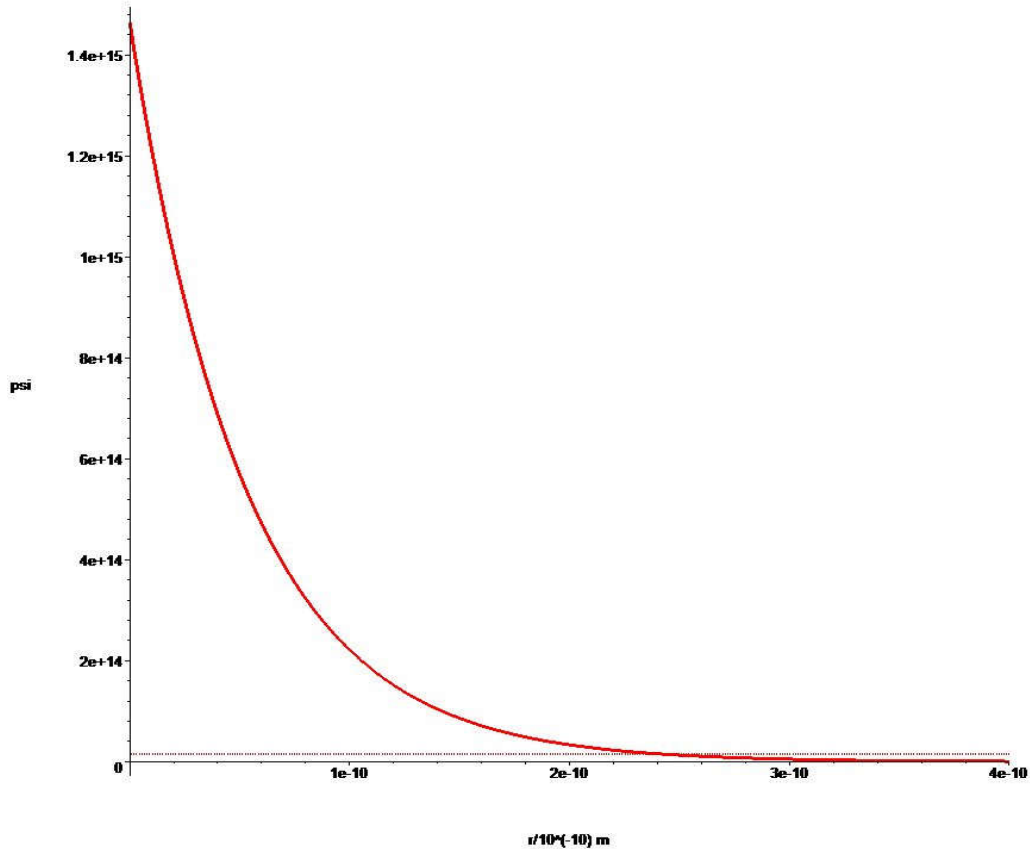

**FIGURE 4**.  Radial dependence of $\psi_{0,0,0}$/m$^{-3/2}$; the intersection of the dotted line with the curve indicates the point at which 0.995 of the density of electronic charge is within that distance from the origin of coordinates. Here, and in succeeding plots, notation 2e+14 implies $2 \times 10^{14}$ ; 2e−10 implies $2 \times 10^{-10}$, and analogously for other values.

In figure 4, a dotted line intersects the curve depicting the radial dependence of $\psi_{0,0,0}$ at a point $r_c$ chosen such that 0.995 of the total density of electronic charge according to an integral,

$$\int_0^{r_c} \psi(0,0,0)^2 \, 4\pi r^2 \, dr$$

is within a domain $0...r_c$; for $\psi_{0,0,0}$, that distance is $r_c = 2.45 \times 10^{-10}$ m.  When we plot a surface of constant $\psi_{0,0,0} = 1.43 \times 10^{13}$ m$^{-3/2}$, which is 0.01 times the maximum value of $\psi_{0,0,0}$, at $r = 0$, we obtain necessarily a sphere, having the stated radius $r_c$, because $\psi_{0,0,0}$ exhibits no angular dependence. We must adopt such a strategy to derive a meaningful idea of any *size* or *shape* of an amplitude function, or the corresponding density of electronic charge that is proportional to $|\psi^2|$ or $\psi^*\psi$ according to Born's interpretation, because amplitude function $\psi_{0,0,0}$ has zero value only as $r \to \infty$; i.e. according to Schroedinger's amplitude function, a single H atom in its ground state occupies the entire universe, but not to the practical exclusion of other atoms!

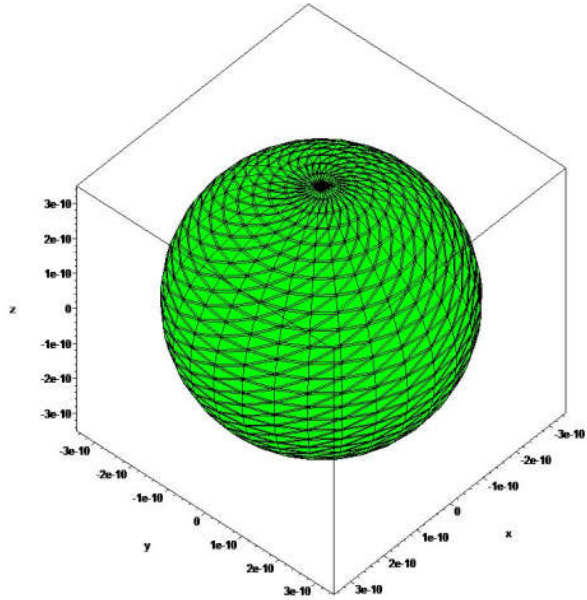

**FIGURE 5.** Surface of $\psi_{0,0,0} = 1.43 \times 10^{13}$ m$^{-3/2}$. Here, and in succeeding plots in three pseudo-dimensions, the unit of length along each coordinate axis is m.

We next consider amplitude function $\psi_{1,0,0}$ that has this explicit formula,

$$\psi_{1,0,0} = -\frac{1}{8} \frac{\sqrt{2}\, e^3\, \mu^{(3/2)}\, Z^{(3/2)}\, \pi\, (\pi\, Z\, e^2\, \mu\, r - 2\, h^2\, \varepsilon_0)\, e^{\left(-1/2\, \frac{\pi\, \mu\, Z\, e^2\, r}{h^2\, \varepsilon_0}\right)}}{\varepsilon_0^{(5/2)}\, h^5}$$

first as its radial profile, shown in figure 6 with two dotted lines at $\pm 3.17 \times 10^{12}$ m$^{-3/2}$. The intersection of the line at $-3.17 \times 10^{12}$ m$^{-3/2}$ with the curve at $r_c = 7.2 \times 10^{-10}$ m indicates the distance within which 0.995 of the total electronic charge is contained. Two other intersections of the curve with those two lines occur, both near $1.07 \times 10^{-10}$ m; the latter location marks the presence of a nodal surface, the single spherical surface between $r = 0$ and $r \to \infty$ at which function $\psi_{1,0,0}$ has zero amplitude and $\psi_{1,0,0}^2 = 0$, consistent with radial quantum number $k = 1$. An explanation of the appearance of that surface is that, relative to the phase convention set on assuming $c = 1$ in the solution of the temporally dependent Schroedinger equation, the phase of the amplitude function changes from positive to negative at that distance, $1.07 \times 10^{-10}$ m, as depicted in figure 6. For that reason we find three concentric spheres, of which the radii correspond to the distances at which the dotted lines intersect the curve in figure 6; at all those three points, the magnitude of $\psi_{1,0,0}$ is equal to that at the third point, at $r_c$, at which 0.995 of the total electronic charge is contained within that radius. Figure 7 shows a complete inner sphere that has a positive phase, and two spheres of negative phase cut open to show that inner sphere. These plotted surfaces of $\psi_{0,0,0}$ and $\psi_{1,0,0}$ and the explanation underlying their generation provide a basis for further plots to illustrate salient features of selected amplitude functions.

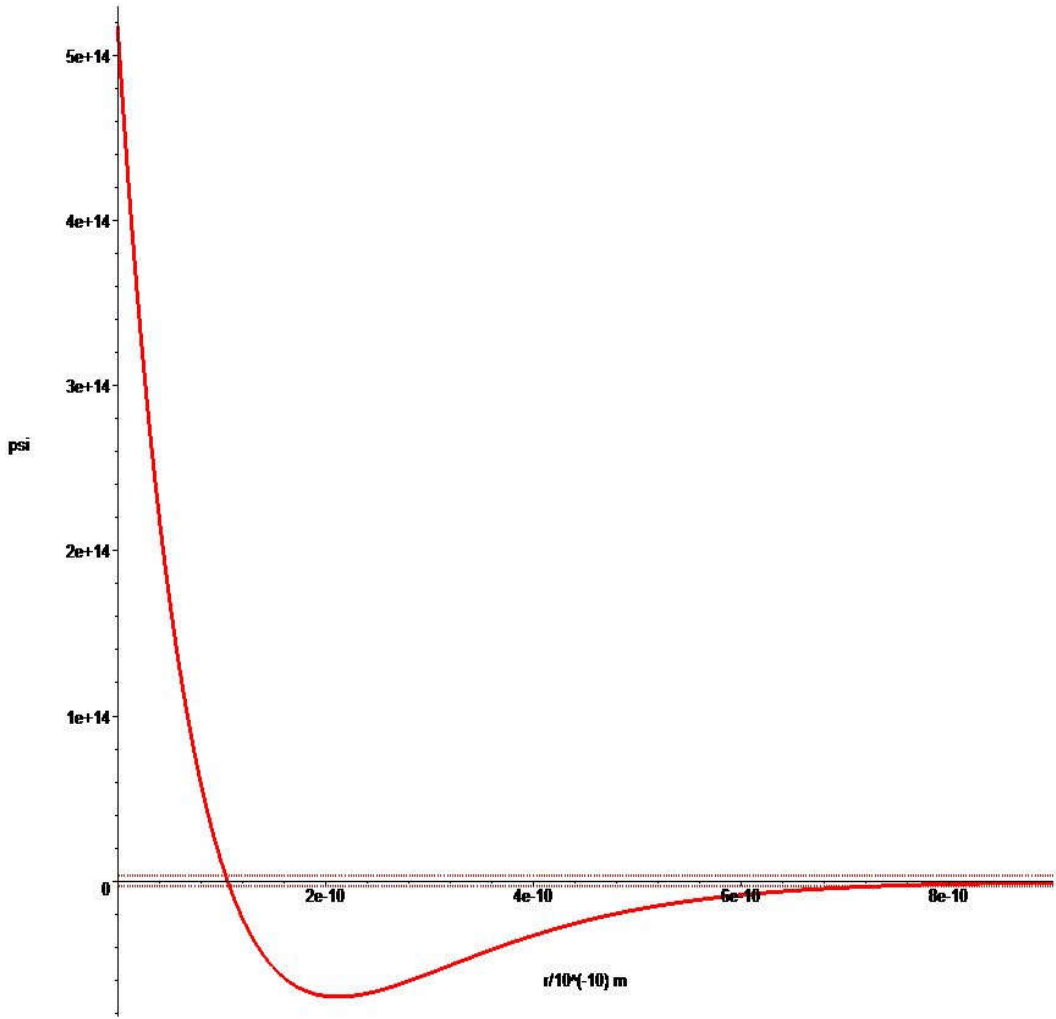

**FIGURE 6.** Radial dependence of $\psi_{1,0,0}$; the intersection, farthest from the origin, of the lower dotted line with the curve indicates the point at which 0.995 of the electronic charge occurs within that distance from the origin of coordinates.

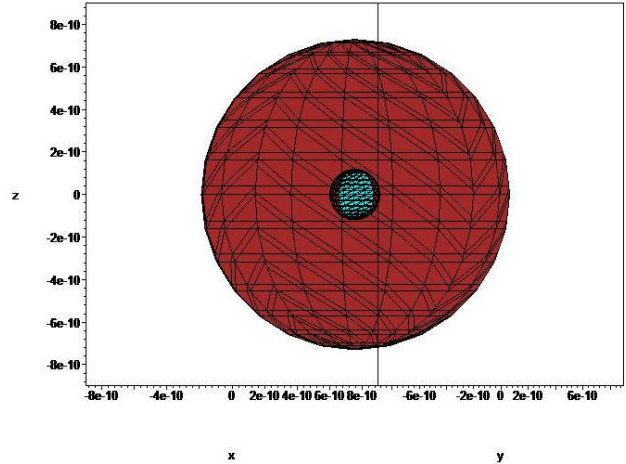

**FIGURE 7.** Surfaces of amplitude function $\psi_{1,0,0} = \pm3.17 \times 10^{12}$ m$^{-3/2}$; of three spheres, the inner one (cyan) contains the amplitude function in its positive phase; in a thick shell between an intermediate sphere (brown), of radius only slightly larger than that of the inner sphere, and the outer sphere (also brown), both shown cut open with $\phi = 0 .. 3\pi/2$ rad, the amplitude function has a negative phase corresponding to the profile in figure 6.

We consider next $\psi_{0,1,0}$, according to this formula:

$$\psi_{0,1,0} = \frac{\sqrt{2}\, e^5\, \mu^{\left(\frac{5}{2}\right)} Z^{\left(\frac{5}{2}\right)} \pi^2\, r\, \mathbf{e}^{\left(-\frac{\pi \mu Z e^2 r}{2 h^2 \varepsilon_0}\right)} \cos(\theta)}{8\, h^5\, \varepsilon_0^{\left(\frac{5}{2}\right)}}$$

Consistent with azimuthal quantum number $l = 1$, we expect one angular node in the domain $\theta = 0 .. \pi$, which is shown in figure 8 to lie in plane $z = 0$ or $\theta = \pi/2$. This figure exhibits two lobes, each almost hemispherical but with rounded edges: the positive lobe for $\psi_{0,1,0} > 0$ is axially symmetric about axis $z$ with $z > 0$; the negative lobe for $\psi_{0,1,0} < 0$ is also axially symmetric about axis $z$ but with $z < 0$; a narrow gap between those surfaces contains a nodal plane between these two lobes at which $\psi_{0,1,0} = 0$. At the selected value $|\psi_{0,1,0}| = 3.17 \times 10^{12}$ m$^{-3/2}$ for this surface, the same as for $\psi_{1,0,0}$, the maximum extent of the surface along axis $z$, about $1.5 \times 10^{-9}$ m, is slightly larger than the maximum extent perpendicular to this direction, about $1.2 \times 10^{-9}$ m, contrary to what one might expect from published plots of only the angular parts or of qualitative sketches based mostly on wishful thinking. The overall shape, nearly spherical or slightly prolate spheroidal, is consistent with a coulombic attraction between a proton and an electron that has no angular dependence. The square of this amplitude function has essentially the same *relative size* and *shape*.

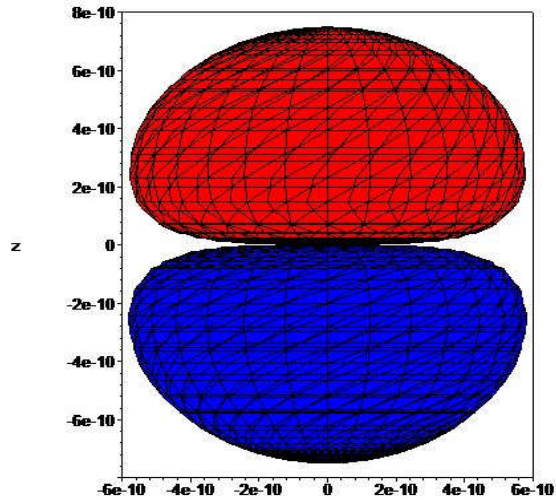

**FIGURE 8.** Surface of amplitude function ψ(0,1,0) = ±3.17x10$^{12}$ m$^{-3/2}$; the upper lobe (red) has a positive phase, the lower lobe (blue) negative.

Like most amplitude functions expressed in spherical polar coordinates, amplitude functions ψ$_{0,1,1}$ and ψ$_{0,1,-1}$ are complex: for a given value of energy quantum number *n*, explicitly $n^2 - n$ functions have both real and imaginary parts for the choice of coefficient *c* = 1. For this function,

$$\psi_{0,1,1} = \frac{e^5 \mu^{\left(\frac{5}{2}\right)} Z^{\left(\frac{5}{2}\right)} \pi^2 r \, \mathrm{e}^{\left(-\frac{\pi \mu Z e^2 r}{2 h^2 \varepsilon_0}\right)} \sin(\theta)(\cos(\phi) + i \sin(\phi))}{8 h^5 \varepsilon_0^{\left(\frac{5}{2}\right)}}$$

of which the real part contains the cosine function, its plot generates a figure identical with that in figure 8 except that it is axially symmetric about axis *y* instead of axis *z*; the plot of the imaginary part, containing the sine function, is, analogously, axially symmetric about axis *x*; i.e. each such surface is the same as that in figure 8 but rotated about axis *x* or *y*. Their positive lobes extend along the respective positive axes *y* and *x*, and their negative lobes along the same negative axes. A more intriguing aspect is the square of either ψ$_{0,1,1}$ or ψ$_{0,1,-1}$, which are identical according to this formula because the dependence on ϕ is lost in the square,

$$\psi_{0,1,1}^2 = \frac{1}{64} \frac{\mu^5 Z^5 e^{10} \pi^4 r^2 \, \mathrm{e}^{\left(-\frac{\pi \mu Z e^2 r}{h^2 \varepsilon_0}\right)} \sin(\theta)^2}{h^{10} \varepsilon_0^5}$$

and which, cut open to reveal the inner structure, is depicted accurately in figure 9; there is explicitly zero density of ψ$_{0,1,1}^2$ or electronic charge along polar axis *z* at which *r* sin(θ) = θ = 0 or π rad. Whereas both ψ$_{0,1,0}$ and its square display an overall shape of slightly prolate spheroid with a plane of zero amplitude or density between two nearly hemispherical lobes, the square of either $\psi_{0,1,1}$ or $\psi_{0,1,-1}$, calculated as $|\psi_{0,1,1}|^2 = \psi_{0,1,1}^* \psi_{0,1,1}$ or analogously, displays as an oblate toroid with zero density along polar axis $z$. The extent of this toroid at the given value of $\psi_{0,1,1}^2$ is about 1.15x$10^{-9}$ m along axis $z$ and 1.3x$10^{-9}$ m perpendicular to this axis. The surface of sum $\psi_{0,1,0}^2 + |\psi_{0,1,1}|^2 + |\psi_{0,1,-1}|^2 = 1.0\text{x}10^{25}$ m$^{-3}$ displays as a perfect sphere, of diameter 1.55x$10^9$ m with the same criterion of enclosed electronic charge.

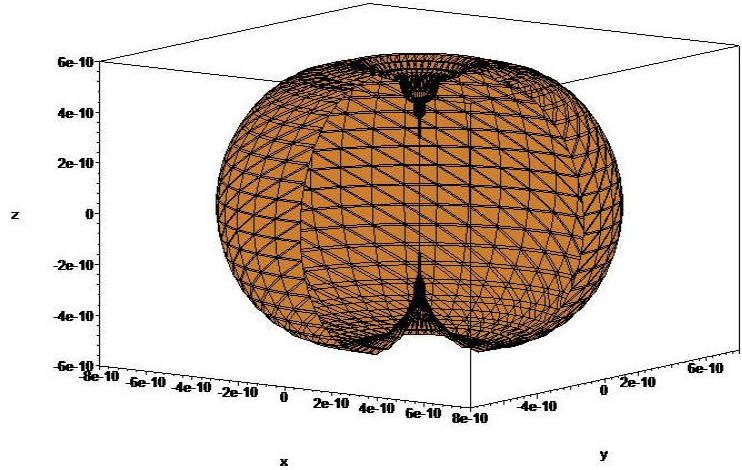

**FIGURE 9.** Surface of squared amplitude function $\psi_{0,1,1}^2 = 1.0\text{x}10^{25}$ m$^{-3}$, shown cut open with $\phi = 0 .. 3\pi/2$ rad to reveal the zero density along polar axis $z$.

A plot of amplitude function $\psi_{2,0,0}$ exhibits two nodal concentric spherical surfaces, consistent with $k = 2$, between one inner sphere of positive phase and two larger spherical shells, of successively negative and positive phases, analogously to $\psi_{1,0,0}$ in figure 6 that displays only one inner spherical nodal surface; with the surfaces set for $|\psi_{2,0,0}| = 1.23\text{x}10^{12}$ m$^{-3/2}$, the diameters/$10^{-10}$ m of the nodal surfaces are equal to about 2 and 8. According to our stated criterion of enclosed electronic charge being 0.995 of the total negative charge, the diameter of the outer sphere is 2.8x$10^{-9}$ m, increased from 0.49x$10^{-9}$ m for $\psi_{0,0,0}$ and 1.44x$10^{-9}$ m for $\psi_{1,0,0}$.

We show in figure 10 the surface of $\psi_{1,1,0}$ that has this formula,

$$\psi_{1,1,0} = -\frac{1}{81} \frac{\sqrt{2}\, e^5\, \mu^{(5/2)}\, Z^{(5/2)}\, \pi^2\, (\pi Z e^2 \mu r - 6 h^2 \varepsilon_0)\, r\, e^{\left(-1/3 \frac{\pi \mu Z e^2 r}{h^2 \varepsilon_0}\right)} \cos(\theta)}{\varepsilon_0^{(7/2)}\, h^7}$$

Like figure 8 that shows two nearly hemispherical lobes, figure 10 shows two large nearly hemispherical lobes, one of positive and one of negative phase, each with an embedded small nearly hemispherical lobe of opposite phase near the origin, all symmetric about axis $z$. For $\psi_{1,1,0} = \pm 1.23\text{x}10^{12}$ m$^{-3/2}$, the maximum extent of the surface along axis $z$ is about 2.8x$10^{-9}$ m, and 2.4x$10^{-9}$ m perpendicular to this direction, so slightly prolate spheroidal in overall shape. For $\psi_{1,1,1}$ or $\psi_{1,1,-1}$ that has both real and imaginary parts, those parts exhibit an identical shape and size axially symmetric about axes $y$ or $x$, respectively, analogous to the relation between $\psi_{0,1,0}$ depicted in figure 8 and $\psi_{0,1,1}$ or $\psi_{0,1,-1}$, but rotated about axis $y$ or $x$. The surface representing sum $\psi_{1,1,0}^2 + \psi_{1,1,1}^*$ $\psi_{1,1,1} + \psi_{1,1,-1}{}^{*}\, \psi_{1,1,-1} = 1.25 \times 10^{24}$ m$^{-3}$ is again a perfect sphere, of diameter about $3 \times 10^{-9}$ m, with no internal structure.

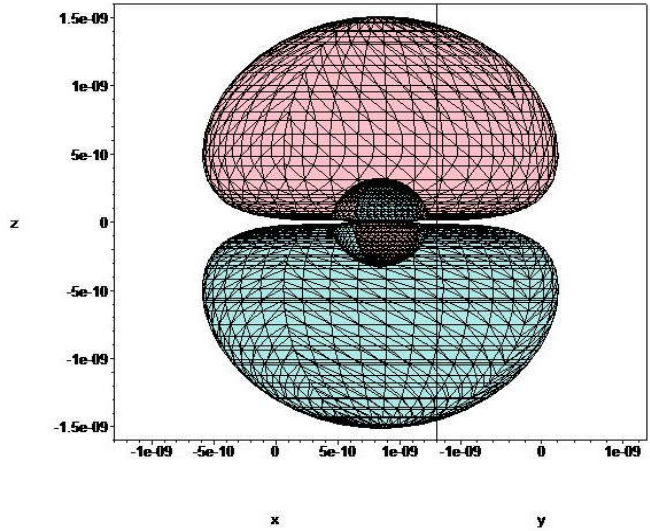

**FIGURE 10.** Surface of amplitude function $\psi_{1,1,0} = \pm 1.23 \times 10^{12}$ m$^{-3/2}$ shown cut open with $\phi = 0$ .. $3\pi/2$ rad to exhibit the inner structure; phases positive (pink) and negative (light blue) are indicated.

As an amplitude function that presents further features, we consider $\psi_{0,2,0}$ that implies $l = 2$, according to this formula:

$$\psi_{0,2,0} = \frac{1}{486} \frac{\sqrt{6}\, e^7\, \mu^{(7/2)}\, Z^{(7/2)}\, \pi^3\, r^2\, e^{\left(-1/3\, \frac{\pi \mu Z e^2 r}{h^2 \varepsilon_0}\right)} (3\cos(\theta)^2 - 1)}{\varepsilon_0^{(7/2)}\, h^7}$$

Its surface with $\psi_{0,2,0} = \pm 1.23 \times 10^{12}$ m$^{-3/2}$ appears in figure 11. In accordance with $k = 0$, there is no radial node – along a line in any direction from the origin, the amplitude function does not change sign, but from polar angle $\theta = 0$ to $\theta = \pi$ rad there are two changes of sign, from positive to negative and back to positive, consistent with $l = 2$. The shape might be described as a torus of positive phase separating two, somewhat conical, negative lobes. The figure is axially symmetric about axis $z$; its maximum extents/$10^{-9}$ m are 2.9 along polar axis $z$ and 2.6 perpendicular to that axis, so again slightly prolate spheroidal in overall shape. The square of this amplitude function has approximately the same *relative size* and *shape*.

For contrast we show surfaces of the real and imaginary parts of complex amplitude function $\psi_{0,2,2}$ expressed with $\phi$ in trigonometric form as

$$\psi_{0,2,2} = -\frac{1}{162} \frac{e^7\, \mu^{(7/2)}\, Z^{(7/2)}\, \pi^3\, r^2\, e^{\left(-\frac{\pi \mu Z e^2 r}{3 h^2 \varepsilon_0}\right)} \sin(\theta)^2\, (\cos(2\phi) + i\sin(2\phi))}{h^7\, \varepsilon_0^{(7/2)}}$$

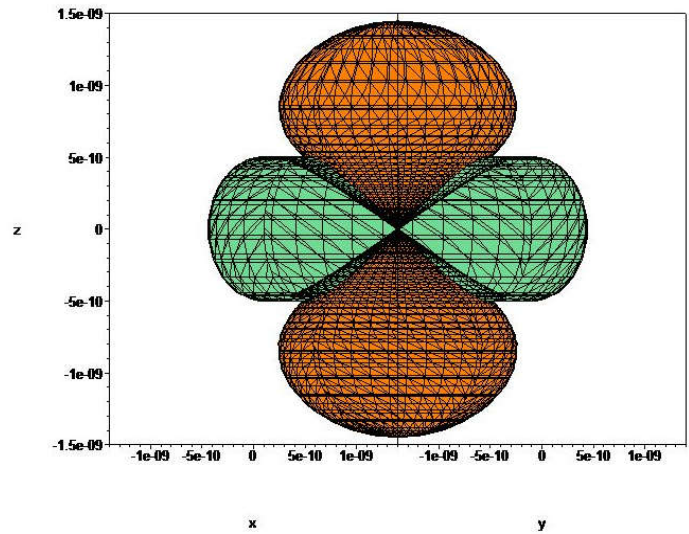

**FIGURE 11.** Surface of amplitude function $\psi_{0,2,0} = \pm 1.23 \times 10^{12}$ m$^{-3/2}$, shown cut open with $\phi = 0$ .. $3\pi/2$ rad to reveal the angular nodes and lack of radial node; coral colour indicates a positive phase, aquamarine negative.

The real part of the surface of this amplitude function $\psi_{0,2,2}$, containing $\cos(2\phi)$ and shown in figure 12, and the imaginary part, containing $\sin(2\phi)$ and shown in figure 13, reveal a four-fold symmetry about axis $z$, apart from their phases.

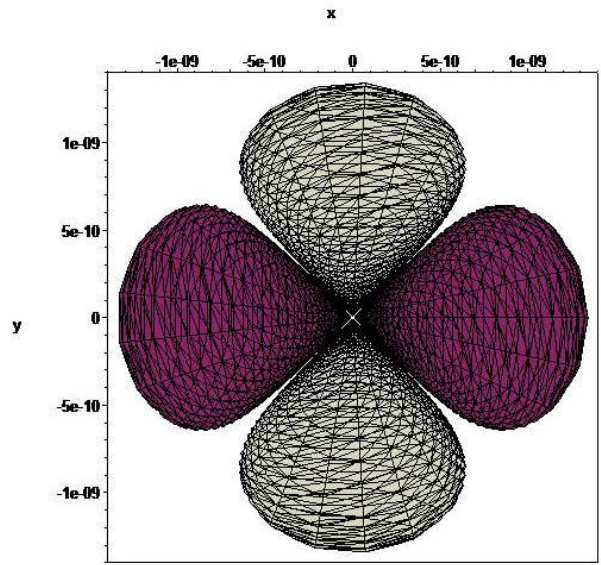

**FIGURE 12.** Surface of real part of amplitude function $\psi_{0,2,2} = \pm 1.23 \times 10^{12}$ m$^{-3/2}$; the wheat lobes have positive phase, the maroon lobes negative.

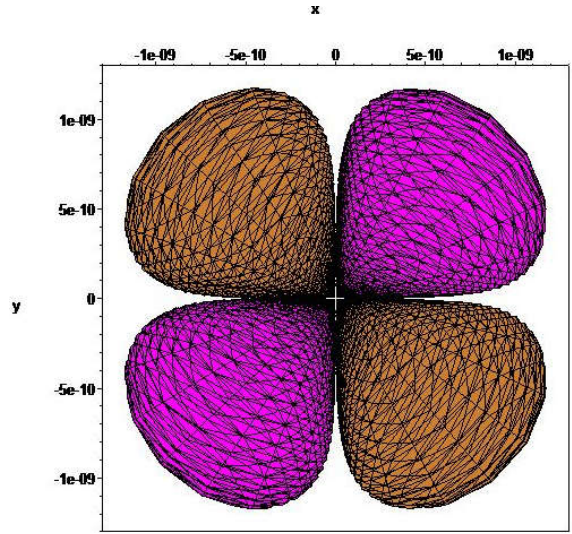

**FIGURE 13.** Surface of imaginary part of amplitude function ψ(0,2,2) = ±1.23x10$^{12}$ m$^{-3/2}$; the golden lobes have positive phase, the magenta lobes negative.

The maximum extent of the displayed surface of the real part of ψ$_{0,2,2}$= ±1.23x10$^{12}$ m$^{-3/2}$ along axes $x$ and $y$ is 2.7x10$^{-9}$ m, but the thickness of the body, parallel to axis $z$, is only about 1.8x10$^{-9}$ m; the imaginary part has similar dimensions but its orientation is rotated about polar axis $z$ by $\pi/8$ rad. The surface of the square of ψ$_{0,2,2}$, calculated as ψ$_{0,2,2}$* ψ$_{0,2,2}$, resembles a torus with a narrow core along axis $z$ that is empty of electronic charge, similar to that in figure 9. The surfaces of ψ$_{0,2,-2}$ = ±1.23x10$^{12}$ m$^{-3/2}$ in their real and imaginary parts are practically identical to those of the corresponding parts of ψ$_{0,2,2}$= ±1.23x10$^{12}$ m$^{-3/2}$. The surfaces of ψ$_{0,2,1}$ and ψ$_{0,2,-1}$ are analogous to those of the imaginary parts of ψ$_{0,2,2}$ and ψ$_{0,2,-2}$ except that they display a nearly four-fold symmetry about axes $x$ and $y$, respectively, rather than about polar axis $z$; their lobes extend only between the cartesian axes, as in figure 13, rather than along axes as in figure 12. The surface of the square of ψ$_{0,2,1}$, calculated as ψ$_{0,2,1}$* ψ$_{0,2,1}$ = 1.5x10$^{24}$ m$^{-3}$, is shown in figure 14; it exhibits a narrow core of zero electronic charge density along polar axis $z$ and a nodal plane at $z = 0$ at which the electronic charge density is also zero. Its maximum extent parallel to axis $z$ is 2.1x10$^{-9}$ m and perpendicular to polar axis $z$ is 2.3x10$^{-9}$ m, making its overall shape approximately slightly oblate spheroidal.

According to the same criterion of enclosed electronic charge, the surface of sum

$$|\psi_{0,2,0}|^2 + |\psi_{0,2,1}|^2 + |\psi_{0,2,-1}|^2 + |\psi_{0,2,2}|^2 + |\psi_{0,2,-2}|^2 = 1.5 \times 10^{24} \text{ m}^{-3}$$

of the squared amplitude functions for $k = 0$ and $l = 2$ is a perfect sphere of diameter 2.9x10$^{-9}$ m.

As examples of the general features of surfaces of spherical polar amplitude functions with further values of azimuthal quantum number $l$, we show in figure 15 the surface of ψ$_{0,3,0}$,

$$\psi_{0,3,0} = \frac{1}{15360} \frac{\sqrt{5}\, e^9\, \mu^{(9/2)}\, Z^{(9/2)}\, \pi^4\, r^3\, \mathrm{e}^{\left(-1/4 \frac{\pi \mu Z e^2 r}{h^2 \varepsilon_0}\right)} \cos(\theta)\, (5\cos(\theta)^2 - 3)}{h^9\, \varepsilon_0^{(9/2)}}$$

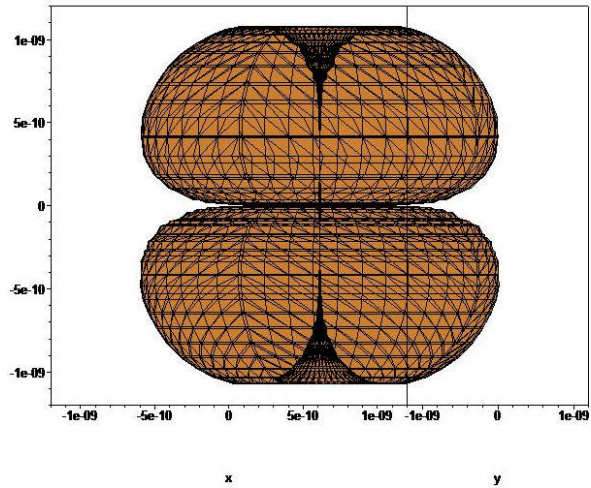

**FIGURE 14**. Surface of $|\psi_{0,2,1}|^2 = \psi_{0,2,1}^* \, \psi_{0,2,1} = 1.5 \times 10^{24}$ m$^{-3}$, shown cut open with $\phi = 0 \; .. \; 3\pi/2$ rad to expose zero density along polar axis $z$ and in plane $z = 0$.

which exhibits the corresponding numbers of angular nodes, $l = 3$, with two tori of roughly conical cross section separating two roughly conical lobes along the polar axis, but no radial node.

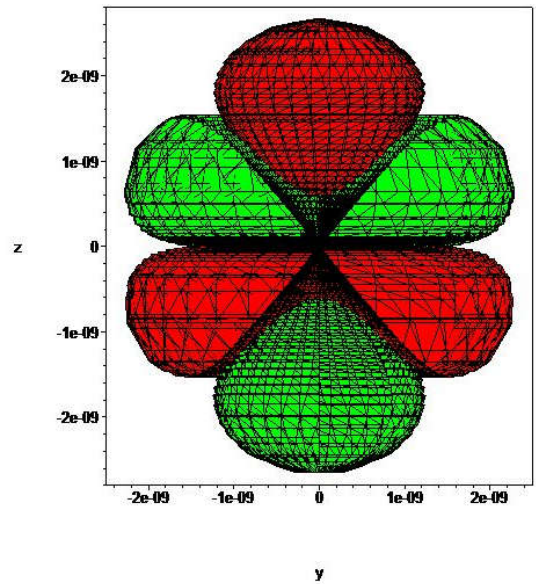

**FIGURE 15.** Surface of spherical polar amplitude function $\psi_{0,3,0} = \pm 1.85 \times 10^{11}$ m$^{-3/2}$, shown cut open with $x = -2.5 \times 10^{-9} ... 0.5 \times 10^{-9}$ m to expose the nodal surfaces; the red lobes have positive phase, the green lobes negative.

Figure 16 displays a surface of $\psi_{0,4,0}$,

$$\psi_{0,4,0} = \frac{1}{13125000} \frac{\sqrt{70} \, e^{11} \, \mu^{(11/2)} \, Z^{(11/2)} \, \pi^5 \, r^4 \, e^{\left(-1/5 \, \frac{\pi \mu Z e^2 r}{h^2 \varepsilon_0}\right)} (35 \cos(\theta)^4 - 30 \cos(\theta)^2 + 3)}{\varepsilon_0^{(11/2)} \, h^{11}}$$

with *l* = 4, which exhibits three tori of roughly conical cross section separating two lobes of roughly conical shape.

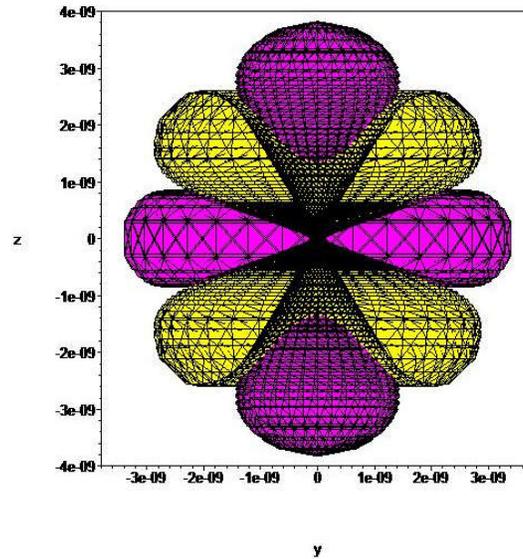

**FIGURE 16.** Surface of spherical polar amplitude function $\psi_{0,4,0}$ = ±1.1x10$^{11}$ m$^{-3/2}$, shown cut open with $x$ = −2.8x10$^{-9}$... 0.8x10$^{-9}$ m to expose the nodal surfaces; the violet lobes have positive phase, the yellow lobes negative.

Many other plots of surfaces of amplitude functions or their squares might be presented, but the examples provided above likely demonstrate both the qualitative and the quantitative features of these functions in spherical polar coordinates as solutions of Schroedinger's temporally independent equation that is applicable to states of the hydrogen atom with discrete energies.

## IV.    DISCUSSION

Although the treatment of the hydrogen atom, or a hydrogenic atom with one electron or the Kepler problem or the central-field problem as alternative descriptions, with Schroedinger's equation temporally dependent or independent in spherical polar coordinates is much discussed elsewhere, a principal objective here is to emphasize the particular quantitative aspects of this system of coordinates for comparison with the results in other systems in succeeding parts of this series. A treatment of this system of spatial coordinates comprising $r$, $\theta$ and $\phi$ yields results in the form of amplitude functions involving quantum numbers $k$, $l$ and $m$; we summarize this condition according to notation $\psi_{k,l,m}(r,\theta,\phi)$ for the amplitude function. Including the temporal variable to specify the wave function as $\Psi_{k,l,m}(r,\theta,\phi,t)$ produces no advantage because the temporal part of the solution of Schroedinger's equation is, naturally, common to all pertinent systems of coordinates in a non-relativistic sense; energy quantum number $n$ results from an analysis of experiments in the form of optical spectra, and the form of its dependence is duly reproduced in the theoretical derivation incorporating a temporal dependence as summarized above for spherical polar coordinates when we associate $k + l + 1$ with $n$.

With each plot presented here is specified an explicit formula of the particular associated amplitude function to emphasize that these plots depict mathematical functions; for the same reason, these functions include explicitly the pertinent fundamental physical constants, evaluated for all purposes rigorously in SI units. *A shape of a surface illustrated in any such plot implies an* *associated formula, and vice versa*; such a shape is meaningless in the absence of both such a formula and the specified system of coordinates, and even a specific constant value of that formula. An accurate surface of each such amplitude function displayed within a preceding plot hence pertains to a particular value of that function $\psi_{k,l,m}$ or its square; such an amplitude function has an accompanying unit in terms of a unit of length to the appropriate power, specifically m$^{-3/2}$ in SI units for $\psi$, and accordingly m$^{-3}$ for $\psi^2$; integration over all space effectively multiplies the units of the squared amplitude function by m$^3$, generating a dimensionless value of that integral that we might associate with unit probability according to Born's interpretation. The shape of the surface depends also on the chosen value of $\psi$ or $\psi^2$ for that surface to some extent; the criterion for the value of $\psi$ might be chosen to differ from that of $\psi^2$ containing 0.995 of the total electronic charge, which would accordingly affect the shape, and size, of the surface.

The important results of the preceding treatment are that spatial coordinates $r$, $\theta$, $\phi$ in this spherical polar system lead irrevocably, on solution of Schroedinger's equations, to formulae expressed in terms of quantum numbers $k$, $l$, $m$. Unlike energy quantum number $n$ that is independent of any system of coordinates and that arises indisputably from experiment, quantum numbers $k$ and $l$ have only a parochial significance: they are artifacts of this particular system and can be accordingly expected to have no meaning for the formulae and shapes of surfaces of amplitude functions apart from this system. Whereas equatorial quantum number $m$ is directly related to the angular-momentum properties of the atom in states that are independent of coordinates in a chosen system, $l$ is only indirectly related, because the squared total angular momentum $L^2$, hence a scalar quantity with no directional property, is equal to $l(l+1)h^2/4\pi^2$. For each value of $k + l + 1$, there are $2l + 1$ values of $m$; the values of $l$ run from 0 to $n - 1 = k + l$. There are hence quantum numbers in

$$\sum_{l=0}^{n-1} (2l+1) = n^2$$

sets that specify a particular amplitude function for each value of $n$; among these orthogonal functions, for coefficient $c = 1$, $n^2 - n$ functions are complex, so having real and imaginary parts; apart from any other consideration, the fact of these imaginary parts means that these functions have no direct physical reality. There are also $n^2$ spectrometric states for each value of $n$. We reiterate that there is, in general, no direct relation between a spectrometric state, as denoted below, and a particular amplitude function $\psi_{k,l,m}$ corresponding to quantum numbers $k$, $l$, $m$ in a specific set; the ground state is an exception to this condition. The importance of this reasoning becomes incontestable when one proceeds to contemplate solutions of Schroedinger's equation temporally independent with coordinates in other systems and their associated quantum numbers, as discussed in succeeding parts of articles in this series. The designation of a spectrometric state of the hydrogen atom is based conventionally on a value of azimuthal quantum number $l$ – S states for $l = 0$, P states for $l = 1$, D states for $l = 2$ et cetera, originating in the terminology of Liveing and Dewar for <u>s</u>harp, <u>p</u>rincipal and <u>d</u>iffuse series of lines in atomic spectra; the quantum number for energy might be included in such a designation as $n\,l$, and all states are *doublet* states, indicated as a prefixed superscript, when the purported intrinsic angular momentum of the electron is taken also into account, to yield a term symbol such as 1 $^2$S, 2 $^2$S, 2 $^2$P et cetera. Such an intrinsic angular momentum is beyond the purview of Schroedinger's treatment. The line marked α in the discrete absorption spectrum in figure 3, of least energy of the photon, hence represents a transition that might be denoted 2 $^2$P ← 1 $^2$S; all such spectral lines have fine structure, not indicated in figure 3

J. F. OGILVIE

because its calculation is beyond the scope of pioneer quantum mechanics and is related to the coupling of intrinsic electronic and nuclear angular momenta.

The Zeeman effect, whereby spectral lines split and have their frequencies or wave lengths altered when a sample of atoms is subjected to an external magnetic field, is related to component $L_z$ of electronic angular momentum due to its motion, which yields equatorial quantum number $m$. Although a treatment of this Zeeman effect is practicable with amplitude functions in spherical polar coordinates, to produce the *normal* Zeeman effect, taking into consideration also the intrinsic angular momentum of the electron -- its purported spin -- produces the *anomalous* Zeeman effect. Because such an electron spin is beyond the scope of the Schroedinger equation as presented above, we omit this discussion.

The *shapes* of surfaces of these amplitude functions $\psi_{k,l,m}(r,\theta,\phi)$ are likewise parochial; they are hence artifacts of this particular system, and accordingly lack meaning beyond this system, as is demonstrated on comparison with shapes of surfaces of amplitude functions in other systems of coordinates, presented in further articles in this series. The *sizes* of the surfaces of these amplitude functions at the selected values according to a particular criterion of enclosed electronic charge are, in contrast, a result of the coulombic potential energy, and are seen to be approximately common to amplitude functions, expressed in various coordinates, corresponding to the same energy quantum number $n$. In the preceding figures, the shapes of the nodal surfaces that lie between lobes of opposite phase are spheres centred at the origin, or cones centred about the polar axis, or planes containing the cartesian axes, reflecting the nature of the system of spherical polar coordinates $r$, $\theta$, $\phi$, respectively.

Apart from any fine structure that results from the purported electronic spin, the energy of a discrete state of a H atom is commonly stated to depend only on $n$; as $n$ must be regarded as the energy quantum number, such a statement appears tautological. For only a hydrogenic atom, as defined above, is the energy synonymous with an energy quantum number. As $n$ represents, for only spherical polar (and spheroconical) coordinates, a sum $k + l + 1$, the energy of a discrete state hence depends equivalently on quantum numbers $k$ and $l$, but not on $m$ in the absence of an external field applied to the atom. Whereas $l$ has an indirect connexion with angular momentum, as explained above, an interpretation of $k$ other than as signifying the number of radial nodes is challenging. For instance, the mean distance of the electron from the origin near the atomic nucleus for atomic number $Z$, or expectation value <$r$>, as a function of quantum numbers is, expressed in terms of Bohr radius $a_0$,

$$<r> = \frac{a_0}{Z} \left( \frac{3}{2} k^2 + 3\,k\,l + l^2 + 3\,k + \frac{5}{2} l + \frac{3}{2} \right),$$

which depends, to comparable extents, on both $k$ and $l$, which can vary independently. In the absence of an externally applied electric or magnetic field, the independence of the energy on equatorial quantum number $m$ is attributed to the isotropic nature of space: there is no preferred axis or direction.

One might seek to overcome the complex nature of amplitude functions on forming linear combinations of two or more functions with the same values of quantum numbers $k$ and $l$, but different $m$, such as in the following examples. This sum of two particular amplitude functions,

$$\frac{\psi_{0,1,1} + \psi_{0,1,-1}}{\sqrt{2}} = \frac{\sqrt{2}\, e^5\, Z^{\left(\frac{5}{2}\right)} \mu^{\left(\frac{5}{2}\right)} \pi^2\, r\, e^{\left(-\frac{Z e^2 \pi \mu r}{2 h^2 \varepsilon_0}\right)} \sin(\theta) \cos(\phi)}{8\, \varepsilon_0^{\left(\frac{5}{2}\right)} h^5}$$

produces a plot exactly like that in figure 8 except that the axis of cylindrical symmetry is *x* instead of *z*. The difference of the same two particular amplitude functions, divided by **i**,

$$\frac{\psi_{0,1,1} - \psi_{0,1,-1}}{\sqrt{2}\, i} = \frac{\sqrt{2}\, e^5\, Z^{\left(\frac{5}{2}\right)} \mu^{\left(\frac{5}{2}\right)} \pi^2\, r\, e^{\left(-\frac{Z e^2 \pi \mu r}{2 h^2 \varepsilon_0}\right)} \sin(\theta) \sin(\phi)}{8\, \varepsilon_0^{\left(\frac{5}{2}\right)} h^5}$$

likewise produces a plot exactly like that in figure 8 except that the axis of cylindrical symmetry is *y* instead of *z*, but both such plots depend on an arbitrary choice of coefficient *c* = 1. The value of *l* for the sum and difference appears to remain unity, but the value of equatorial quantum number *m* is indeterminate – until one recognizes that those results simply correspond to a rotation of the coordinate axes [13] to redirect $\psi_{0,1,0}$. Such a rotation hence implies again no direct physical reality.

    Another notable aspect of the figures showing surfaces of constant amplitude is the dominance of the polar axis, to which polar angle θ is referred. Although an amplitude function such as $\psi_{0,1,1}(r,\theta,\phi)$ is not axially symmetric with respect to this polar axis, its square acquires that property. Despite this apparent special spatial feature, in the absence of an externally applied electric or magnetic field a hydrogen atom is *spherically symmetric* – there is no preferred axis of symmetry, as mentioned above. The solution of Schroedinger's equation, temporally dependent or independent, in spherical polar coordinates is valid only under conditions of rigorously spherical symmetry -- no other matter in the system, no externally applied electric field. A related aspect is the small extent of a deviation from spherical symmetry demonstrated by the only slightly prolate or oblate spheroidal overall shapes of the surfaces of amplitude functions and their squares. A major distinction between the surfaces of amplitude functions $\psi_{0,0,0}$ and $\psi_{1,0,0}$ in figures 5 and 7 or between $\psi_{0,1,0}$ and $\psi_{1,1,0}$ in figures 8 and 10 is the appearance of an inner sphere, in the functions with *l* = 0, or two hemispheres, for *l* = 1. This feature is common to all further amplitude functions in spherical polar coordinates in which radial quantum number *k* increases by one unit between two instances; when *k* increases by two units, the distinction amounts to two further inner spheres or their parts or two further radial or spherical nodal surfaces. This property is a direct result of the fundamental significance of this quantum number *k*: it specifies the number of radial nodes. Likewise, as is evident from figures 5, 8, 11, 15 and 16, quantum number *l* specifies the number of angular nodes. As amplitude functions for *l* > 4 have no practical application in a chemical or physical context, further plots of surfaces to show the shapes are of negligible interest.

    A claim [14] to have obtained an additional solution of Schroedinger's equation in rectangular or cartesian coordinates is misleading because those four coordinates (*x,y,z,r*) include also the radial distance *r*. The other coordinates provide, as ratios of cartesian coordinates and their combinations, an alternative to spherical harmonics, and allow eigenfunctions of angular momentum to avoid reference to polar angles. This system of coordinates must be considered to be merely a variant of spherical polar coordinates.

In a subsequent part of this sequence of articles, we form an overview of all solutions of Schroedinger's temporally dependent equation in the four systems of coordinates, the quantum numbers in their corresponding sets and the related properties.

# THE HYDROGEN ATOM ACCORDING TO WAVE MECHANICS – II. PARABOLOIDAL COORDINATES

*J. F. Ogilvie**

Centre for Experimental and Constructive Mathematics, Department of Mathematics, Simon Fraser University, Burnaby, British Columbia V5A 1S6 Canada

Escuela de Química, Universidad de Costa Rica, Ciudad Universitaria Rodrigo Facio, San Pedro de Montes de Oca, San José, 11501-2050 Costa Rica

**Abstract**

In the second of five parts in a series, the Schroedinger equation is solved in paraboloidal coordinates to yield amplitude functions that enable accurate plots of their surfaces to illustrate the variation of shapes and sizes with quantum numbers $n_1$, $n_2$, $m$, for comparison with the corresponding plots of amplitude functions in coordinates of other systems. A useful property of these functions in paraboloidal coordinates is their application to treat the Stark effect, when a hydrogen atom is placed in an isotropic electric field.

**Resumen**

En el segundo de cinco artículos de esta serie, la ecuación de Schrödinger se resuelve en coordenadas parabólicos para producir funciones de amplitud que permiten gráficos exactos de superficies, para ilustrar la variación de formas y tamaños con números cuánticos $n_1$, $n_2$ y $m$, y comparar con los gráficos correspondientes de funciones de amplitud en coordenadas de otros sistemas. Una propiedad útil de estas funciones en coordenadas paraboloides es su aplicación para tratar el efecto Stark, cuando un átomo de hidrógeno se ubica en un campo eléctrico isotrópico.

**Key words:** hydrogen atom, wave mechanics, paraboloidal coordinates, orbitals, atomic spectra

**Palabras clave**: átomo de hidrógeno, mecánica cuántica, coordenadas parabólicas, orbitales, espectro atómico.

## I. INTRODUCTION

In Schroedinger's third article of four in a series of title *Quantisation as a Problem of Proper Values* [*1,2,3,4*] with which he introduced wave mechanics, he applied his differential equation to the solution of the hydrogen atom in circular paraboloidal coordinates, and specified a method to calculate the intensities of spectral lines [*3*]. As the type of central field of force in the hydrogen atom is coulombic, the variables in the partial-differential equation are separable in paraboloidal coordinates, to yield three ordinary-differential equations, one for each spatial variable in the definition of a space of three dimensions. In this part II of a series of articles devoted to the hydrogen atom with its coordinates of the governing partial-differential equation separable in four systems, we state the temporally independent partial-differential equation and its solution in paraboloidal coordinates, and provide plots of selected amplitude functions as surfaces corresponding to a chosen value of amplitude. As the dependence on time occurs in the same manner in all systems of coordinates in which the Schroedinger equation is separable, we accept the results from part I [5], and avoid that repetition.  Although the equations governing the form of the amplitude functions are here, of necessity, defined in coordinates according to a paraboloidal system, we view the surfaces of these amplitude functions invariably in rectangular cartesian coordinates: a computer procedure (in *Maple*) translates effectively from the original system of coordinates in which the algebra and calculus are performed to the system to which the human eye is accustomed.

---

* Corresponding author: ogilvie@cecm.sfu.ca

## II.    SCHROEDINGER'S EQUATION IN PARABOLOIDAL COORDINATES

We relate these mutually orthogonal circular paraboloidal coordinates $u$, $v$, $\phi$ to cartesian $x$, $y$, $z$, and spherical polar $r$, $\theta$, $\phi$ coordinates as algebraic formulae in both direct and inverse relations according to an established convention [6].

$$x = u\, v \cos(\phi), \quad y = u\, v \sin(\phi), \quad z = \tfrac{1}{2}(u^2 - v^2), \quad r = \tfrac{1}{2}(u^2 + v^2)$$
$$u^2 = r + z = r\,(1 + \cos(\theta)), \quad v^2 = r - z = r\,(1 - \cos(\theta)), \quad \phi = \arctan(y/x)$$

Surfaces of $u$, $v$ and $\phi$ as constant quantities are exhibited in figure 1. Although this system of coordinates might be described elsewhere as parabolic, the surfaces of two defining coordinates in three dimensions are clearly paraboloids, or parabolas of revolution, that have circular cross sections, which thus dictate the most informative name of the system. The surfaces of constant $u$ describe confocal paraboloids about the polar axis, $z$ in cartesian coordinates, that open in the direction of negative $z$ or $\theta = \pi$ rad and have a focus at the origin; the surfaces of constant $v$ analogously describe confocal paraboloids that open in the direction of positive $z$, or $\theta = 0$, and have also a focus at the origin.  The limiting cases of $u$ and $v$ tend to a line along axis $z$ as $u \to 0$ or $v \to 0$, with $z < 0$ and $z > 0$, respectively, and to a plane perpendicular to axis $z$ as $u$ or $v$ becomes large with $z \gg 0$ or $z \ll 0$, respectively.  The surfaces of constant equatorial angle $\phi$ have the same property as those in spherical polar coordinates – half-planes extending from the polar axis. With appropriate values of $u$, $v$ and $\phi$, a point can clearly locate anywhere in the coordinate space.  For use within the volume element in subsequent integrals, the jacobian of the transformation between cartesian and paraboloidal coordinates, as defined above, is $u\, v\, (u^2 + v^2)$.

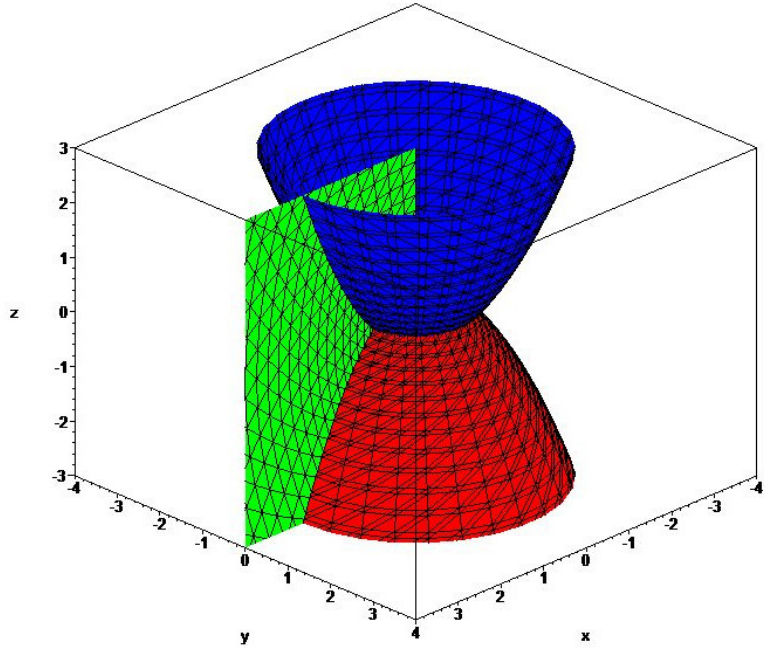

**FIGURE 1.** Definition of paraboloidal coordinates $u$, $v$, $\phi$: a paraboloid opening along negative $z$ (red) has $u = 1$ unit and a focus at the origin; another paraboloid (blue), opening along positive $z$, has $v = 1$ unit and a focus at the origin; a half-plane has equatorial angle $\phi = \pi$ rad with respect to plane $xz$.

A separation of the coordinates of the centre of mass of the H atom produces reduced mass $\mu$ of the system that is distant $r$ from the origin, to supplant distance $r$ between the electron and the atomic nucleus. Schroedinger's temporally independent equation in explicit SI units then contains within terms on the left side of the equality an electrostatic potential energy and first and second partial derivatives of an assumed amplitude function $\psi(u, v, \phi)$ with respect to spatial coordinates $u$, $v$, $\phi$ within an hamiltonian operator $H$ to take into account the kinetic and potential energies of the system; the right side of the equality comprises a product of energy $E$, as a variable parameter that has no dependence on coordinates, with the same amplitude function. The resultant form, as $H(u, v, \phi) \, \psi(u, v, \phi) = E \, \psi(u, v, \phi)$, resembles an eigenvalue relation.

$$-\frac{1}{8}\frac{h^2 \left(\frac{\partial}{\partial u}\psi(u,v,\phi)\right)}{\pi^2 \mu (u^2+v^2) u} - \frac{1}{8}\frac{h^2 \left(\frac{\partial^2}{\partial u^2}\psi(u,v,\phi)\right)}{\pi^2 \mu (u^2+v^2)} - \frac{1}{8}\frac{h^2 \left(\frac{\partial}{\partial v}\psi(u,v,\phi)\right)}{\pi^2 \mu (u^2+v^2) v}$$
$$-\frac{1}{8}\frac{h^2 \left(\frac{\partial^2}{\partial v^2}\psi(u,v,\phi)\right)}{\pi^2 \mu (u^2+v^2)} - \frac{1}{8}\frac{h^2 \left(\frac{\partial^2}{\partial \phi^2}\psi(u,v,\phi)\right)}{\pi^2 \mu (u^2+v^2) v^2} - \frac{1}{8}\frac{h^2 \left(\frac{\partial^2}{\partial \phi^2}\psi(u,v,\phi)\right)}{\pi^2 \mu (u^2+v^2) u^2}$$
$$-\frac{1}{4}\frac{Z e^2 \psi(u,v,\phi)}{\pi \varepsilon_0 \left(\frac{u^2}{2} + \frac{v^2}{2}\right)} = E \, \psi(u,v,\phi)$$

Apart from fundamental physical constants electric permittivity of free space $\varepsilon_0$, Planck constant $h$ and protonic charge $e$, there appear parameters $Z$ for atomic number – $Z = 1$ for H – and $\mu$ for the reduced mass of the atomic system, practically equal to the electronic rest mass $m_e$. After separation of the variables and solution of the three consequent ordinary-differential equations including definition of the separation parameters or integration constants, the full solution of the above equation has exactly this formula [7].

$$\psi(u, v, \phi) = c\,(-1)^{|m|} \sqrt{\frac{Z\pi\mu e^2}{\varepsilon_0 h^2}} \sqrt{\frac{2\,n_1!\,n_2!}{(n_1+|m|)!\,(n_2+|m|)!}}$$

$$\left(\frac{\pi Z e^2 \mu}{h^2 \varepsilon_0\,(|m|+n_1+n_2+1)}\right)^{(1+|m|)} (u v)^{|m|} e^{\left(-\frac{\pi Z e^2 \mu (u^2+v^2)}{2 h^2 \varepsilon_0\,(|m|+n_1+n_2+1)}\right)} e^{(i m \phi)}$$

$$\mathrm{LaguerreL}\left(n_1, |m|, \frac{\pi Z e^2 \mu u^2}{h^2 \varepsilon_0\,(|m|+n_1+n_2+1)}\right)$$

$$\mathrm{LaguerreL}\left(n_2, |m|, \frac{\pi Z e^2 \mu v^2}{h^2 \varepsilon_0\,(|m|+n_1+n_2+1)}\right) \Big/ (\sqrt{2\pi}\,(|m|+n_1+n_2+1))$$

This formula is accurately normalized such that

$$\int \psi(u, v, \phi)^* \,\psi(u, v, \phi)\,\mathrm{d}vol = 1,$$

in which d*vol* is a volume element containing the jacobian specified above; the implied triple integration is over all space; an asterisk as raised suffix, so ψ∗ of an amplitude function implies a complex conjugate of ψ such that, wherever $\mathbf{i} = \sqrt{-1}$ appears in ψ, $-\mathbf{i}$ appears in ψ*. A normalizing factor stated elsewhere [8] is incorrect. The presence of **i** in an exponential factor as product with φ signifies that this formula is complex, thus containing real and imaginary parts. Coefficient *c* that equals any complex number of modulus unity such as a fourth root of unity – i.e. ±1, ±√−1, occurs because Schroedinger's equation is a linear homogeneous partial-differential equation, or equally because the temporally independent Schroedinger equation has the form of an eigenvalue relation, as shown above. The conventional choice *c* = 1, which is arbitrary and lacks physical justification, signifies that some solutions ψ(*u*,*v*,φ) as amplitude functions from the temporally independent Schroedinger equation appear in a purely real form, whereas most are complex; with a mathematically valid alternative choice *c* = i, some amplitude functions would be entirely imaginary, but most would still be complex and thus alien to physical space. Choosing instead *c* = −1 or −**i** merely reverses the phase of an amplitude function or its constituent parts. Parameters that appear in the solution but not in the partial-differential equation take discrete values, imposed by boundary conditions, as follows: *m* is called the equatorial, or magnetic, quantum number that assumes only integer values and that arises in the solution of the angular equation to define Φ(φ), as in spherical polar coordinates; the first arguments of the associated Laguerre functions, $n_1$ and $n_2$, like radial quantum number *k* among the three quantum numbers pertaining to spherical polar coordinates, must be non-negative integers so that for bound states of the hydrogen atom the Laguerre functions in U(*u*) and V(*v*) terminate at finite powers of variable *u* or *v*, and remain finite for *u* or *v* taking large values, respectively. The sum $n_1 + n_2$ of paraboloidal quantum numbers plays a role similar to that of radial quantum number *k* among the quantum numbers for spherical polar coordinates [8]; the difference $n_1 - n_2$, or its reverse, might be called an *electric quantum number* [8], because the energy of the linear Stark effect, whereby the H atom interacts with an external electric field, depends on that difference; *vide infra*.

Whereas the solution of the temporally independent Schroedinger equation in spherical polar coordinates comprises a product of one function of distance R($r$), involving radial variable $r$, and two angular functions $\Theta(\theta)$ and $\Phi(\phi)$ of which the product Y($\theta,\phi$) = $\Theta(\theta)\,\Phi(\phi)$ constitutes spherical harmonics involving angles polar $\theta$ and equatorial $\phi$, the analogous solution in paraboloidal coordinates comprises a product of two functions of distance variables, U($u$) and V($v$), and one and the same equatorial angular function $\Phi(\phi)$:

$$\psi(u,v,\phi) = U(u)\,V(v)\,\Phi(\phi)$$

each variable $u$ and $v$ has physical dimension of square root of length, so SI unit m$^{1/2}$; $\psi(u,v,\phi)$ has a physical dimension consistent with SI unit m$^{-3/2}$. Since Schroedinger himself [3], both functions U($u$) and V($v$) are expressed traditionally in terms of Laguerre polynomials for the discrete states, although Kummer and Whittaker functions serve the purpose just as satisfactorily. Just as a Laguerre polynomial in R($r$) in spherical polar coordinates contains a sum of quantum numbers that occurs also in the exponent of the temporal factor, so both U($u$) and V($v$) contain, in the third arguments of their Laguerre functions, a sum $n_1 + n_2 + |m| + 1$; that sum that must take values of a positive integer likewise occurs in the temporal factor, omitted above; we associate that sum with $n$, an integer quantum number for energy that was originally defined from experiment. For a particular value of energy quantum number $n$ and magnetic quantum number $m = 0$, quantum numbers $n_1$ and $n_2$ can be chosen in $n$ distinct ways; for $|m| > 0$, there are two ways of choosing $m$ as $\pm|m|$, which yields a total degeneracy $n^2$ of amplitude functions, or the corresponding sets of quantum numbers, for a particular value of $n$ and hence the energy associated with a particular amplitude function $\psi(u,v,\phi)$. With coefficient $c = 1$, of $n^2$ amplitude functions for a given value of $n$, $n$ functions are real and $n^2 - n$ are complex, hence containing both real and imaginary parts that defy direct plots in less than six spatial dimensions. Whereas in spherical polar coordinates the energy of a H atom not subject to an external electric or magnetic field is formally independent of $m$, in paraboloidal coordinates under the same conditions the energy depends directly on its absolute value, $|m|$, in combination with quantum numbers $n_1$ and $n_2$, according to the formula above; as $|m|$ is simply equivalent to a lower limit of $l$ in spherical polar coordinates, the same sense of dependence remains in paraboloidal coordinates.

Regarding the frequencies and intensities of spectral lines as the principal observable properties of an atom, as Heisenberg recognised, independent of parochial quantum numbers $n_1$, $n_2$ and $m$ in the case of paraboloidal coordinates, the frequency of a spectral line depends on only the difference of energies of spectrometric states. The energy of each state depends on the inverse square of energy quantum number $n = n_1 + n_2 + |m| + 1$. The intensities are just as readily calculated with paraboloidal amplitude functions [7] as with spherical polar amplitude functions, being proportional to the squares of matrix elements of cartesian coordinate $z$, or the spherical polar product $r\cos(\theta)$, or the corresponding paraboloidal coordinate $\tfrac{1}{2}(u^2 - v^2)$ as defined above. The absorption spectrum thus maintains a form exactly as calculated with spherical polar coordinates and depicted in part I of articles in this series [5].

### III. GRAPHICAL REPRESENTATIONS OF AMPLITUDE FUNCTION $\Psi(u,v,\phi)$

Not only for comparison with graphical representations of amplitude functions calculated in coordinates of other systems but also to present quantitatively accurate shapes and sizes of these functions, we exhibit here some selected examples. As a plot involving three independent variables – spatial coordinates $u,v,\phi$ – and dependent variable $\psi(u,v,\phi)$ would require at least four dimensions, the best way to proceed with two dimensions, or three pseudo-dimensions, is to exhibit a surface of constant ψ at a value selected to display the overall spatial properties of a particular real amplitude function in an satisfactory manner, as explained elsewhere [5]. Whereas some textbooks of quantum mechanics in physics discuss amplitude functions of H in paraboloidal coordinates, for instance that by Schiff [9], almost invariably without plots, no known textbook of chemistry even mentions this topic; we hence present here accurate plots of exemplary functions, emphasizing a comparison with related functions expressed in spherical polar coordinates.

The formula of amplitude function $\psi_{0,0,0}$ associated with the state of least energy,

$$\psi_{0,0,0} = \frac{e^3 Z^{\left(\frac{3}{2}\right)} \mu^{\left(\frac{3}{2}\right)} \pi \, \mathbf{e}^{\left(-\frac{Z e^2 \pi \mu (u^2 + v^2)}{2 h^2 \varepsilon_0}\right)}}{\varepsilon_0^{\left(\frac{3}{2}\right)} h^3},$$

shows no dependence on the angular variable, merely an exponential decay with distance from the origin of the system of coordinates that is effectively at the atomic nucleus, because $u^2 + v^2$ in the exponent is equivalent to distance $2\,r$ from the nucleus. With $Z = 1$, we plot a surface of this amplitude function that hence exhibits a spherical shape; its radius is about $2.45 \times 10^{-10}$ m. The value $\psi_{0,0,0} = 1.46 \times 10^{13}$ m$^{-3/2}$ that is chosen for this surface corresponds to both 1/100 of the maximum value of $\psi_{0,0,0}$ at $u = v = 0$ and the volume of $\psi_{0,0,0}^2$ that encloses about 0.995 electronic charge, as explained in part I of this series of papers [5]. The shape and size of this surface of $\psi_{0,0,0}(u,v,\phi)$ in paraboloidal coordinates coincide exactly with the surface of $\psi_{0,0,0}(r,\theta,\phi)$ in spherical polar coordinates. The related surface of $\psi_{0,0,0}^2 = 1.46 \times 10^{26}$ m$^{-3}$ is necessarily also spherical and has a radius about $2.55 \times 10^{-10}$ m; for both this amplitude function and its square, the spherical shape reflects the lack of angular dependence of the electrostatic attraction between the atomic nucleus and the electron in the absence of an electromagnetic field.

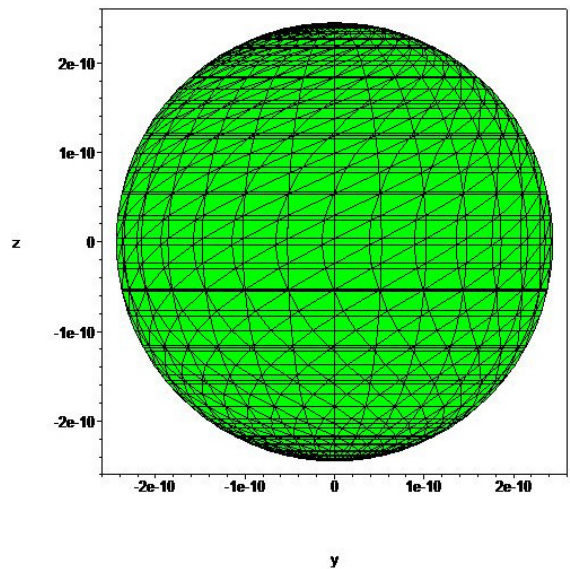

**FIGURE 2.** Surface of real paraboloidal amplitude function $\psi_{0,0,0} = 1.46 \times 10^{13}$ m$^{-3/2}$. Here, and in succeeding plots in three pseudo-dimensions, the unit of length along each coordinate axis is m; notation 2e−10 implies $2 \times 10^{-10}$, and analogously, in this and succeeding figures.

The surface of amplitude function $\psi_{1,0,0}$, which incorporates $Z = 1$ here and henceforth,

$$\psi_{1,0,0} = -\frac{e^3 \mu^{(3/2)} \pi \, e^{\left(-\frac{e^2 \pi \mu (u^2 + v^2)}{4 h^2 \varepsilon_0}\right)} (\pi e^2 \mu u^2 - 2 h^2 \varepsilon_0)}{8 \varepsilon_0^{(5/2)} h^5},$$

in figure 3 exhibits a novel shape, unlike that of any amplitude function directly derived in spherical polar coordinates; the domain of $\phi$ is curtailed at $3\pi/2$ rad to reveal the interior so as to emphasize the ostensibly peculiar structure. Although the overall shape is axially symmetric about axis $z$ and is roughly spherical, the centre of that sphere is displaced from the origin by about $2 \times 10^{-10}$ m along positive axis $z$; a nodal surface of zero amplitude, of paraboloidal shape, exists between a large positive lobe, extending mostly along positive axis $z$, and a small negative lobe, extending along negative axis $z$. The surface of the square of this amplitude function, plotted for $\psi_{1,0,0}^2 = 5 \times 10^{25}$ m$^{-3}$, has a similar size and shape.

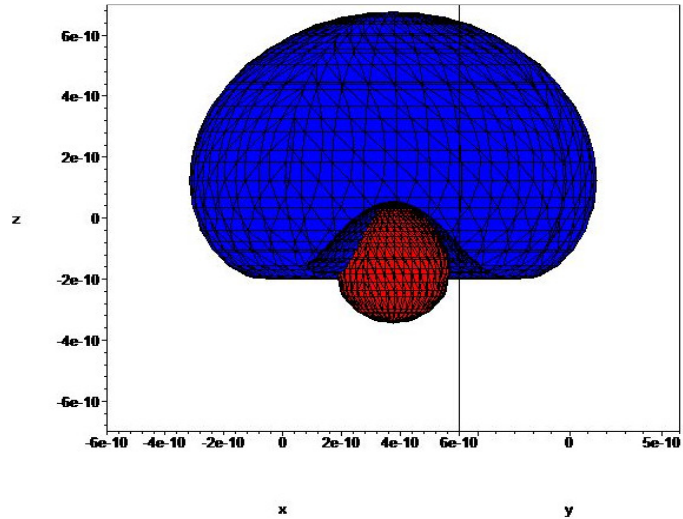

**FIGURE 3.** Surface of real paraboloidal amplitude function $\psi_{1,0,0} = 1.46 \times 10^{13}$ m$^{-3/2}$; the surface is cut open to reveal the structure of the positive lobe (blue) extending mostly above plane $z=0$ and the negative lobe (red) extending along negative axis $z$.

Consistent with the complementary shapes of surfaces of $u$ and $v$ of the same value as seen in figure 1, the shape of the surface of $\psi_{0,1,0} = 1.46 \times 10^{13}$ m$^{-3/2}$ is exactly the reflection of $\psi_{1,0,0}$ across the plane $z = 0$: a small positive lobe hence extends along positive axis $z$ and a large negative lobe along negative axis $z$; the centre of the surface of $\psi_{0,1,0}$ is located about $2 \times 10^{-10}$ m along negative axis $z$ from the origin. For the real and imaginary parts of complex paraboloidal amplitude functions $\psi_{0,0,1}$ and $\psi_{0,0,-1}$ with equatorial quantum number $m$ different from zero, the shapes and sizes of the surfaces of amplitude functions are essentially identical to those of the corresponding parts of $\psi_{0,1,1}(r,\theta,\phi)$ and $\psi_{0,1,-1}(r,\theta,\phi)$ in spherical polar coordinates, which are in turn identical to $\psi_{0,1,0}(r,\theta,\phi)$ apart from their orientation, except that the real part of $\psi_{0,1,1}(r,\theta,\phi)$ is symmetric about axis $y$ whereas the real part of $\psi_{0,0,1}(u,v,\phi)$ is symmetric about axis $x$, and vice versa for the imaginary parts. The sum $\psi_{0,1,0}^2 + \psi_{1,0,0}^2 + |\psi_{0,0,1}|^2 + |\psi_{0,0,-1}|^2$ in paraboloidal coordinates plots as a perfect sphere, of radius about $6.4 \times 10^{-10}$ m.

In figure 4, we show next the surface of paraboloidal amplitude function $\psi_{2,0,0}$, according to this formula.

$$\psi_{2,0,0} = \frac{e^3 \mu^{(3/2)} \pi \, \mathbf{e}^{\left(-\frac{e^2 \pi \mu (u^2 + v^2)}{6 h^2 \varepsilon_0}\right)} (\pi^2 e^4 \mu^2 u^4 - 12 \pi e^2 h^2 \mu u^2 \varepsilon_0 + 18 h^4 \varepsilon_0^2)}{162 \, \varepsilon_0^{(7/2)} h^7}$$

Its surface exhibits three lobes, two of positive phase and one of negative phase between those two, with paraboloidal nodal surfaces between these lobes. Apart from that lobal structure, the overall shape is roughly spherical, of radius $8 \times 10^{-10}$ m, but the centre of the sphere is located along positive axis $z$ about $4 \times 10^{-10}$ m from the origin. Amplitude function $\psi_{0,2,0}$ is the reverse of $\psi_{2,0,0}$, roughly spherical in shape but extending mostly along negative axis $z$ and centred near $z = -4 \times 10^{-10}$ m.

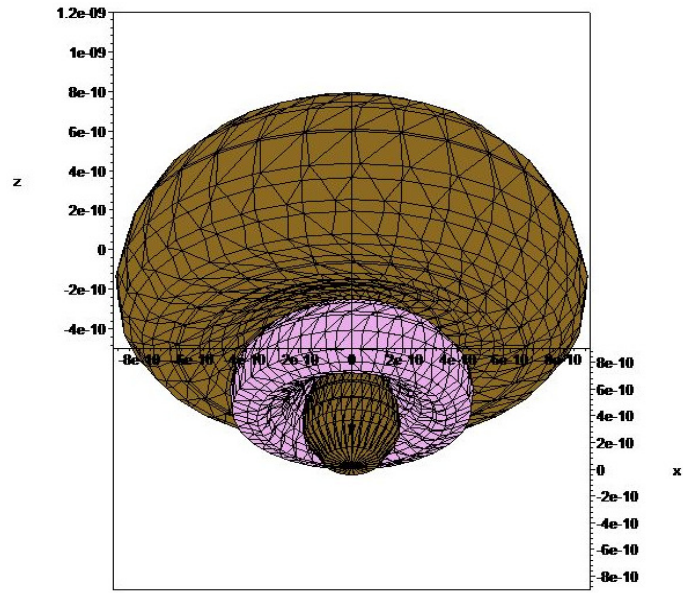

**FIGURE 4.** Surface of real paraboloidal amplitude function $\psi_{2,0,0} = 1.46 \times 10^{13}$ m$^{-3/2}$; the largest and smallest lobes (sienna) have positive phase, the intermediate lobe (plum) has a negative phase.

The surface of amplitude function $\psi_{1,1,0}$,

$$\psi_{1,1,0} = \frac{e^3 \mu^{(3/2)} \pi \, \mathbf{e}^{\left(-\frac{e^2 \pi \mu (u^2 + v^2)}{6 h^2 \varepsilon_0}\right)} (\pi e^2 \mu u^2 - 3 h^2 \varepsilon_0)(\pi e^2 \mu v^2 - 3 h^2 \varepsilon_0)}{81 \, \varepsilon_0^{(7/2)} h^7},$$

which, like the preceding two functions, $\psi_{2,0,0}$ and $\psi_{0,2,0}$, corresponds to energy quantum number $n$ = 3, is shown in figure 5.  In this case a small positive lobe of slightly oblate spheroidal shape is centred at the origin; a large positive lobe is a torus, with two pronouncedly spheroidal negative lobes directed along positive and negative axis $z$ separated by the small positive lobe.  The overall shape is roughly oblate spheroidal; its square exhibits a similar shape.

As an example of a novel complex amplitude function, we show in figure 6 the surface of the real part of $\psi_{1,0,1}$ that conforms to this formula:

$$\psi_{1,0,1} = \frac{e^5 \mu^{\left(\frac{5}{2}\right)} 2^{\left(\frac{1}{2}\right)} \pi^2 u\, v\, \mathbf{e}^{\left(-\frac{e^2 \pi \mu u^2 + e^2 \pi \mu v^2 - 6\, i\, \phi\, h^2\, \varepsilon_0}{6\, h^2\, \varepsilon_0}\right)} (\pi\, e^2\, \mu\, u^2 - 6\, h^2\, \varepsilon_0)}{162\, \varepsilon_0^{\left(\frac{7}{2}\right)} h^7}$$

Of four lobes, two are large and two are small, one each of each phase.  The shapes of lobes in this figure are common to $\psi_{1,0,1}$, $\psi_{1,0,-1}$, $\psi_{0,1,1}$ and $\psi_{0,1,-1}$ in their real or imaginary parts; they are symmetric across planes $y$ = 0 or $x$ = 0. For $\psi_{1,0,1}$ or $\psi_{1,0,-1}$, the large lobes lie mostly above plane $z$ = 0, as for $\psi_{1,0,0}$, whereas for $\psi_{0,1,1}$ or $\psi_{0,1,-1}$ the large lobes lie mostly on the negative side of plane $z$= 0, as for $\psi_{0,1,0}$. The surfaces of the squares of $\psi_{1,0,1}$ and $\psi_{1,0,-1}$ are identical, and resemble the shapes of the surfaces of $\psi_{1,0,0}$, shown in figure 3, and $\psi_{0,1,0}$, except that there is a tunnel of zero electronic density along the polar axis, making the shapes toroidal.

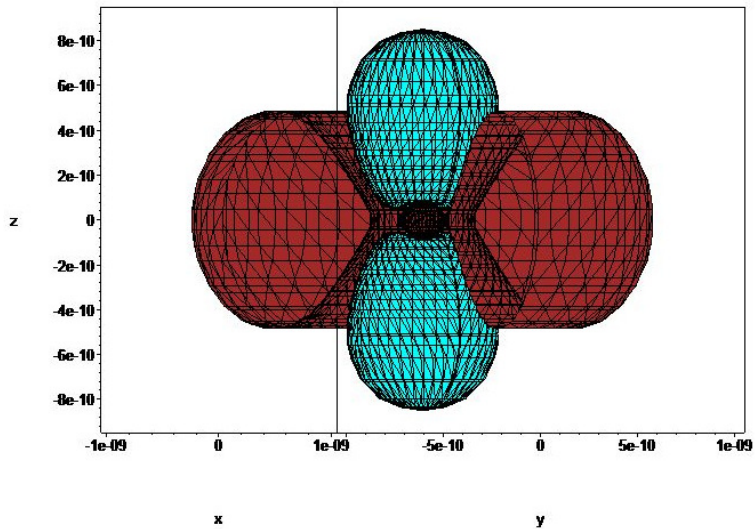

**FIGURE 5.** Surface of real paraboloidal amplitude function $\psi_{1,1,0}$ = 1.46x10$^{13}$ m$^{-3/2}$, cut open to reveal the interior structure; two negative lobes (cyan) are separated by a small positive oblate spheroidal lobe (brown) around the origin, all partially surrounded with a large positive torus (brown).

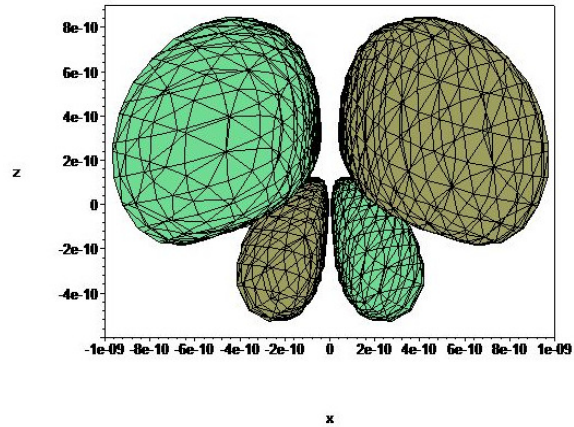

**FIGURE 6.** Surface of the real part of complex paraboloidal amplitude function $\psi_{0,1,1}$ = $1.46 \times 10^{13}$ m$^{-3/2}$; the positive lobes (khaki) and negative lobes (aquamarine) are symmetric with respect to plane $xz$.

In contrast, for complex paraboloidal amplitude functions $\psi_{0,0,2}$ and $\psi_{0,0,-2}$ in their real or imaginary parts, the surfaces have four lobes of equal size, extending along axes $x$ and $y$ for $\psi_{0,0,2}$ and $\psi_{0,0,-2}$ in their real parts and between these axes for their imaginary parts, so rotated by $\pi/8$ rad from one another. As an example in figure 7, we display the shape of the surface of the real part of $\psi_{0,0,2}$ that conforms to this formula.

$$\psi_{0,0,2} = \frac{e^7 \mu^{\left(\frac{7}{2}\right)} \pi^3 u^2 v^2 \, e^{\left(-\frac{e^2 \pi \mu (u^2+v^2)}{6 h^2 \varepsilon_0}\right)} \cos(2\phi)}{162 \, h^7 \varepsilon_0^{\left(\frac{7}{2}\right)}}$$

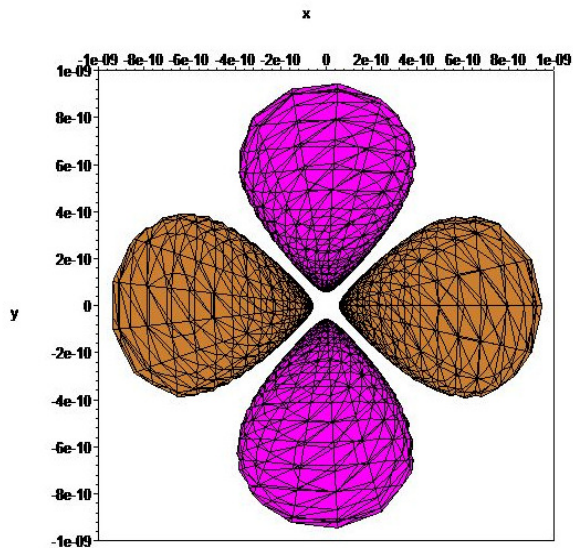

**FIGURE 7.** Surface of the real part of complex paraboloidal amplitude function $\psi_{0,0,2}$ = $1.46 \times 10^{13}$ m$^{-3/2}$; the lobes of positive phase (golden) lie along axis $y$ and those of negative phase (magenta) along axis $x$.

These surfaces of complex paraboloidal amplitude functions $\psi_{0,0,\pm2}$ in their real and imaginary parts thus resemble the surfaces of spherical polar amplitude functions $\psi_{0,2,\pm2}(r,\theta,\phi)$, analogously to the respective surfaces of $\psi_{0,0,\pm1}$ and $\psi_{0,1,\pm1}(r,\theta,\phi)$. The shapes of the surfaces of the squares of $\psi_{0,0,2}$ and $\psi_{0,0,-2}$ are identical to each other, and constitute oblate tori surrounding the polar axis.

As a further example, figure 8 displays the surface of paraboloidal amplitude function $\psi_{1,2,0}$, according to this formula,

$$\psi_{1,2,0} = -e^3 \mu^{\left(\frac{3}{2}\right)} \pi\, e^{\left(-\frac{e^2 \pi \mu (u^2+v^2)}{8 h^2 \varepsilon_0}\right)} (\pi e^2 \mu u^2 - 4 h^2 \varepsilon_0)$$
$$(\pi^2 e^4 \mu^2 v^4 - 16 \pi e^2 h^2 \mu v^2 \varepsilon_0 + 32 h^4 \varepsilon_0^2) \Big/ \left(2048\, \varepsilon_0^{\left(\frac{9}{2}\right)} h^9\right)$$

which exhibits four spheroidal lobes, two of each phase, and two tori, one of each phase, all symmetric about the polar axis. These features of tori and spheroids are typical of the shapes of surfaces of real paraboloidal amplitude functions with energy quantum number $n$ greater than 3.

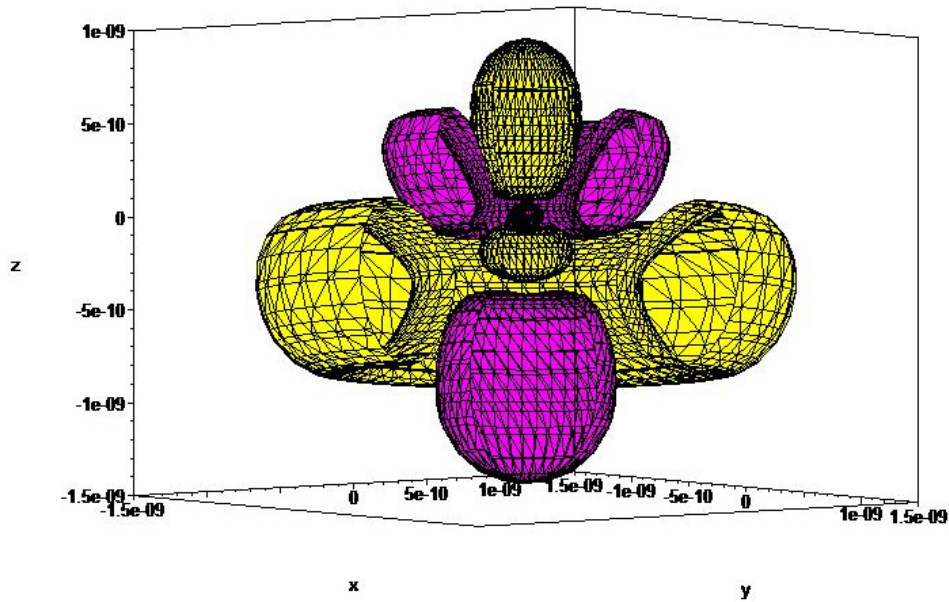

**FIGURE 8.** Surface of real paraboloidal amplitude function $\psi_{1,2,0}$ = 1.46x10$^{13}$ m$^{-3/2}$, cut open to reveal the interior structure; the large torus (yellow) has a negative phase and the small torus (magenta) a positive phase; of the four spheroidal lobes, the largest (magenta) has a positive phase, and the smallest also of positive phase (magenta) is between two larger spheroidal lobes of negative phase.

## IV.   DISCUSSION

Schroedinger developed this solution of his equation in paraboloidal coordinates [3] primarily as a method to treat, with perturbation theory that he concurrently developed, the Stark effect on the hydrogen atom, explicitly the shifting and splitting of spectral lines as a result of hydrogen atoms being subjected to a uniform electric field. To take such account, he added to the hamiltonian a term $e\,F\,z = \tfrac{1}{2}\,e\,F\,(u^2 - v^2)$ for a uniform electric field of strength $F$ in direction $+z$, i.e. so to define the polar axis; the additional term was treated as a perturbation of the system. The consequent energy of a state corresponding to quantum numbers $n_1$, $n_2$ and $m$ becomes

$$E = -\frac{\mu\,Z^2\,e^4}{8\,h^2\,\varepsilon_0^{\,2}\,(n_1 + n_2 + |m| + 1)^2} + \frac{3\,\mu\,e^4\,(n_1 + n_2 + |m| + 1)\,(n_1 - n_2)\,F}{8\,h^2\,\varepsilon_0^{\,2}\,Z}$$

in which the first term is precisely the energy of that state in the absence of the field, $F = 0$; the second term shows the linear dependences on both that mild field and the difference between quantum numbers $n_1$ and $n_2$, i.e. the *electric quantum number*. For $n_1 > n_2$, the electron is predominantly located with $z > 0$, consistent with the raising of the energy through the potential energy of the electron and the external field. Figures 2 – 8 show that, for amplitude functions with $n_1 = n_2$, plane $z = 0$ is a plane of symmetry. With increasing strength of the electric field, an additional term, quadratic in $F$, appears, and eventually, at large fields, further terms become significant. In all cases the magnitude of the Stark effect depends also on the magnitude of equatorial quantum number $m$, as shown in the above formula for the linear Stark effect for instance. Regarding an homogeneous magnetic field acting on a hydrogen atom $^1$H, the problem is complicated because of the presence of magnetic dipole moments associated with the intrinsic angular momenta of both electron and proton, known as spin. Neglect of the effect of *electron spin* yields a variation of the energy of the atom linearly proportional to magnetic quantum number $m$, considered to be the *normal* Zeeman effect, but in practice the energy of interaction of the electron spin with its motion about the atomic nucleus is larger than the energy change due to the external magnetic field, up to flux density $B_z = 10$ T. Taking into account also the purported electron spin involves a treatment that yields the *anomalous* Zeeman effect, but spherical polar coordinates are suitable for this calculation.

    Other contexts in which these paraboloidal coordinates are particularly useful include the photoelectric effect, the Compton effect and a collision of an electron with a H atom; in each case a particular direction in space is distinguished according to some external force [*8*]: that direction becomes the polar axis about which equatorial angle $\phi$ is measured with respect to a reference plane. Although paraboloidal coordinates might appear to obscure the innate spherical symmetry of an isolated H atom, in nearly any practical experiment on these atoms, apparently apart from a measurement of the simple spectrum of an unperturbed system, that spherical symmetry is lost. Any chemical application of the amplitude functions of a H atom, in particular, inevitably involves an interaction with another chemical species, which defines a particular direction to become prospectively the polar axis. From a chemical or physical point of view, a consideration of the H atom in paraboloidal coordinates seems more important than in spherical polar coordinates, but ellipsoidal coordinates are more practical than either spherical polar or paraboloidal. We consider this matter further in Part III on ellipsoidal coordinates.

    Quantum numbers in each set $n_1$, $n_2$, $m$ and each amplitude function that they designate, with their corresponding plots, are all parochial to this system of paraboloidal coordinates, just as quantum numbers $k$, $l$, $m$ and their associated amplitude functions are parochial to the system of spherical polar coordinates, and have no meaning beyond the context of the same particular system of coordinates; equatorial quantum number fortuitously happens to be common to both systems because equatorial angle $\phi$ is likewise a common coordinate. With $n = n_1 + n_2 + |m| + 1$, the

total number of states or amplitude functions for *n* of given value is equal to that number *n* with *m* = 0 plus twice the sum with *m* ≠ 0,

$$n + 2\left(\sum_{m=1}^{n-1}(n-m)\right) = n^2,$$

the same as for spherical polar coordinates.

The preceding plots of surfaces of amplitude functions in paraboloidal coordinates show that, in general, these functions are asymmetric to the plane $z = 0$ or for which $u^2 = v^2$ unless $n_1 = n_2$; for $n_1 > n_2$ most electronic charge is located in the half space in which $z > 0$, and conversely for $n_1 < n_2$. A comparison of plots of surfaces of amplitude functions in spherical polar and paraboloidal coordinates makes clear that the shapes and sizes of the surfaces of the real and imaginary parts of complex paraboloidal amplitude functions $\psi_{0,0,1}(u,v,\phi)$ and $\psi_{0,0,-1}(u,v,\phi)$ are essentially identical with those of the corresponding parts of $\psi_{0,1,1}(r,\theta,\phi)$ and $\psi_{0,1,-1}(r,\theta,\phi)$ in spherical polar coordinates, except that the real part of $\psi_{0,1,1}(r,\theta,\phi)$ is symmetric about axis *y* whereas the real part of $\psi_{0,0,1}$ is symmetric about axis *x*, and vice versa. For the paraboloidal and spherical polar amplitude functions, there are relations, not one to one as for the complex functions specified above, but as a sum or difference of amplitude functions in one system of coordinates to generate a particular amplitude function in another system with a common value of energy, and hence energy quantum number *n*. This condition necessarily follows from the solution of the same hydrogen atom under the same conditions, for instance, a non-relativistic treatment in three spatial dimensions in the absence of an external field. As instances of formulae for interconversion of these functions between $\psi_{n_1,n_2,m}(u,v,\phi)$ and $\psi_{k,l,m}(r,\theta,\phi)$, which correspond merely to transformations of coordinates, we state the following results for the sum,

$$\psi_{0,1,0}(r,\theta,\phi) + \psi_{1,0,0}(r,\theta,\phi) = \sqrt{2}\ \psi_{1,0,0}(u,v,\phi)$$

or the inverse relation,

$$\sqrt{2}\ \psi_{1,0,0}(r,\theta,\phi) = \psi_{0,1,0}(u,v,\phi) + \psi_{1,0,0}(u,v,\phi)$$

and for the difference,

$$\psi_{0,1,0}(r,\theta,\phi) - \psi_{1,0,0}(r,\theta,\phi) = \sqrt{2}\ \psi_{0,1,0}(u,v,\phi)$$

and its inverse relation.

$$\sqrt{2}\ \psi_{0,1,0}(r,\theta,\phi) = \psi_{0,1,0}(u,v,\phi) - \psi_{1,0,0}(u,v,\phi)$$

Analogous sums and differences – linear combinations of multiple functions in general – connect any amplitude function in spherical polar coordinates with appropriately selected functions in paraboloidal coordinates, provided that energy quantum number *n* is common to each set; for the above relations $n = 2$. Schroedinger mentioned this interconversion in his second lecture to the Royal Institution in London, 1928 [*10*]. The identity of surfaces of the separate real or imaginary parts of $\psi_{0,0,1}(u,v,\phi)$ and $\psi_{0,0,-1}(u,v,\phi)$ with those of the corresponding parts of $\psi_{0,1,1}(r,\theta,\phi)$ and $\psi_{0,1,-1}(r,\theta,\phi)$ fails to hold directly for amplitude functions with $n > 2$, but remains applicable to the appropriate linear combinations of complex amplitude functions of one coordinate system to generate a particular complex amplitude function in another system. Linear combinations of degenerate amplitude functions with common energy quantum number *n*, which are consequently likewise solutions of the same Schroedinger equation with the same energy, correspond to merely a rotation of the axes in the same system of coordinates or to another choice of system of coordinates [9]. Such degeneracy occurs in general when the amplitude equation is solvable in multiple ways – either for distinct systems of coordinates or within a single coordinate system variously oriented. For *l* = 0 in spherical polar coordinates, the amplitude function is spherically symmetric, so having the same form for any orientation of the polar axis. For paraboloidal coordinates, quantum number *l* for angular momentum is undefined, even though the component of angular momentum along the polar axis is well defined according to equatorial quantum number *m*; the only intrinsically spherically symmetric surface of an amplitude function, or of one of its lobes, occurs for $n_1 = n_2 = m = 0$, applicable to the ground state of this H atom with a plot shown in figure 2; in this case the solutions for the amplitude equation derived in spherical polar and paraboloidal coordinates are equivalent. With the amplitude functions as defined above, we calculate an expectation value of the square of angular momentum $L^2$ in its external motion for any state defined with quantum numbers $n_1$, $n_2$, $m$ as

$$<L^2> = \frac{1}{4} \frac{((n_1+n_2+1)|m|+m^2+n_1+2n_1n_2+n_2)h^2}{\pi^2},$$

which accordingly depends on all three quantum numbers, like the energy of that state. From that formula one might derive an expectation value for *l*,

$$l = -\frac{1}{2} + \sqrt{(|m|+n_1+n_2+1)(|m|+1)+2n_1n_2-|m|-\frac{3}{4}}\ h/2\pi,$$

which clearly assumes no integer or half-integer value except when $n_1 = n_2 = 0$ giving $l = |m|$, and which might otherwise have little meaning as a quantum number; taking $|m|$ as being a lower limit of *l* is clearly preferable. The coefficient of $h/2\pi$ in the above formula is thus effectively an expectation value of *l* for states associated with amplitude functions expressed in paraboloidal coordinates. For comparison, in spherical polar coordinates, the total square of angular momentum, hence a scalar quantity and having no directional dependence, has value $l(l+1)\left(\frac{h}{2\pi}\right)^2$ involving quantum number *l*, of which the numerical coefficient of $\left(\frac{h}{2\pi}\right)^2$ assumes an integer value only for *l* = 0. For both systems of coordinates, the component $L_z$ of angular momentum parallel to the polar axis is given by $m\,h/2\pi$, which is independent of quantum numbers $n_1$ and $n_2$ but limited by *l*.

The preceding results in the form of formulae for amplitude functions ψ(*u*,*v*,φ), their associated quantum numbers $n_1$, $n_2$, *m*, and the surfaces of those functions are all clearly parochial to these paraboloidal coordinates, just as the corresponding quantities for spherical polar coordinates are parochial to those coordinates. An analogous conclusion is inevitable for the solution of Schroedinger's equation in ellipsoidal and spheroconical coordinates, presented in further papers of this series.

# THE HYDROGEN ATOM ACCORDING TO WAVE MECHANICS – III. ELLIPSOIDAL COORDINATES

*J. F. Ogilvie*[*]

Centre for Experimental and Constructive Mathematics, Department of Mathematics, Simon Fraser University, Burnaby, British Columbia V5A 1S6 Canada

Escuela de Química, Universidad de Costa Rica, Ciudad Universitaria Rodrigo Facio, San Pedro de Montes de Oca, San José, 11501-2050 Costa Rica

**Abstract**

Schroedinger's temporally independent partial-differential equation is directly solvable in ellipsoidal coordinates to yield three ordinary-differential equations; with a common factor in equatorial angular coordinate $\phi$ as in spherical polar and paraboloidal coordinates, the product of their solutions contains confluent Heun functions in coordinates $\xi$ and $\eta$ that impede further calculations at present. To provide plots of these functions, we apply published solutions from Kereselidze et al. in series to illustrate the dependence of the shape of the amplitude functions on distance *d* between the foci of the ellipsoids, between limiting cases of amplitude functions in spherical polar coordinates as $d \to 0$ and in paraboloidal coordinates as $d \to \infty$. These ellipsoidal coordinates are most appropriate for a treatment of a hydrogen atom in a diatomic-molecular context.

**Resumen**

La ecuación parcial-diferencial independiente de la temporalidad de Schroedinger es solucionable en coordenadas elipsoidales para producir tres ecuaciones diferenciales ordinarias. Así como en las coordenadas esféricas polares y paraboloidales, ella tiene otro factor en la coordenada angular ecuatorial $\phi$, cuyo producto de su solución contiene funciones Heun confluentes en coordenadas $\xi$ y $\eta$ que impiden cálculos adicionales en la actualidad. Las soluciones publicadas de Kereselidze et al se aplican en serie para proporcionar gráficos de estas funciones e ilustrar la dependencia de la forma de las funciones de amplitud en la distancia *d* entre los focos de los elipsoides, entre casos limitantes de funciones de amplitud en coordenadas polares esféricas cuando $d \to 0$ y en coordenadas paraboloidales cuando $d \to \infty$. Estas coordenadas elipsoidales son las más apropiadas para un tratamiento de un átomo de hidrógeno en un contexto diatómico-molecular.

**Key words:** hydrogen atom, wave mechanics, ellipsoidal coordinates, orbitals, atomic spectra

**Palabras clave:** átomo de hidrógeno, mecánica de onda, coordenadas elipsoidales, orbitales, espectro atómico.

---

[*] Corresponding author: ogilvie@cecm.sfu.ca

**THE HYDROGEN ATOM ACCORDING TO WAVE MECHANICS – III. ELIPSOIDAL COORDINATES**

I.     INTRODUCTION

In Schroedinger's four articles in a series of title *Quantisation as a Problem of Proper Values* [*1,2,3,4*] through which he introduced wave mechanics, he applied his partial-differential equation to the solution of the hydrogen atom in spherical polar and paraboloidal coordinates, and specified a method to calculate the frequencies and intensities of spectral lines. As the type of central field of force in the hydrogen atom is coulombic, the variables in the partial-differential equation are separable in not only spherical polar and paraboloidal coordinates but also ellipsoidal and spheroconical coordinates. In each case, the solution of Schroedinger's partial-differential equation independent of time yields three ordinary-differential equations, one for each spatial variable in the definition of a space of three dimensions. In this part III of a series of articles devoted to the hydrogen atom with its coordinates separable in these four systems, we state the temporally independent partial-differential equation and its solution in ellipsoidal coordinates, and provide plots of selected amplitude functions as surfaces corresponding to a chosen value of amplitude. As the dependence on time occurs in the same manner in all systems of coordinates in which the Schroedinger equation is separable, we accept the results from part I [*5*], and avoid that repetition. Although the equations governing the form of the amplitude functions are here, of necessity, defined in coordinates according to an ellipsoidal system, we view the surfaces of these amplitude functions invariably in rectangular cartesian coordinates: a computer procedure (in *Maple*) translates effectively from the original system of coordinates in which the algebra and calculus are performed to the system to which a human eye is accustomed.

II.    SCHROEDINGER'S TEMPORALLY INDEPENDENT EQUATION IN ELLIPSOIDAL COORDINATES

We first relate these three mutually orthogonal ellipsoidal coordinates $\xi$, $\eta$, $\phi$ to cartesian *x*, *y*, *z* and spherical polar coordinate *r* as algebraic formulae in direct relations. The system of ellipsoidal, or prolate spheroidal, coordinates, which Pauling and Wilson [*6*] called confocal elliptical coordinates, has two centres, corresponding to the foci of a respective ellipsoid; the distance between these two centres we denote *d*. For distances $r_a$ and $r_b$ of an electron from one or other centre, we define two coordinates $\xi$ and $\eta$, which are dimensionless.

$$\xi = \frac{r_a + r_b}{d} \quad \eta = \frac{r_a - r_b}{d}$$

The relations between these dimensionless distance coordinates $\xi$ and $\eta$, with equatorial angle $\phi$, and cartesian and polar coordinates follow.

$$x = \frac{d\sqrt{(\xi^2 - 1)(1 - \eta^2)} \cos(\phi)}{2}, \quad y = \frac{d\sqrt{(\xi^2 - 1)(1 - \eta^2)} \sin(\phi)}{2},$$

$$z = \frac{d(\eta\xi + 1)}{2}, \quad r = \frac{d(\eta + \xi)}{2}, \quad \phi = \arctan\left(\frac{y}{x}\right)$$

The domains of these variables are $1 \leq \xi < \infty$, $-1 \leq \eta \leq +1$, $0 \leq \phi < 2\pi$. We take an atomic nucleus of electric charge $+Z > 0$ as being located at a centre of coordinates with $\xi = 1$ and $\eta = -1$; the other focus at distance *d* has coordinates $\xi = 1$ and $\eta = 1$ and is a dummy centre with $Z = 0$, which might, however, become the location of another atomic nucleus in the case of a diatomic molecule, such as H$_2^+$. Surfaces of constant values of these coordinates are for ξ an ellipsoid, for η an hyperboloid of one sheet, and for equatorial angle φ a half-plane extending from polar axis *z*, the latter as in both spherical polar and paraboloidal coordinates. Although coordinates ξ and η are both dimensionless, one might consider ξ to be quasi-radial and η to be quasi-angular. For use within the volume element in subsequent integrals, the jacobian of the transformation between cartesian and ellipsoidal coordinates, as defined above, is $\frac{(\eta^2 - \xi^2)\, d^3}{8}$. Whereas these coordinates have been described as prolate spheroidal, the presence of a characteristic ellipsoid with its two foci makes preferable a description as ellipsoidal coordinates. These coordinates are illustrated in figure 1.

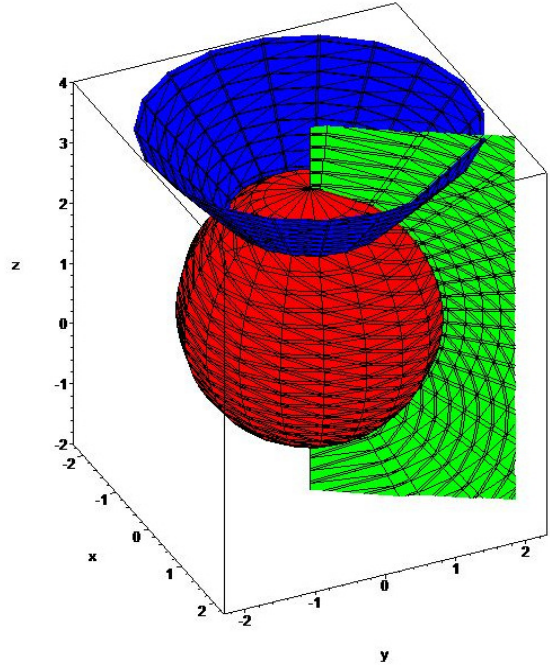

**FIGURE 1.** Surfaces of constant ellipsoidal coordinates: for the ellipsoid, ξ = 1.2 units, red; for the hyperboloid, η = π/4 units, blue; for the half-plane, equatorial angle φ = π/3 rad green.

A separation of the coordinates of the centre of mass of the H atom produces reduced mass μ of the system that is distant *r* from the origin, to supplant distance *r* between the electron and the atomic nucleus at one centre of the ellipsoid in the limit of infinite nuclear mass. Schroedinger's temporally independent equation in explicit SI units then contains within terms on the left side of the equality an electrostatic potential energy and first and second partial derivatives of an assumed amplitude function ψ(ξ,η,φ) with respect to spatial coordinates ξ, η, φ within an hamiltonian operator *H* ψ; the right side of the equality comprises a product of parameter energy, *E*, which is independent of coordinates, with the same amplitude function. The resultant form, as *H*(ξ,η,φ) ψ(ξ,η,φ) = *E* ψ(ξ,η,φ), resembles an eigenvalue relation:

$$\frac{1}{2} h^2 \Bigg( \Bigg( \Bigg( \frac{\partial^2}{\partial \xi^2} \psi(\xi, \eta, \phi) \Bigg) \eta^2 \xi^4 - \Bigg( \frac{\partial^2}{\partial \eta^2} \psi(\xi, \eta, \phi) \Bigg) \eta^4 \xi^2 + 2 \Bigg( \frac{\partial}{\partial \xi} \psi(\xi, \eta, \phi) \Bigg) \eta^2 \xi^3$$

$$- 2 \Bigg( \frac{\partial}{\partial \eta} \psi(\xi, \eta, \phi) \Bigg) \eta^3 \xi^2 - 2 \Bigg( \frac{\partial^2}{\partial \xi^2} \psi(\xi, \eta, \phi) \Bigg) \eta^2 \xi^2 - \Bigg( \frac{\partial^2}{\partial \xi^2} \psi(\xi, \eta, \phi) \Bigg) \xi^4$$

$$+ \Bigg( \frac{\partial^2}{\partial \eta^2} \psi(\xi, \eta, \phi) \Bigg) \eta^4 + 2 \Bigg( \frac{\partial^2}{\partial \eta^2} \psi(\xi, \eta, \phi) \Bigg) \eta^2 \xi^2 - 2 \Bigg( \frac{\partial}{\partial \xi} \psi(\xi, \eta, \phi) \Bigg) \eta^2 \xi$$

$$- 2 \Bigg( \frac{\partial}{\partial \xi} \psi(\xi, \eta, \phi) \Bigg) \xi^3 + 2 \Bigg( \frac{\partial}{\partial \eta} \psi(\xi, \eta, \phi) \Bigg) \eta^3 + 2 \Bigg( \frac{\partial}{\partial \eta} \psi(\xi, \eta, \phi) \Bigg) \eta \xi^2$$

$$+ \Bigg( \frac{\partial^2}{\partial \xi^2} \psi(\xi, \eta, \phi) \Bigg) \eta^2 + 2 \Bigg( \frac{\partial^2}{\partial \xi^2} \psi(\xi, \eta, \phi) \Bigg) \xi^2 - 2 \Bigg( \frac{\partial^2}{\partial \eta^2} \psi(\xi, \eta, \phi) \Bigg) \eta^2$$

$$- \Bigg( \frac{\partial^2}{\partial \eta^2} \psi(\xi, \eta, \phi) \Bigg) \xi^2 + \Bigg( \frac{\partial^2}{\partial \phi^2} \psi(\xi, \eta, \phi) \Bigg) \eta^2 - \Bigg( \frac{\partial^2}{\partial \phi^2} \psi(\xi, \eta, \phi) \Bigg) \xi^2$$

$$+ 2 \Bigg( \frac{\partial}{\partial \xi} \psi(\xi, \eta, \phi) \Bigg) \xi - 2 \Bigg( \frac{\partial}{\partial \eta} \psi(\xi, \eta, \phi) \Bigg) \eta - \Bigg( \frac{\partial^2}{\partial \xi^2} \psi(\xi, \eta, \phi) \Bigg) + \Bigg( \frac{\partial^2}{\partial \eta^2} \psi(\xi, \eta, \phi) \Bigg)$$

$$\Bigg) \Bigg/ (\pi^2 \mu (\eta^4 \xi^2 - \eta^2 \xi^4 - \eta^4 + \xi^4 + \eta^2 - \xi^2) d^2) - \frac{1}{2} \frac{Z e^2 \psi(\xi, \eta, \phi)}{\pi \varepsilon_0 d (\eta + \xi)} = E \psi(\xi, \eta, \phi)$$

Apart from fundamental physical constants electric permittivity of free space $\varepsilon_0$, Planck constant $h$ and protonic charge $e$, there appear parameters $Z$ for atomic number – $Z = 1$ for H – and $\mu$ for the reduced mass of the atomic system, practically equal to the electronic rest mass $m_e$. After separation of the variables and solution of the three consequent ordinary-differential equations including definition of the separation parameters or integration constants, the eventual full solution of the above equation has exactly this form [7].

$$\psi(\xi, \eta, \phi) = c\, N\, (\xi^2 - 1)^{\left(\frac{|m|}{2}\right)} (\eta^2 - 1)^{\left(\frac{|m|}{2} + \frac{1}{2}\right)} e^{\left(-\frac{d Z (\xi + \eta + 1)}{2 a_0 n} + i\, m\, \phi\right)}$$

$$\text{HeunC}\Bigg( \frac{2 d Z}{a_0 n}, |m|, |m|, \frac{2 d}{a_0}, -\frac{Z^2 d^2}{4 a_0^2 n^2} + \frac{m^2}{2} + \lambda - \frac{d}{a_0}, \frac{\xi}{2} + \frac{1}{2} \Bigg)$$

$$\text{HeunC}\Bigg( \frac{2 d Z}{a_0 n}, |m|, |m|, \frac{2 d}{a_0}, -\frac{Z^2 d^2}{4 a_0^2 n^2} + \frac{m^2}{2} + \lambda - \frac{d}{a_0}, \frac{\eta}{2} + \frac{1}{2} \Bigg) \Bigg/ \sqrt{2 \pi}$$

In this formula some clusters of fundamental constants have been replaced with Bohr radius $a_0$,

$$a_0 = \frac{\varepsilon_0 h^2}{\pi m_e e^2}$$

that invariably appears as a ratio with distance $d$ to maintain the correct dimensions; normalizing factor $N$ ensures that

$$\int \psi(\xi, \eta, \phi) * \psi(\xi, \eta, \phi)\, \mathrm{d}vol = 1,$$

in which d*vol* is a volume element containing the jacobian specified above; the implied triple integration is over all space. An asterisk as raised suffix of an amplitude function, so $\psi *$, implies a complex conjugate of $\psi$ such that, wherever $\mathbf{i} = \sqrt{-1}$ appears in $\psi$, $-\mathbf{i}$ appears in $\psi *$. The presence of **i** in an exponential factor as product with equatorial angle φ signifies that this formula is complex, thus containing real and imaginary parts. Coefficient *c* that equals any complex number of magnitude unity, such as a fourth root of unity – i.e. ±1, ±√−1, occurs because the Schroedinger equation independent of time is a linear homogeneous partial-differential equation, or equally because the temporally independent Schroedinger equation has the form of an eigenvalue relation, as shown above. The conventional choice *c* = 1, which is arbitrary and lacks physical justification, signifies that some solutions ψ(ξ,η,φ) as amplitude functions from the temporally independent Schroedinger equation might appear in a purely real form; with a mathematically valid alternative choice *c* = **i**, some amplitude functions might be entirely imaginary, but most would still be complex and thus alien to physical space. Choosing *c* = −1 or −**i** merely reverses the phase of an amplitude function or its constituent parts. Parameters that appear in the solution but not the partial-differential equation take discrete values, imposed by boundary conditions, as follows: *m* is called the equatorial, or magnetic, quantum number that assumes values of only negative and positive integers and zero, and that arises in the solution of the equatorial angular equation to define Φ(φ); *n* denotes the energy quantum number, which is incorporated in the formula at several locations on comparison with the corresponding solutions in spherical polar coordinates. The third quantum parameter λ that arises in this solution is unique to these ellipsoidal coordinates; it occurs only in the fifth argument of each confluent Heun function, so for these functions of ξ and η separately but in the same form. Apart from factor $(\eta^2 - 1)^{\left(\frac{|m|}{2} + \frac{1}{2}\right)}$ that also generally contributes a complex character to H(η), the confluent Heun function containing variable η might assume only real values, whereas the Heun function containing variable ξ definitely assumes complex values, so having real and imaginary parts, in addition to the real and imaginary parts resulting from i *m* φ in the exponential term.

## III. GRAPHICAL REPRESENTATIONS OF AMPLITUDE FUNCTION ψ(ξ, η, φ)

Not only for comparison with graphical representations of amplitude functions calculated in coordinates of other systems but also to present quantitatively accurate shapes and sizes of these functions, we display here some selected examples. As a plot involving three independent variables – spatial coordinates ξ, η, φ – and dependent variable ψ(ξ,η,φ) would require four dimensions, the best way to proceed with two dimensions, or three pseudo-dimensions, is to exhibit a surface of constant ψ at a value selected to display the overall spatial properties of a particular amplitude function in a satisfactory manner. As our following figures demonstrate plainly, the shapes of these ellipsoidal amplitude functions depend markedly on distance *d* between the centres of the ellipsoids. In the two limiting cases, as *d* → 0, the shape of an amplitude function tends to a shape of a corresponding function in spherical polar coordinates, whereas, as *d* → ∞, the shape tends to a shape of a respective function in paraboloidal coordinates. With the direct confluent Heun functions as specified above, these features are depicted with difficulty because the confluent Heun function of at least ξ, in the general formula above, has complex values, in addition to the complex character dictated in relation to equatorial angular coordinate φ. Working with these confluent Heun functions is hence at present difficult; for this reason we present figures prepared with ellipsoidal amplitude functions indirectly obtained through solution of Schroedinger's equation in series [8], rather than our direct solution stated above. These functions ψ$_{n_\xi,n_\eta,m}$ are characterised with three quantum numbers -- $n_\xi$, $n_\eta$, *m*, of which equatorial quantum number *m* has the same significance as for amplitude functions in spherical polar and paraboloidal coordinates; the values of $n_\xi$ and $n_\eta$ take non-negative integers, and *m* positive and negative integers and zero, as before. The relation for energy quantum number is $n = n_\xi + n_\eta + |m| + 1$. When $d \to 0$, $n_\xi \to$ radial quantum number $k$ for spherical polar coordinates, $n_\eta \to l - |m|$; when $d \to \infty$, $n_\xi \to$ paraboloidal quantum number $n_1$ and $n_\eta \to n_2$ [8]. According to the nature of the preparation of these amplitude functions, their algebraic form contains long expressions in their algebraic normalizing factors involving polynomials in $d$ to inverse powers, except $\psi_{0,0,0}(\xi,\eta,\phi)$. In this formula and all succeeding formulae and plots, atomic number is set to $Z = 1$, so appropriate directly to the hydrogen atom.

$$\psi_{0,0,0} := \frac{e^{\left(-\frac{d(\xi+\eta)}{2a_0}\right)}}{\sqrt{\pi}}$$

The plot of this surface, in figure 2, has the form of a perfect sphere.

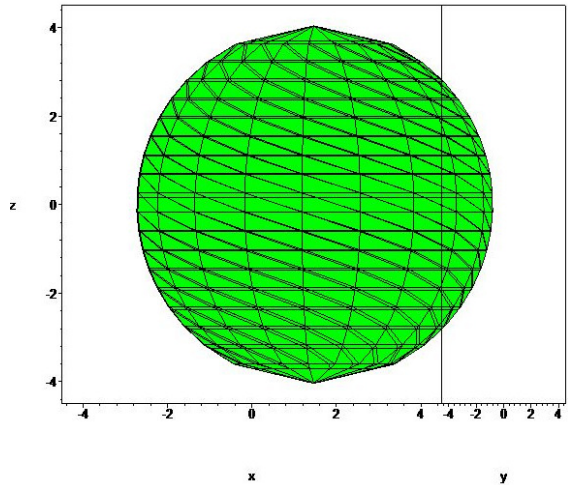

**FIGURE 2.** Surface of $\psi_{0,0,0} = 1/100\ a_0^{-3/2}$; as parameter $d$ is taken in unit $a_0$, the distance along each axis is expressed also in this unit, in this and succeeding plots of surfaces.

As in figures of surfaces of amplitude functions in spherical polar and paraboloidal coordinates in preceding parts of this series [5,9], this surface of an amplitude function at a stated value of that function is chosen such that the square of the amplitude function contains about 0.995 of the total electronic charge; this criterion is applicable to all further plots of surfaces presented in this article. The plot in figure 2 is actually formed with $d = 1/10\ a_0$, because a numerical value of $d$ must be provided to make such a plot. The shape, spherical, of this surface of constant $\psi$ is invariant with $d$; the size decreases only slightly, in unit $a_0$, as $d$ increases to 200 $a_0$; the reason is that, although variable $\eta$ has a fixed domain -1..1, the same for each plot, the extent of variable $\xi$ to make the plot varies greatly. For instance, for $d = a_0/10$, the necessary extent of $\xi$ is 1..90, whereas for $d = 200\ a_0$, the domain is only 1..1.041. Although all surfaces appear to have apparently pointed extremities along polar axis $z$, this effect is likely a distortion, due to a numerical artifact of the plotting routine with finite numerical accuracy for poorly behaved functions in terms of awkward arguments and of the conversion between ellipsoidal and cartesian coordinates. This amplitude function $\psi_{0,0,0}$ is the only one corresponding to energy quantum number $n = 1$.

For further amplitude functions of which the shape varies markedly with distance *d*, we exhibit a few surfaces at varied values of *d*, to illustrate the limiting and intermediate cases. As noted above, the limiting case as $d \to 0$ is a respective surface of an amplitude function in spherical polar coordinates; as $d \to \infty$, the respective surface is that in paraboloidal coordinates.

For the first instance, at distances $d/a_0$ = 1/10, 1.55 and 20 between the foci of the ellipsoids, in figures 3a, 3b and 3c respectively we exhibit $\psi_{0,1,0}$; this formula is expressed with symbolic normalizing factor $N_{0,1,0}$ because its explicit algebraic formula [8], which was used in the calculations to form the plot, is too long for practical presentation here.

$$\psi_{0,1,0} = N_{0,1,0}\, e^{\left(-\frac{d(\xi+\eta)}{4 a_0}\right)} \left(\xi - \frac{2 a_0 \left(1 - \sqrt{1 + \frac{d^2}{4 a_0^2}}\right)}{d}\right)\left(\eta - \frac{2 a_0 \left(1 - \sqrt{1 + \frac{d^2}{4 a_0^2}}\right)}{d}\right)$$

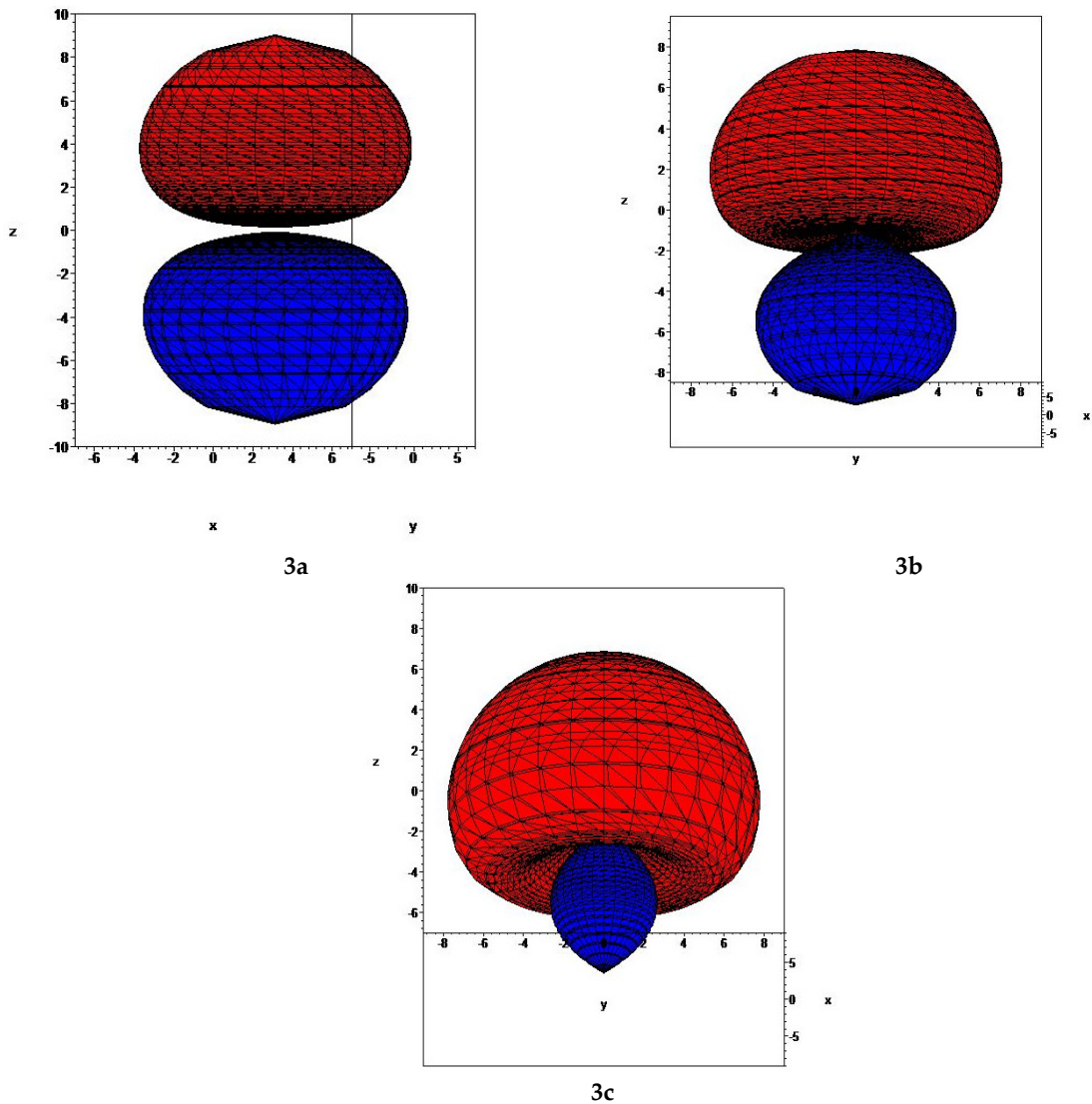

**FIGURE 3**. Surfaces of $\psi_{0,1,0}$ = 1/100 $a_0^{-3/2}$ at, from top down, a) $d$ = 1/10 $a_0$, b) $d$ = 1.55 $a_0$ and c) $d$ = 20 $a_0$; the positive lobe is red and the negative lobe is blue.

The shape of $\psi_{0,1,0}$ at $d = 1/10$ $a_0$ is practically indistinguishable from that of $\psi_{0,1,0}(r,\theta,\phi)$ in spherical polar coordinates as shown in figure 8 of part I [5]: the two lobes are nearly hemispherical with rounded edges, and have nearly the same size with a nodal plane of zero amplitude between them. At $d = 20$ $a_0$ the shape is nearly that of $\psi_{1,0,0}(u,v,\phi)$, as shown in figure 3 of part II [9]. The intermediate case at $d = 1.55$ $a_0$ simply shows that the negative lobe contracts gradually and the volume of the positive lobe expands with increasing $d$.

The next amplitude function, $\psi_{1,0,0}$,

$$\psi_{1,0,0} = N_{1,0,0}\, e^{\left(-\frac{d(\xi+\eta)}{4 a_0}\right)} \left(\xi - \frac{2 a_0 \left(1+\sqrt{1+\frac{d^2}{4 a_0^2}}\right)}{d}\right)\left(\eta - \frac{2 a_0 \left(1+\sqrt{1+\frac{d^2}{4 a_0^2}}\right)}{d}\right)$$

again expressed with symbolic normalizing factor of algebraic form that becomes converted to the appropriate numerical value, displays a different transition from one limit to the other, as shown in figures 4a, 4b, 4c at distances $d/a_0 = 1/10$, 1.55 and 20 respectively.

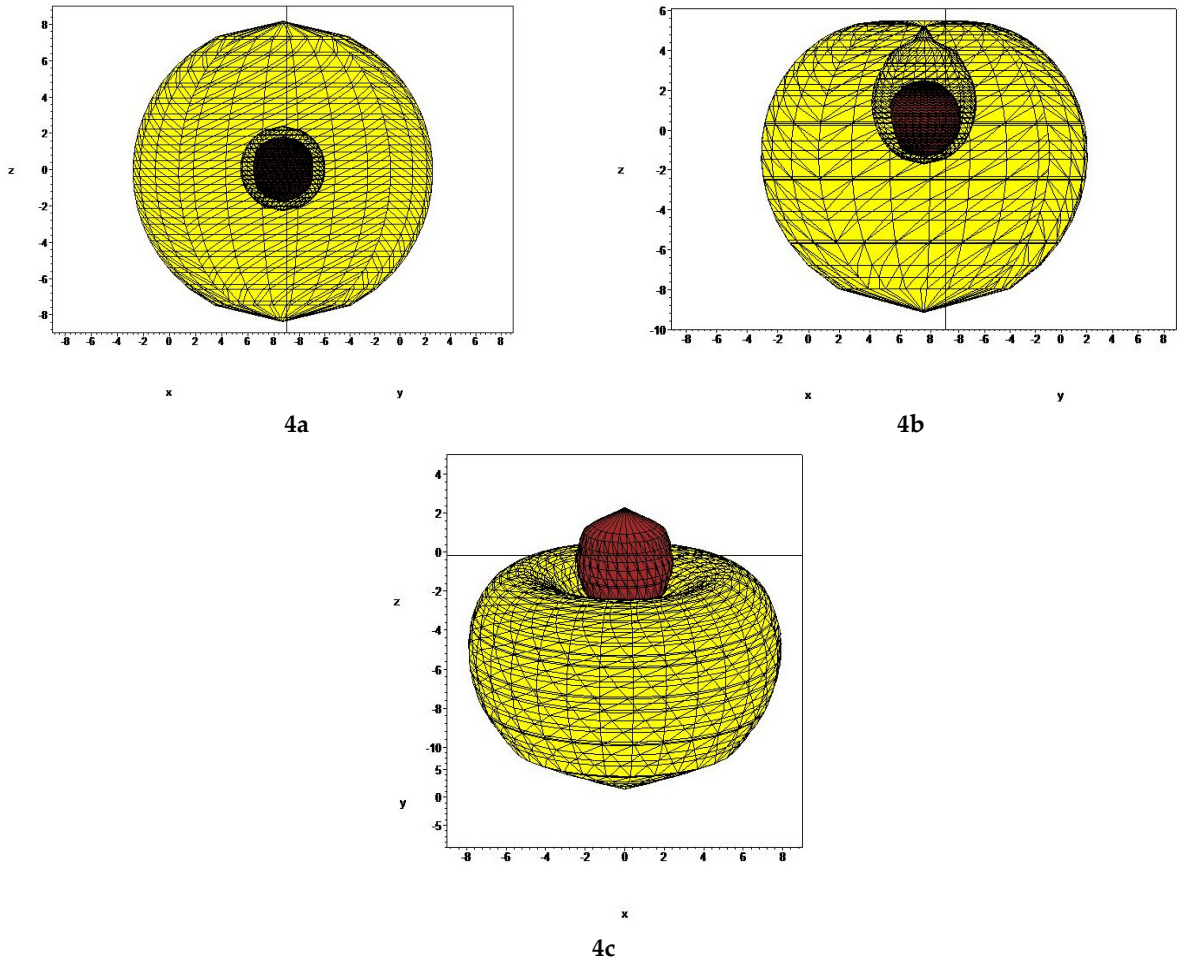

4a

4b

4c

**FIGURE 4.** Surfaces of $\psi_{1,0,0} = 1/100\ a_0^{-3/2}$ at, from top down, a) $d = 1/10\ a_0$, b) $d = 1.55\ a_0$ and c) $d = 20\ a_0$; the positive lobe is brown and the negative lobe is yellow. The surfaces in figures 4a and 4b are cut open to reveal the internal structure.

The shape of the surface at $d = 1/10\ a_0$ in figure 4a is essentially identical with that of $\psi_{1,0,0}(r,\theta,\phi)$ in figure 7 [5], having one small and nearly spherical positive lobe totally enclosed within a second, nearly spherical, shell of negative phase. At $d = 20\ a_0$, the shape is nearly that of $\psi_{0,1,0}(u,v,\phi)$, just the reverse of that shape in figure 3c along axis $z$. The intermediate shape in figure 4b is chosen at $d = 1.55\ a_0$ because at that distance a minute gap opens in the outer shell at the top of the negative lobe, through which the inner positive lobe eventually emerges to be become entirely external, as in figure 4c. These two amplitude functions $\psi_{0,1,0}$ and $\psi_{1,0,0}$, with $\psi_{0,0,1}$ depicted below, have common energy quantum number $n = 2$.

According to the solution in series [8], the next three amplitude functions involve the roots of a cubic equation, which, according to Viète's method, are expressed algebraically explicitly in terms of trigonometric functions and their inverses, specifically cosine and arc cosine; their forms [8] become consequently too voluminous for convenient presentation here, even with symbolic normalizing factors that would be even more extensive in explicit algebraic presentation. We hence merely present appropriate figures depicting the surfaces of these functions under the same conditions as above, first $\psi_{0,2,0}$ in figures 5a for $d = a_0/10$, 5b for $d = 5\ a_0$ and 5c for $d = 20\ a_0$.

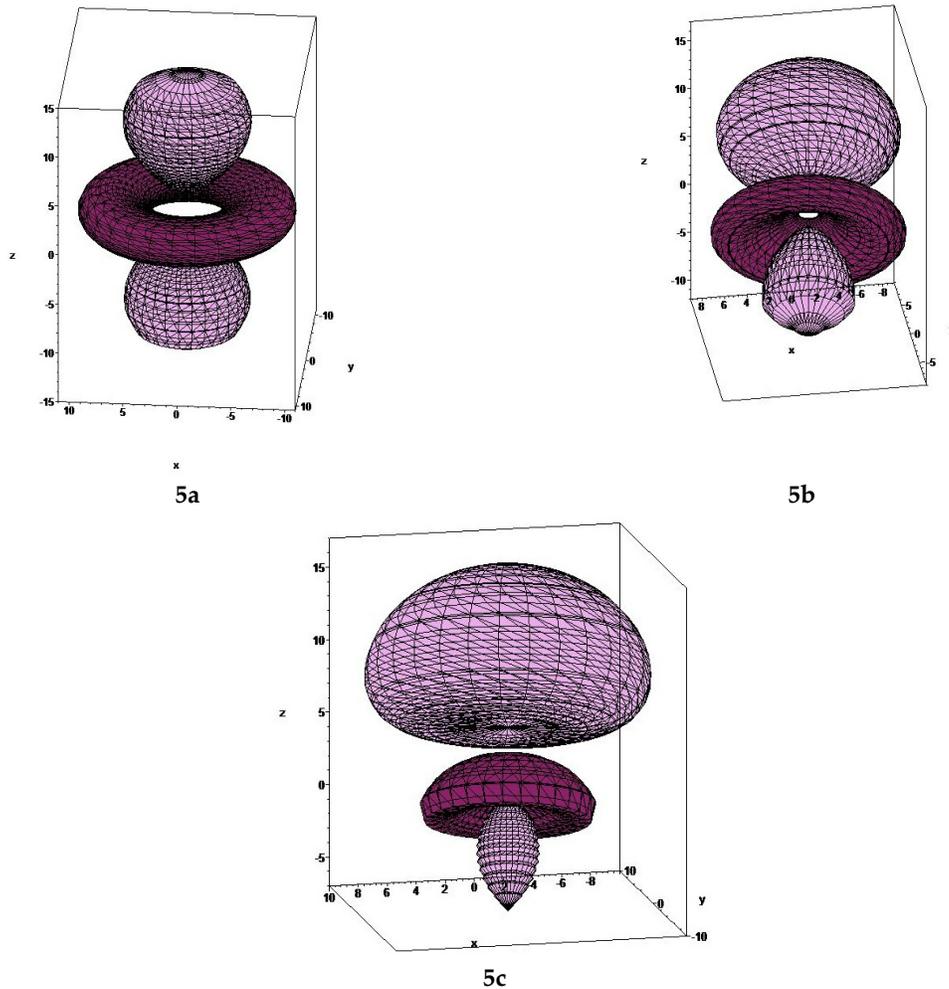

**FIGURE 5**. Surfaces of $\psi_{0,2,0} = 1/100\ a_0^{-3/2}$ at, from top down, a) $d = 1/10\ a_0$, b) $d = 5\ a_0$ and c) $d = 20\ a_0$; the positive lobe is plum colour and the negative lobe is maroon.

Whereas, at $d = 1/10$ $a_0$, the two positive lobes have practically equal size on either side of a negative torus, similar to $\psi_{0,2,0}(r,\theta,\phi)$ in figure 11 [5], at $d = 20$ $a_0$, the torus becomes transformed into an inverted bowl, with a small prolate spheroidal lobe below it and above it a large lobe of somewhat oblate spheroidal shape, as in figure 4 [9]. Here the intermediate case is chosen at $d = 5$ $a_0$, because in that condition the negative lobe exhibits only a small orifice that becomes closed completely at $d = 20$ $a_0$. Most surface at $d = 20$ $a_0$ lies above plane $xy$ at $z = 0$.

The next amplitude function is $\psi_{2,0,0}$ in figure 6a, b, c, with surfaces plotted at $d/a_0 = 1/10$, 1 and 20.

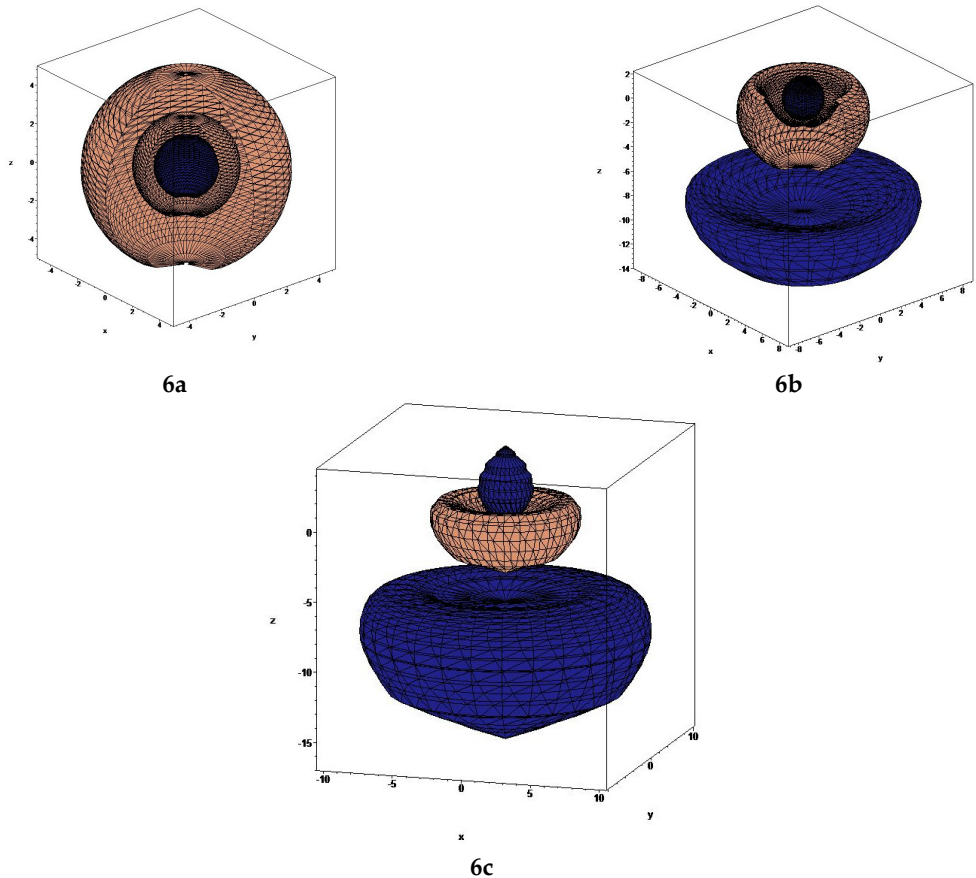

**6a**   **6b**

**6c**

**FIGURE 6.** Surfaces of $\psi_{2,0,0} = 1/100$ $a_0^{-3/2}$ at, from top down, a) $d = a_0/10$, b) $d = a_0$ and c) $d = 20a_0$; the positive lobes are navy blue and the negative lobe is tan. In figure 6a, all surfaces are cut open to reveal the interior structure; in figure 6b, only the negative lobe is cut open.

At distance $d = 1/10$ $a_0$, figure 6a shows a nearly spherical inner lobe of positive phase surrounded by a nearly spherical shell of negative phase; their centres are at approximately the origin of the system of cartesian coordinates. At $d = a_0$, the inner and positive sphere has just emerged from its negative surrounding lobe at the top, but a second positive lobe of bowl shape appears below the negative lobe. At $d = 20$ $a_0$, the upper positive lobe has a prolate spheroidal shape, partially below the rim of a negative lobe of bowl shape, and the lower positive lobe has almost a hemispherical shape, just the reverse of figure 5c along polar axis $z$. Most surface of this amplitude function is below plane $xy$ at $z = 0$.

The next amplitude function is $\psi_{1,1,0}$, of which surfaces in figures 7a,b,c,d appear at $d = a_0/10$, $a_0$, 5 $a_0$ and 20 $a_0$, respectively.

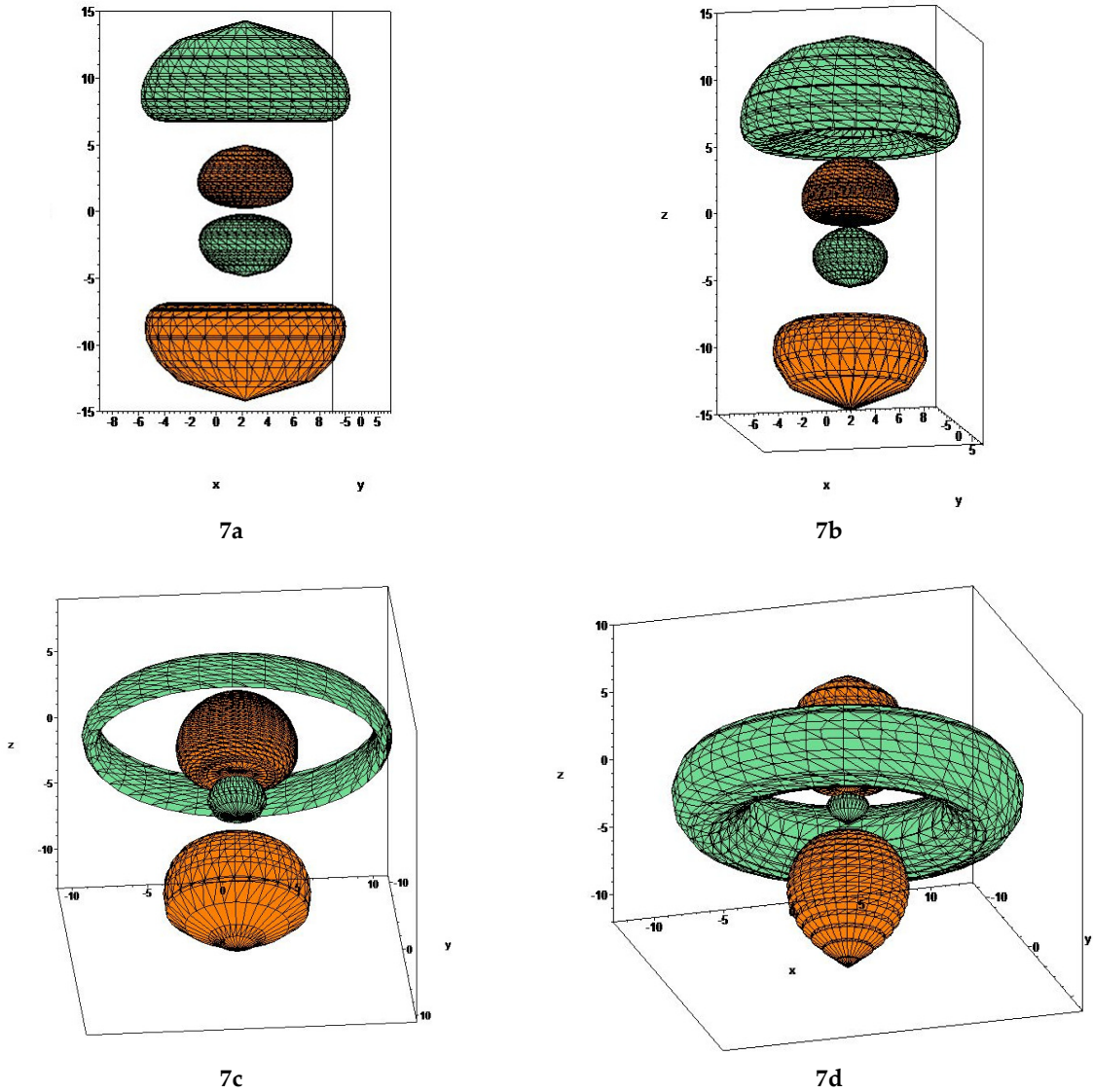

7a  7b

7c  7d

**FIGURE 7.** Surfaces of $\psi_{1,1,0} = 1/100\ a_0^{-3/2}$ at, from top down, a) $d = 1/10\ a_0$, b) $d = a_0$, c) $d = 5\ a_0$ and d) $d = 20\ a_0$; the positive lobes have coral colour, the negative lobes aquamarine.

The total surface of $\psi_{1,1,0} = 1/100\ a_0^{-3/2}$ at $d = 1/10\ a_0$ comprises four lobes, two of each phase, one small lobe of each phase between the other small lobe and a large lobe; this pattern correlates with $\psi_{1,1,0}(r,\theta,\phi)$. At $d = a_0$, the upper negative lobe increases in breadth, but at $d = 5\ a_0$ it becomes a torus surrounding the upper part of a positive lobe; at $d = 20\ a_0$ that torus has moved to plane $xy$ nearly at $z = 0$, at which it surrounds the other small negative lobe that is nearly spherical at the origin. The two positive lobes have prolate spheroidal shapes and are nearly symmetrically disposed across plane $xy$ at $z = 0$. The latter shape correlates with $\psi_{1,1,0}(u,v,\phi)$ in paraboloidal coordinates. The latter three amplitude functions have energy quantum number $n = 3$.

Amplitude function $\psi_{0,0,1}$, or equivalently $\psi_{0,0,-1}$, has both real and imaginary parts, according to the presence of an exponential factor containing $i\,m\,\phi$ with equatorial quantum number $m = 1$, in which $\mathbf{i} = \sqrt{-1}$.

$$\psi_{0,0,1} = N_{0,0,1}\, e^{\left(-\frac{d(\xi+\eta)}{4a_0}\right)} \sqrt{(\xi^2-1)(-\eta^2+1)}\; e^{(i\phi)}$$

Figure 8 presents the imaginary part of this amplitude function, of which the shape and size of the surface are practically invariant with varying distance $d$.

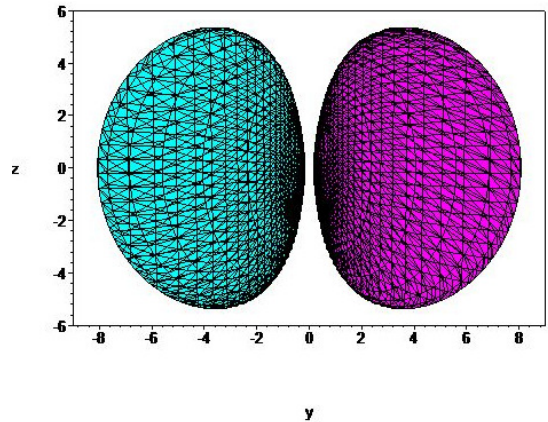

**FIGURE 8.** Surface of the imaginary part of $\psi_{0,0,1} = 1/100\; a_0^{-3/2}$ at $d = 1/10\; a_0$; the positive lobe is magenta and the negative lobe is cyan.

This surface is characteristic of $\psi_{0,0,1}(r,\theta,\phi)$ or $\psi_{0,0,-1}(r,\theta,\phi)$, and equivalently $\psi_{0,0,1}(u,v,\phi)$ or $\psi_{0,0,-1}(u,v,\phi)$, in either real or imaginary part. Although all preceding figures show surfaces of the depicted amplitude functions of which the surfaces are axially symmetric about the polar axis $z$, this surface is axially symmetric about cartesian axis $y$; its real counterpart is analogously axially symmetric about axis $x$, and the real and imaginary parts of $\psi_{0,0,-1}$ have analogous spatial dispositions axially symmetric about the same axes $x$ and $y$, respectively.

## IV. DISCUSSION

The three or four plots in each composite figure 3 - 7 demonstrate clearly how the familiar shape of the surface of an ellipsoidal amplitude function [5] for distance $d$ between foci nearly zero, which resembles a surface of an amplitude function in spherical polar coordinates, transforms into a much less familiar shape resembling the surface of a corresponding amplitude function in paraboloidal coordinates [9], as $d$ increases to a large value. The number of such amplitude functions that are available for this purpose is limited at present because of the problem of the complex nature of confluent Heun functions in the directly derived amplitude functions and because of the intractability of explicit algebraic solutions of quartic polynomials in the indirectly derived functions [8]. When one is able to plot real amplitude functions in both limiting cases and deduces the appropriate correlation, it is not difficult to imagine the course of the gradual transformation between those two limiting cases with increasing or decreasing distance $d$ between the foci of the ellipsoids; the two limiting cases must conform to the same value of energy quantum number $n$. As all these amplitude functions are common to the hydrogen atom, they must be convertible from one form, in one coordinate system, to another form in a separate coordinate system; this property would enable further explicit formulae to be generated through a transformation of coordinates, but the resulting expressions likely have a complicated algebraic form.

Much interest in the hydrogen atom treated in ellipsoidal coordinates arises because of the features pertinent to a system with two coulombic centres, such as $H_2^+$, for which ellipsoidal coordinates have long been applied [*6*]. Following the pioneering work of Burrau [*10*] and Wilson [*11,12*], Teller recognised that the amplitude functions from the solution of Schroedinger's equation in ellipsoidal coordinates to treat interatomic interactions in a diatomic molecule were applicable also to the hydrogen atom itself [*13*]. These ellipsoidal amplitude functions that result from the solution of Schroedinger's equation, either directly containing confluent Heun functions as presented above or indirectly through solution in series [*8*], are hence most appropriate for the treatment of a hydrogen atom, or analogous atomic ion with one electron, interacting with a point charge, located at the dummy focus; in such a case the atomic amplitude function, or atomic orbital, becomes *de facto* a molecular orbital. Other applications arise in an investigation of diatomic molecules or their ions in Rydberg states, or in other excited states [*14*]. As these confluent Heun functions become developed, these applications will become increasingly practical with the direct solutions of the Schroedinger equation.

**Acknowledgement**

I thank Professor T. Kereselidze for providing helpful information.

# THE HYDROGEN ATOM ACCORDING TO WAVE MECHANICS – IV. SPHEROCONICAL COORDINATES

J. F. Ogilvie[*]

Centre for Experimental and Constructive Mathematics, Department of Mathematics, Simon Fraser University, Burnaby, British Columbia V5A 1S6 Canada

Escuela de Química, Universidad de Costa Rica, Ciudad Universitaria Rodrigo Facio, San Pedro de Montes de Oca, San José, 11501-2060 Costa Rica

**Abstract**

In this fourth of five parts in a series, the Schroedinger equation is solved in spheroconical coordinates to yield amplitude functions that enable accurate plots of their surfaces to illustrate the variation of shapes and sizes with quantum numbers $k$, $l$, κ, for comparison with the corresponding plots of amplitude functions in coordinates of other systems. These amplitude functions directly derived have the unique feature of being prospectively only real, with no imaginary part.

**Resumen**

En este cuarto artículo de una serie de cinco, se resuelve la ecuación de Schrödinger en coordenadas esferocónicas para producer funciones de amplitude que facilitan gráficos exactos de sus superficies para ilustrar la variación de formas y tamaños con los números cuánticos $k$, $l$, *κ*, para comparación con los gráficos correspondientes de funciones de amplitus en coordenadas de otros sistemas. Estas funciones de amplitude derivadas directamente tiene la característica única de ser solo prospectivamente real, sin ninguna parte imaginaria.

**Key words:** hydrogen atom, wave mechanics, spheroconical coordinates, orbitals, atomic spectra

**Palabras clave:** átomo de hidrógeno, mecánica de onda, coordenadas esferocónicas, orbitales, espectro atómico.

## I. INTRODUCTION

Schroedinger founded wave mechanics with four papers under a collective title in English translation [1] *Quantisation as a problem of proper values*, with auxiliary essays and lectures. In the first and third papers of that sequence, Schroedinger calculated the energies of the hydrogen atom in discrete states according to the solution of his partial-differential equation in coordinates in two systems -- spherical polar and paraboloidal, respectively. The former might be primarily appropriate to an isolated hydrogen atom subject to no external influence, so without breaking symmetry O₄ (also written as O(4)) of that atom, whereas the primary purpose of the latter coordinates was to facilitate the calculation of the influence of an externally applied electric field according to the linear Stark effect, which breaks that O₄ symmetry. Although a confirmation that a separation of coordinates in a molecular context is practicable also in ellipsoidal coordinates in an application to H$_2^+$ followed shortly [2] after Schroedinger's original work, a half century passed before the analogous recognition of spheroconical coodinates [3]. Of coordinates in those four systems in which Schroedinger's temporally independent equation is separable, the amplitude functions in ellipsoidal coordinates have as limiting cases the corresponding amplitude functions in either spherical polar coordinates, as distance $d$ between the two centres of the ellipsoidal system tends to zero, or paraboloidal coordinates, as $d \to \infty$. In all three systems, one common coordinate $\phi$ is the equatorial angle between a half-plane containing polar axis $z$ and the projection of a given point $(x,y,z)$ and a reference half-plane also containing the polar axis, so as to define a half-plane extending from that polar axis; associated equatorial quantum number $m$ is correspondingly common to these three systems. In the fourth system that we describe as spheroconical coordinates (called also spheroconal), that equatorial angular coordinate $\phi$ is, in contrast, no longer a member of the set; this system is hence distinct from the other three systems in that regard, but retains a radial distance $r$ in common with spherical polar coordinates.

---

[*] Corresponding author: ogilvie@cecm.sfu.ca

In this part IV of a series of articles devoted to the hydrogen atom with its coordinates separable in four systems, we state the temporally independent partial-differential equation in spheroconical coordinates and its direct solution, for the first time, and provide plots of selected amplitude functions as surfaces corresponding to an appropriately chosen value of amplitude; no plot of an explicit spheroconical amplitude function is previously reported. As the dependence on time occurs in the same manner in all systems of coordinates in which the Schroedinger equation is separable, we accept the results from part I [4] and avoid that repetition. Although the equations governing the form of the amplitude functions are here, of necessity, defined in coordinates according to a spheroconical system, we view the surfaces of these amplitude functions invariably in rectangular cartesian coordinates: a computer procedure (in *Maple*) translates effectively from the original system in which the algebra and calculus are performed to the system to which a human eye is accustomed.

## II.    SCHROEDINGER'S EQUATION IN SPHEROCONICAL COORDINATES

Among the three coordinates for three spatial dimensions, we define two right elliptical cones, each with two nappes, orientated about axes $x$ and $z$ that complement a radial distance $r$ from the origin, as presented with surfaces of constant values of these coordinates in figure 1; each nappe must have an elliptical cross section perpendicular to its respective axis. Such elliptical cones might be considered to be limiting geometric cases of paraboloids that occur in the paraboloidal coordinates or the hyperboloid that occurs in the ellipsoidal coordinates. These spheroconical coordinates $\xi, r, \eta$ are related to cartesian coordinates $x, y, z$ as follows:

$$x = \frac{r\sqrt{(b^2+\xi^2)(b^2-\eta^2)}}{b}, \quad y = \frac{r\,\xi\,\eta}{a\,b}, \quad z = \frac{r\sqrt{(a^2-\xi^2)(a^2+\eta^2)}}{a}$$

the use here of $\xi$ and $\eta$ as symbols for dimensionless coordinates is not to be confused with those same symbols to denote coordinates in the ellipsoidal system. The domains of these coordinates are $-a \le \xi \le a$, $0 \le r < \infty$, $-b \le \eta \le b$. To conform to a requirement that $a^2 + b^2 = 1$ that enables the separation of coordinates in Schroedinger's partial-differential equation, we set $a = b = 1/\sqrt{2}$; the square root of the sum of the squares of the coordinates according to the above definitions, i.e. $(x^2 + y^2 + z^2)^{½} = r$, simplifies the form of the electrostatic potential energy. Required in integrals over volume, the jacobian of the transformation of coordinates is

$$\frac{4\, r^2\, (\eta^2 + \xi^2)}{\sqrt{(-4\, \xi^4 + 1)\,(-4\, \eta^4 + 1)}}\,.$$

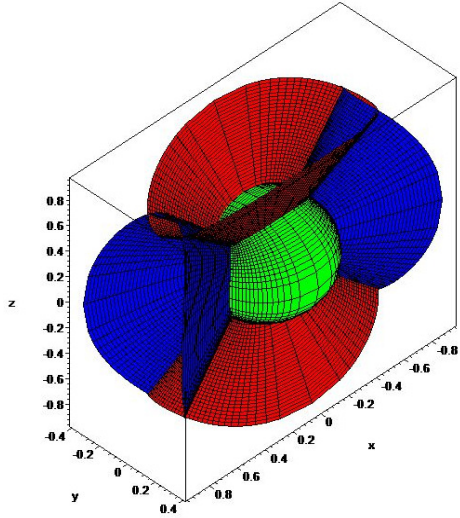

**FIGURE 1.** Definition of spheroconical coordinates $\xi$, $r$, $\eta$: a surface of a double elliptical cone (red), opening along positive and negative axis $z$, has $\xi = ¼$ and its apices at the origin; a surface of a sphere (green) has its centre at the origin and radius $r = 2/5$ units; a surface of another double elliptical cone (blue), opening along positive and negative axis $x$, has $\eta = ¼$ and its apices at the origin.

After separation of the coordinates of the centre of mass of the H atom, Schroedinger's temporally independent equation in explicit SI units contains within terms on the left side an electrostatic potential energy, proportional to $r^{-1}$, and first and second partial derivatives of an assumed amplitude function $\psi(\xi,r,\eta)$ with respect to spatial coordinates $\xi$, $r$, $\eta$ within hamiltonian operator $H(\xi,r,\eta)$; the right side comprises a product of energy as parameter $E$ independent of ccordinates with the same amplitude function, so that the entire equation resembles an eigenvalue relation expressed as $H(\xi,r,\eta)\,\psi(\xi,r,\eta) = E\,\psi(\xi,r,\eta)$.

$$-h^2 \left( -\frac{\xi^4 \left( \frac{\partial^2}{\partial \xi^2} \psi(\xi, r, \eta) \right)}{r^2 (\eta^2 + \xi^2)} + \frac{\eta^2 \left( \frac{\partial^2}{\partial r^2} \psi(\xi, r, \eta) \right)}{\eta^2 + \xi^2} + \frac{\xi^2 \left( \frac{\partial^2}{\partial r^2} \psi(\xi, r, \eta) \right)}{\eta^2 + \xi^2} \right.$$

$$\left. -\frac{\eta^4 \left( \frac{\partial^2}{\partial \eta^2} \psi(\xi, r, \eta) \right)}{r^2 (\eta^2 + \xi^2)} - \frac{2 \left( \frac{\partial}{\partial \xi} \psi(\xi, r, \eta) \right) \xi^3}{r^2 (\eta^2 + \xi^2)} + \frac{2 \left( \frac{\partial}{\partial r} \psi(\xi, r, \eta) \right) \eta^2}{r (\eta^2 + \xi^2)} \right.$$

$$+\frac{2\left(\frac{\partial}{\partial r}\psi(\xi,r,\eta)\right)\xi^2}{r(\eta^2+\xi^2)} - \frac{2\left(\frac{\partial}{\partial \eta}\psi(\xi,r,\eta)\right)\eta^3}{r^2(\eta^2+\xi^2)} + \frac{\frac{\partial^2}{\partial \xi^2}\psi(\xi,r,\eta)}{4\,r^2(\eta^2+\xi^2)} + \frac{\frac{\partial^2}{\partial \eta^2}\psi(\xi,r,\eta)}{4\,r^2(\eta^2+\xi^2)}\Bigg)$$

$$\Big/\,(8\,\pi^2\,\mu) - \frac{Z\,e^2\,\psi(\xi,r,\eta)}{4\,\pi\,\varepsilon_0\,r} = E\,\psi(\xi,r,\eta)$$

Apart from fundamental physical constants electric permittivity of free space $\varepsilon_0$, Planck constant $h$ and protonic charge $e$, there appear parameters $Z$ for atomic number – $Z = 1$ for H – and $\mu$ for the reduced mass of the atomic system, practically equal to the electronic rest mass $m_e$, apart from energy $E$ that is absent from the temporally dependent Schroedinger equation in these same coordinates. After separation of the variables in the partial-differential equation to produce a product of functions of each a single variable,

$$\psi(\xi,r,\eta) = \Xi(\xi)\,R(r)\,H(\eta)$$

and solution of the three consequent ordinary-differential equations including definition of the integration constants, the full solution of the above equation has this form [5],

$$\psi(\xi,r,\eta) = c\,N\sqrt{\frac{Z\,k!}{(1+2\,l+k)!\,a_0}}\left(\frac{2\,Z}{a_0\,(k+l+1)}\right)^{(l+1)} r^l\,e^{\left(-\frac{Z\,r}{a_0\,(k+l+1)}\right)}$$

$$\mathrm{LaguerreL}\left(k, 2\,l+1, \frac{2\,Z\,r}{a_0\,(k+l+1)}\right)\sqrt{1-2\,\xi^2}$$

$$\mathrm{HeunG}\left(-1, \kappa+\frac{1}{4}, 1+\frac{l}{2}, -\frac{l}{2}+\frac{1}{2}, \frac{1}{2}, \frac{1}{2}, -2\,\xi^2\right)\sqrt{1-2\,\eta^2}$$

$$\mathrm{HeunG}\left(-1, -\kappa+\frac{1}{4}, 1+\frac{l}{2}, -\frac{l}{2}+\frac{1}{2}, \frac{1}{2}, \frac{1}{2}, -2\,\eta^2\right)\bigg/(k+l+1)$$

that has become simplified on incorporating Bohr radius $a_0$,

$$a_0 = \frac{\varepsilon_0\,h^2}{\pi\,m_e\,e^2},$$

to contain other constants and parameters in a compact manner. That solution is formally normalized such that

$$\int \psi(\xi,r,\eta)^*\,\psi(\xi,r,\eta)\,\mathrm{d}vol = 1,$$

in which d*vol* is a volume element incorporating the jacobian specified above. Coefficient $N$ is a normalizing factor, to be evaluated, to take into account that factors $\Xi(\xi)$ and $H(\eta)$ of $\psi(\xi,r,\eta)$ are not separately normalized, unlike factor $R(r)$. Coefficient $c$ that equals a complex number of modulus unity such as a fourth root of unity – $c = \pm 1, \pm\sqrt{-1}$, appears because Schroedinger's equation is linear and homogeneous, or equally because that temporally independent equation has the form of an eigenvalue relation, as shown above. The conventional choice $c = 1$ – a choice that is arbitrary and lacks physical justification – imposes that solutions $\psi(\xi,r,\eta)$ as amplitude functions from Schroedinger's temporally independent equation appear in an entirely real form because a general Heun function, denoted HeunG, includes here no imaginary part; with a mathematically valid alternative choice $c = \pm\,\mathbf{i}$, amplitude functions would be entirely imaginary, thus alien to physical space, or $c = -1$ would merely reverse the phase of the amplitude function. This direct solution of the differential equations hence contains general Heun functions of coordinates both $\xi$ and $\eta$ appearing as their squares, not previously suggested to be applicable in this context; Lamé

J. F. OGILVIEJ. F. OGILVIE

polynomials, of which products form ellipsoidal harmonics, have been instead mentioned [*3*], although no explicit formula was ever provided. The ellipsoidal harmonics in these spheroconical coordinates replace the spherical harmonics of spherical polar coordinates. Lamé's differential equation corresponds to a special case of the Heun differential equation with particular relations between the parameters; in the solution of the Heun differential equation, the fifth and sixth arguments in the general, or non-confluent, Heun function equal ½ for this particular Lamé case, as exhibited above. Parameters that appear in the solution but not the partial-differential equation take discrete variables, imposed by boundary conditions, as follows: radial quantum number *k* and azimuthal quantum number *l* appear in generalized Laguerre function R(*r*) in exactly the same form as in spherical polar coordinates because, after the separation of variables, the ordinary-differential equation that governs R(*r*) is exactly the same in both spherical polar and spheroconical coordinates. There is no constraint on the relative values of quantum numbers *k* and *l*, each of which assumes values of non-negative integers. Another quantity κ occurs in one of seven arguments of the general Heun function of each coordinate ξ and η; although these two coordinates have, by design, the same domain, specifically $-1/\sqrt{2} \ldots 1/\sqrt{2}$, κ occurs in distinct forms in those two arguments: κ + ¼ for coordinate ξ, and κ − ¼ for coordinate η. The energy depends on only *k* and *l*, hence *n* = *k* + *l* + 1 as for spherical polar coordinates; in the absence of an external field imposed on a hydrogen atom, the energy is thus independent of κ in spheroconical coordinates, as proved by calculations with varied κ, similarly to a lack of dependence on *m* in spherical polar coordinates [*4*].

### III. GRAPHICAL REPRESENTATIONS OF AMPLITUDE FUNCTION ψ(ξ,*r*,η)

As these amplitude functions ψ(ξ,*r*,η) in spheroconical coordinates were entirely unknown in an explicit algebraic form before this work, we here provide several instances of their nature and form, represented as surfaces in three spatial dimensions for ψ set equal to a particular value, analogously to the presentation of amplitude functions in other systems of coordinates in three preceding parts of this series of papers. All these functions contain general Heun functions that fail to simplify to an explicit algebraic structure when particular values of parameters are specified, but they might be converted approximately to polynomials through formation of Taylor series. The latter practice is useful primarily because the domain of each of ξ and η is finite; calculations, such as plots or integrations, involving these functions are thus implemented with such expansions within *Maple* to attain a satisfactory accuracy.

The formula for ψ(ξ,*r*,η) for the state of least energy specified with quantum numbers *k* = *l* = κ = 0, with Z = 1 assumed here and in each following formula, is thereby expressed as

$$\psi_{0,0,0} = \frac{58720}{70219} e^{\left(-\frac{r}{a_0}\right)} \sqrt{1 - 2\xi^2} \sqrt{1 - 2\eta^2} \, \text{HeunG}\left(-1, \frac{1}{4}, \frac{1}{2}, 1, \frac{1}{2}, \frac{1}{2}, -2\xi^2\right)$$

$$\text{HeunG}\left(-1, \frac{1}{4}, \frac{1}{2}, 1, \frac{1}{2}, \frac{1}{2}, -2\eta^2\right) \Big/ a_0^{(3/2)}$$

of which a surface for a particular value, $\psi_{0,0,0}$ = 0.008 $a_0^{-3/2}$, yields a shape shown in figure 2. In generating that formula for arguments in a particular set, a simplification is automatically incorporated involving the third and fourth arguments; there is hence no inconsistency between the general formula above and its particular representation here. In the absence of an explicit algebraic formula for the normalizing factor, we apply a numerical method that yields the numerical coefficient specified within this, and other, formula in rational form. In all plots of these surfaces of amplitude functions formed in spheroconical coordinates, the distance scale has unit $a_0$ = 5.2917721067x10$^{-11}$ m; the value of the surface of $\psi(\xi,r,\eta)$ in each figure is 1/100 of the maximum value of $\psi(\xi,r,\eta)/a_0^{3/2}$ so that the corresponding volume of $\psi(\xi,r,\eta)^2$ encloses about 0.995 of the total electronic charge density. Because of the presence of factors $(1 - 2\xi^2)^{1/2}$ and $(1 - 2\eta^2)^{1/2}$ in each amplitude function, each such square root must be accommodated in both its positive and negative signs. The symmetric patterns observable in the plots reflect also the presence of $\xi$ and $\eta$ in the seventh arguments of the Heun functions appearing as squares. The surface of each spheroconical amplitude function $\psi(k,l,\kappa)$ must hence be plotted as four separate segments; a gap between each two segments results from the fact that calculation of the general Heun functions becomes slow when $\xi$ or $\eta$ is near either bound at $\pm 1/\sqrt{2}$, necessitating making the magnitudes of the bounds of these variables in the plot slightly less than $1/\sqrt{2}$. Despite the presence of Bohr radius $a_0$ that is an atomic unit, the use of SI units is maintained throughout: $a_0$ serves as merely a scaling factor.

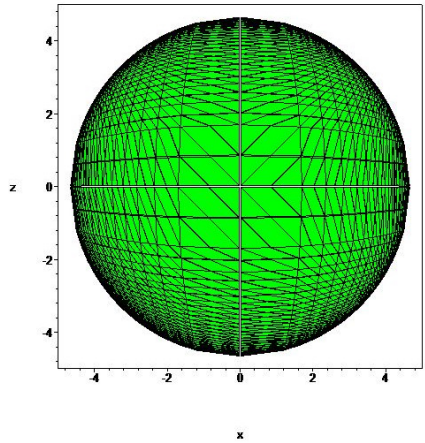

**FIGURE 2.** Surface of real spheroconical amplitude function $\psi_{0,0,0}$ = 0.008 $a_0^{-3/2}$; the distance scale here and in succeeding plots has unit $a_0$ = 5.2917721067x10$^{-10}$ m.

The surface of $\psi_{0,0,0}$ = 0.008 $a_0^{-3/2}$ has hence a rigorously spherical shape; its radius is 4.72 $a_0$ = 2.47x10$^{-10}$ m, exactly the same as for the surface of spherical polar amplitude function $\psi_{0,0,0}(r,\theta,\phi)$ or paraboloidal amplitude function $\psi_{0,0,0}(u,v,\phi)$ according to a corresponding criterion.

The variation of the shape of the surface of $\psi(k,l,\kappa)$ with $\kappa$ is of particular interest because of the novelty of the present solution of the Schroedinger equation in spheroconical coordinates that uniquely contains this particular quantum parameter. For these amplitude functions with varied $\kappa$, the numerical normalizing factor for $\psi_{k,l,-\kappa}$ is the same as that of $\psi_{k,l,+\kappa}$ and is independent of the value of $k$. This formula for $\psi_{0,0,1}$,

$$\psi_{0,0,1} = \frac{12441}{30725} e^{\left(-\frac{r}{a_0}\right)} \sqrt{1-2\xi^2} \sqrt{1-2\eta^2} \, \text{HeunG}\left(-1, \frac{5}{4}, \frac{1}{2}, 1, \frac{1}{2}, \frac{1}{2}, -2\xi^2\right)$$
$$\text{HeunG}\left(-1, \frac{-3}{4}, \frac{1}{2}, 1, \frac{1}{2}, \frac{1}{2}, -2\eta^2\right) \Big/ a_0^{(3/2)}$$

yields a surface presented in figure 3; this surface is symmetric across planes $x = 0$, $y = 0$ and $z = 0$, and has planar nodal surfaces containing axis $z$. The positive lobes along axis $x$ have the shapes of two spheroids that become pointed at the origin. The negative lobe resembles an elliptical torus around axis $x$, extending almost to the origin to separate the two positive lobes. The extent of the lobes parallel to axes $x$ and $y$ is about 13 $a_0$, but only 9 $a_0$ along axis $z$. The surface of the square of this amplitude function, which has accordingly only a positive phase, has a similar shape and size. The corresponding surface of $\psi_{0,0,-1}$ has a form similar to that of $\psi_{0,0,1}$, but its extent perpendicular to plane $x = 0$ is less than across planes $y = 0$ and $z = 0$; its positive lobes lie along axis $z$ and its negative lobe resembles an elliptical torus around $z$.

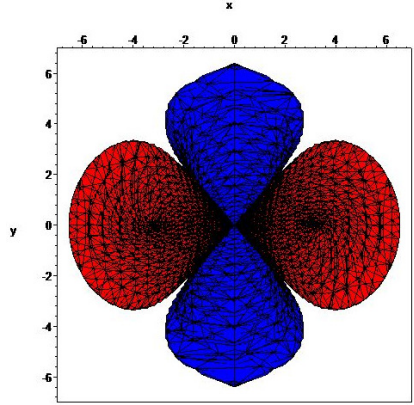

**FIGURE 3.** Surface of real spheroconical amplitude function $\psi(0,0,1) = 0.0041$ $a_0^{-3/2}$, cut open to reveal the diagonal nodal surfaces; the positive lobes (red) extend along axis $y$ and the negative lobe like an elliptical torus (blue) is perpendicular to axis $y$.

Spheroconical amplitude function $\psi_{0,0,2}$ conforms to this formula,

$$\psi_{0,0,2} = \frac{2783}{21438} e^{\left(-\frac{r}{a_0}\right)} \sqrt{1 - 2\xi^2} \sqrt{1 - 2\eta^2} \; \text{HeunG}\left(-1, \frac{9}{4}, \frac{1}{2}, 1, \frac{1}{2}, \frac{1}{2}, -2\xi^2\right) \text{HeunG}\left(-1, \frac{-7}{4}, \frac{1}{2}, 1, \frac{1}{2}, \frac{1}{2}, -2\eta^2\right) \Big/ a_0^{(3/2)}$$

and presents a surface in figure 4 in which there are again planar nodal surfaces through the origin that separate the positive and negative lobes; these lobes are symmetric across that origin, but the negative lobe is an elliptical torus around axis $y$; its cross section in plane $z = 0$ is larger than the cross section of the positive lobes, in contrast with the respective lobes of $\psi_{0,0,1}$. The corresponding surface of $\psi_{0,0,-2}$ has a similar form but with its positive lobes along axis $z$; the negative lobe resembles a flattened torus also about axis $y$, separating the positive lobes along axis $z$; its thickness parallel to axis $x$ is thus less than that parallel to axes $y$ and $z$.

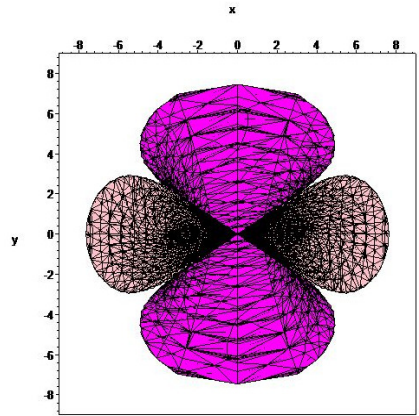

**FIGURE 4.** Surface of real spheroconical amplitude function $\psi_{0,0,2}$ = 0.0013 $a_0^{-3/2}$, cut open to reveal the diagonal nodal surfaces; the positive lobes (pink) extend along axis $x$ and the negative lobe (magenta) is a torus about axis $y$.

Figure 5 shows a surface of real spheroconical amplitude function $\psi_{0,0,3}$ that conforms to this formula.

$$\psi_{0,0,3} = \frac{36998}{561969} e^{\left(-\frac{r}{a_0}\right)} \sqrt{1-2\xi^2} \sqrt{1-2\eta^2} \, \text{HeunG}\left(-1, \frac{13}{4}, \frac{1}{2}, 1, \frac{1}{2}, \frac{1}{2}, -2\xi^2\right)$$
$$\text{HeunG}\left(-1, \frac{-11}{4}, \frac{1}{2}, 1, \frac{1}{2}, \frac{1}{2}, -2\eta^2\right) \Big/ a_0^{(3/2)}$$

This surface exhibits three nodal surfaces: two such surfaces resemble hyperboloids oriented separating the spheroidal positive lobes, and plane $x = 0$ constitutes a third nodal surface. The single torus of $\psi(0,0,1)$ or $\psi(0,0,2)$ here is split into positive and negative lobes with the planar nodal surface between them. The surface of $\psi_{0,0,3}^2$ has a similar shape and size. The corresponding surface of real spheroconical amplitude function $\psi_{0,0,-3}$ has three analogous nodal surfaces and each lobe is symmetric across plane $x = 0$.

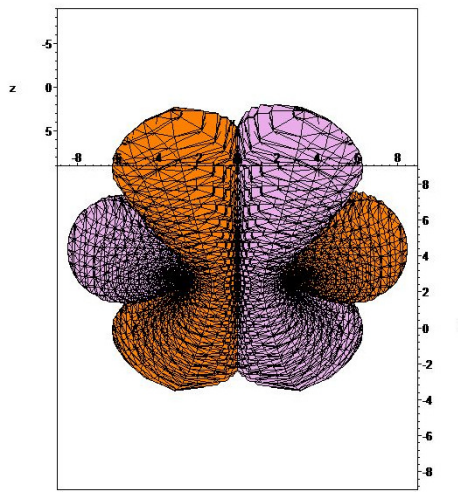

**FIGURE 5.** Surface of real spheroconical amplitude function $\psi_{0,0,3}$ = 0.00065 $a_0^{-3/2}$; the positive lobes (coral) are symmetrically related to the negative lobes (plum) across plane $x$=0 with reversed phase.

Figure 6 presents a surface of real spheroconical amplitude function $\psi_{0,0,4}$ that conforms to this formula.

$$\psi_{0,0,4} = \frac{2249}{57376} e^{\left(-\frac{r}{a_0}\right)} \sqrt{1 - 2\xi^2} \sqrt{1 - 2\eta^2} \, \text{HeunG}\left(-1, \frac{17}{4}, \frac{1}{2}, 1, \frac{1}{2}, \frac{1}{2}, -2\xi^2\right)$$
$$\text{HeunG}\left(-1, \frac{-15}{4}, \frac{1}{2}, 1, \frac{1}{2}, \frac{1}{2}, -2\eta^2\right) \Big/ a_0^{(3/2)}$$

There are two positive lobes that extend along axis $x$ from a point at the origin, and three lobes resembling tori about axis $y$, of which a positive toroidal lobe separates two negative toroidal lobes; all lobes are symmetric with respect to plane $z = 0$. The surface has an extent greater parallel to axes $x$ and $y$ than parallel to axis $z$. Real spheroconical amplitude function $\psi_{0,0,-4}$ has a similar shape and size; its toroidal lobes are also perpendicular to axis $y$, but its extension parallel to axis $x$ is less than in the other two directions.

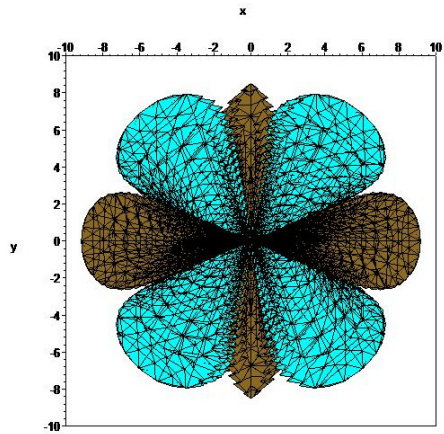

**FIGURE 6.** Surface of real spheroconical amplitude function $\psi_{0,0,4} = 0.00039 \, a_0^{-3/2}$, cut open to reveal the four nodal surfaces that all cross the origin; the positive lobes (sienna) extend along axis $x$; two negative tori (cyan) with one positive torus (sienna) between them surround axis $y$.

As a first exhibit of the shape of a surface of a spheroconical amplitude function with quantum number $l > 0$, figure 7 shows first a surface of $\psi_{0,1,0}$, which conforms to this formula.

$$\psi_{0,1,0} = \frac{112337}{2139648} \sqrt{6} \, r \, e^{\left(-\frac{r}{2a_0}\right)} \sqrt{1 - 2\xi^2} \sqrt{1 - 2\eta^2} \, \text{HeunG}\left(-1, \frac{1}{4}, 0, \frac{3}{2}, \frac{1}{2}, \frac{1}{2}, -2\xi^2\right)$$
$$\text{HeunG}\left(-1, \frac{1}{4}, 0, \frac{3}{2}, \frac{1}{2}, \frac{1}{2}, -2\eta^2\right) \Big/ a_0^{(5/2)}$$

For the particular surface of spheroconical amplitude function $\psi_{0,1,0}$ depicted in figure 7, the shape is roughly an oblate spheroid; the maximum extent in direction $x$ or $z$ is about 10.8 $a_0$, but only 9.4 $a_0$ in direction $y$. Only one lobe is discernible, corresponding to a positive phase of $\psi_{0,1,0}$; there is no nodal plane. As the amplitude function contains a factor $r$, the function has zero amplitude at the origin of the coordinate system and hence formally an inner surface at which $\psi_{0,1,0} = 0.00093 \, a_0^{-3/2}$, but its radius is too small to appear even when the surface is cut open. The surface of $\psi_{0,1,0}^2$ resembles that of $\psi_{0,1,0}$ in figure 7, also having an oblate spheroidal shape with minor axis $y$.

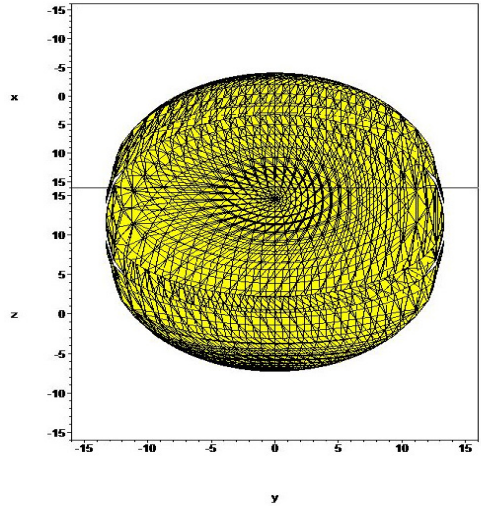

**FIGURE 7.** Surface of real spheroconical amplitude function $\psi_{0,1,0} = 0.00093\ a_0^{-3/2}$

Figure 8 shows the surface of $\psi_{0,2,0}$, which has this algebraic form.

$$\psi_{0,2,0} = \frac{8393}{6598665} \sqrt{30}\ r^2\ e^{\left(-\frac{r}{3a_0}\right)} \sqrt{1-2\xi^2}\ \sqrt{1-2\eta^2}\ \mathrm{HeunG}\!\left(-1, \frac{1}{4}, \frac{-1}{2}, 2, \frac{1}{2}, \frac{1}{2}, -2\xi^2\right)$$
$$\mathrm{HeunG}\!\left(-1, \frac{1}{4}, \frac{-1}{2}, 2, \frac{1}{2}, \frac{1}{2}, -2\eta^2\right) \Big/ a_0^{(7/2)}$$

This surface shows four lobes, of alternating positive and negative phase around axis *y*, directed parallel to axis *y* between nodal planes *x* = 0 and *z* = 0. Its maximum extent parallel to axes *x* and *z* is about 14.3 $a_0$ but parallel to axis *y* only 11.1 $a_0$, so exhibiting an overall roughly oblate spheroidal shape. The shape of this surface strongly resembles the corresponding surface of the imaginary part of $\psi_{0,2,1}(r,\theta,\phi)$ in spherical polar coordinates.

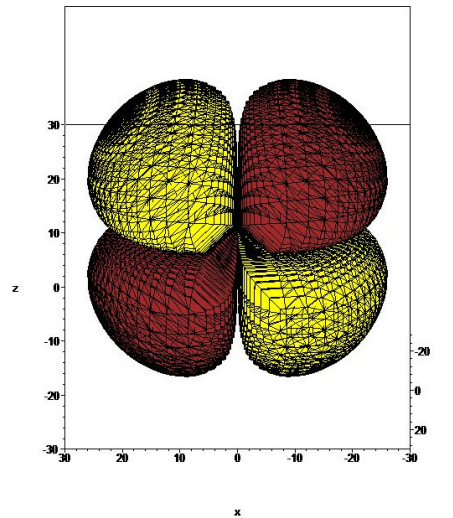

**FIGURE 8**. Surface of real spheroconical amplitude function $\psi_{0,2,0} = 0.00034\ a_0^{-3/2}$ with two positive lobes (yellow) and two negative lobes (brown)

Figure 9 shows the surface of amplitude function $\psi_{0,3,0}$, which has this algebraic form.

$$\psi_{0,3,0} = \frac{7135}{186267648} \sqrt{35}\, r^3\, e^{\left(-\frac{r}{4 a_0}\right)} \sqrt{-2\xi^2 + 1}\, \sqrt{-2\eta^2 + 1}$$

$$\mathrm{HeunG}\left(-1, \frac{1}{4}, -1, \frac{5}{2}, \frac{1}{2}, \frac{1}{2}, -2\xi^2\right) \mathrm{HeunG}\left(-1, \frac{1}{4}, -1, \frac{5}{2}, \frac{1}{2}, \frac{1}{2}, -2\eta^2\right) \Big/ a_0^{(9/2)}$$

Whereas the surfaces of spheroconical amplitude functions $\psi_{0,0,0}$ and $\psi_{0,1,0}$ both have only one lobe and the surface of $\psi_{0,2,0}$ has four lobes, two of positive phase and two of negative phase, according to a phase convention with coefficient $c = 1$, amplitude function $\psi_{0,3,0}$ has ten lobes, four of negative phase along axes $x$ and $z$; of six lobes of positive phase, four lie between planes $xy$ and $yz$ but two are located along axis $y$ on either side of, and remote from, the origin.

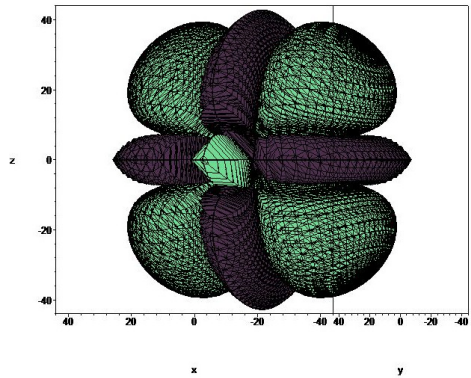

**FIGURE 9.** Surface of real spheroconical amplitude function $\psi_{0,3,0} = 0.00020\, a_0^{-3/2}$. The negative lobes (aquamarine) are directed along axes $x$ and $z$; the positive lobes (violet) lie between those axes with additional small lobes located along axis $y$ on either side of the origin.

Figure 10 shows the surface of amplitude function $\psi_{0,4,0}$, which has this algebraic form.

$$\psi_{0,4,0} = \frac{1226}{2877459375} \sqrt{70}\, r^4\, e^{\left(-\frac{r}{5 a_0}\right)} \sqrt{1 - 2\xi^2}\, \sqrt{1 - 2\eta^2}$$

$$\mathrm{HeunG}\left(-1, \frac{1}{4}, \frac{-3}{2}, 3, \frac{1}{2}, \frac{1}{2}, -2\xi^2\right) \mathrm{HeunG}\left(-1, \frac{1}{4}, \frac{-3}{2}, 3, \frac{1}{2}, \frac{1}{2}, -2\eta^2\right) \Big/ a_0^{(11/2)}$$

Like spheroconical amplitude function $\psi_{0,3,0}$ and unlike function $\psi_{0,2,0}$ that exhibits only four lobes, of its ten lobes function $\psi_{0,4,0}$ has four lobes of positive phase between planes $xy$ and $yz$, and two further and smaller lobes along axis $y$ remote from the origin, but the positive lobes along axis $y$ are larger than the corresponding features of function $\psi_{0,3,0}$; four negative lobes lie along axes $x$ and $z$. The four positive lobes between the axes might appear to be connected across the origin, so separating the negative lobes, but factor $r^4$ in the formula for the amplitude function above imposes zero amplitude at the origin, independent of direction. For amplitude function $\psi_{0,5,0}$ that has no chemical or physical interest and is hence not shown here, the positive lobes along axis $y$ become still larger relative to the other lobes.

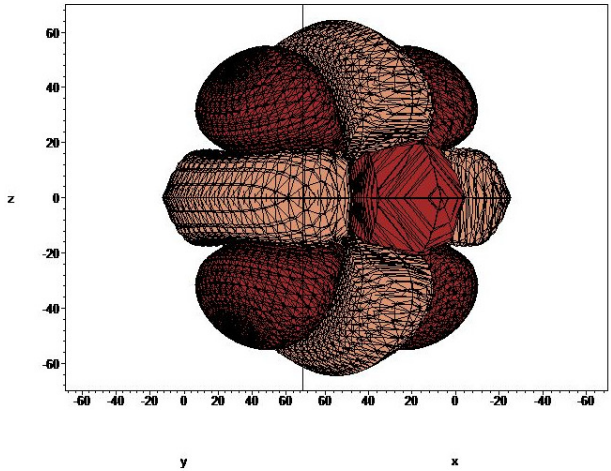

**FIGURE 10.** Surface of real spheroconical amplitude function $\psi_{0,4,0} = 0.000105\ a_0^{-3/2}$. The positive lobes (brown) are directed between axes $x$ and $z$ and the negative lobes (tan) lie along those axes, with additional positive lobes (brown) located along axis $y$ farther from the origin.

Spheroconical amplitude function $\psi_{1,0,0}$ has this formula;

$$\psi_{1,0,0} = \frac{7340}{70219}\sqrt{2}\ e^{\left(-\frac{r}{2a_0}\right)} \sqrt{1-2\xi^2}\ \sqrt{1-2\eta^2}\ (2a_0 - r)$$
$$\text{HeunG}\left(-1, \frac{1}{4}, \frac{1}{2}, 1, \frac{1}{2}, \frac{1}{2}, -2\xi^2\right) \text{HeunG}\left(-1, \frac{1}{4}, \frac{1}{2}, 1, \frac{1}{2}, \frac{1}{2}, -2\eta^2\right) \Big/ a_0^{(5/2)}$$

its surfaces have the shape shown in figure 11. Three concentric spheres display their centres at the origin: one innermost sphere has a positive phase, and an only slightly larger sphere has a negative phase; the latter sphere and the outer sphere demarcate a spherical shell of negative phase. These surfaces strongly resemble those of $\psi_{1,0,0}(r,\theta,\phi)$ in spherical polar coordinates.

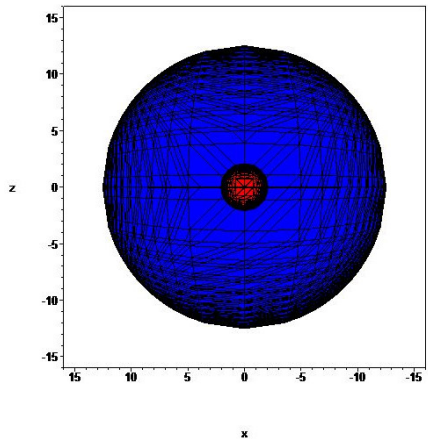

**FIGURE 11.** Surface of real spheroconical amplitude function $\psi_{1,0,0} = 0.0014\ a_0^{-3/2}$ cut open to reveal the interior structure; a small inner positive spherical lobe (red) is surrounded with a thick negative spherical shell (blue).

Spheroconical amplitude function $\psi_{1,1,0}$ has this formula;

$$\psi_{1,1,0} = \frac{112337}{21663936}\sqrt{6}\, r\, e^{\left(-\frac{r}{3a_0}\right)} \sqrt{1-2\xi^2}\sqrt{1-2\eta^2}\,(6a_0-r)$$
$$\text{HeunG}\!\left(-1, \frac{1}{4}, 0, \frac{3}{2}, \frac{1}{2}, \frac{1}{2}, -2\xi^2\right) \text{HeunG}\!\left(-1, \frac{1}{4}, 0, \frac{3}{2}, \frac{1}{2}, \frac{1}{2}, -2\eta^2\right) \bigg/ a_0^{(7/2)}$$

its surfaces of constant $\psi_{1,1,0}$ have the shape exhibited in figure 12. Like the surfaces of amplitude function $\psi_{1,0,0}$ in figure 11, there is a small inner spheroidal surface, nearly spherical and of positive phase, surrounded closely with a surface of negative phase and an outer spheroidal surface, oblate like that of amplitude function $\psi_{0,1,0}$ presented in figure 7, that marks the distance at which the amplitude function decays to the stated value on its way asymptotically to zero.

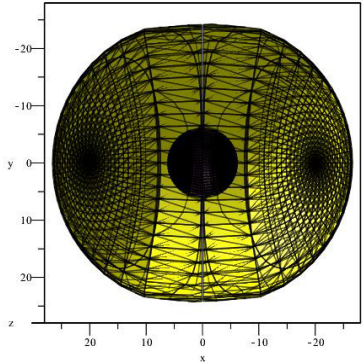

**FIGURE 12.** Surface of real spheroconical amplitude function $\psi_{1,1,0} = 0.00053\, a_0^{-3/2}$, cut open to reveal the interior structure; an inner positive spherical lobe (violet) is surrounded with a thick negative oblate spheroidal shell (yellow).

Surfaces of further real spheroconical amplitude functions $\psi_{k,l,\kappa}$ exhibit features that are predictable on the basis of the figures presented above, specifically that inner spheres numbering $k$ appear within an outer surface resembling that of $\psi_{0,l,\kappa}$.

## IV. DISCUSSION

Like coordinates in three other systems, as specified in three preceding papers in this series, the spheroconical coordinates enable a separation of the variables in the temporally independent Schroedinger equation, with a restriction on variables $\xi$ and $\eta$ that limits each domain to the same interval $-1/\sqrt{2}$ to $+1/\sqrt{2}$ in a unit with no physical dimension. Only radial distance $r$ from the origin is common to another system of coordinates. In conjunction with these three distinct variables, three quantum parameters $k$, $l$ and $\kappa$ characterize these spheroconical amplitude functions and the shapes of their surfaces of $\psi_{k,l,\kappa}$ set at a selected value. Although the amplitude functions in this spheroconical system have uniquely defined algebraic formulae and shapes of their surfaces according to the specified criterion, a transformation of coordinates converts an amplitude function of this system into amplitude functions of a selected other system in an appropriate linear combination, just as illustrated between spherical polar and paraboloidal coordinates in part II of this series [7]. The existence of such a general transformation in no way makes one system, and the amplitude functions expressed therein, objectively superior or preferable to another system and its particular functions. The great advantage of amplitude functions in this spheroconical system is that, with coefficient $c = 1$, all formulae are *real* – i.e. have no imaginary part, so that their full surfaces can be presented directly in real space, as figures 2 – 12 demonstrate emphatically. Amplitude functions $\psi_{k,l,\kappa}(\xi,r,\eta)$ beyond those depicted in the eleven specified figures show inner spheroidal surfaces directly analogous to those of surfaces of $\psi_{k,l,m}(r,\theta,\phi)$ or amplitude functions in the two other systems of coordinates, just as the surfaces in figures 11 and 12 transcend those of figures 2 and 7.

Regarding spheroconical amplitude functions $\psi_{0,0,\kappa}$, their size increases slightly with increasing $\kappa$, and the number of nodal surfaces tends also to increase, although not from $\psi_{0,0,1}$ to $\psi_{0,0,2}$. In all cases there exist axes of symmetry two fold along the coordinate axes, which reflects the dual axes about which the double cones of coordinates $\xi$ and $\eta$ locate; for the same reason, the planes of symmetry for the thinner extents are $z = 0$ for $\psi_{0,0,+\kappa}$ and $x = 0$ for $\psi_{0,0,-\kappa}$. Other planes of symmetry are generally also present. All lobes of $\psi_{0,0,\kappa}$ with $\kappa > 0$ begin at one of these two axes and have zero magnitude at that axis.

For spheroconical amplitude functions $\psi_{0,l,0}$, in contrast their size increases rapidly with increasing quantum number $l$, like the size of functions $\psi_{k,0,0}$, and according to the same property: the energy of such an amplitude function increases proportionally to $-1/(k + l + 1)^2$, in which quantum numbers $k$ and $l$ appear on an equivalent basis.

Because quantum numbers $k$ for radial and $l$ for angular momentum are precisely defined for amplitude functions in spheroconical coordinates, the spectrometric states conventionally expressed in terms of these quantum numbers are defined with respect to these coordinates equally as well as in spherical polar coordinates. Explicitly, the designation of a spectrometric state of the hydrogen atom not subject to an externally applied field is conventionally based on such a value of the quantum number for angular momentum – S states for $l = 0$, P states for $l = 1$, D states for $l = 2$ et cetera; energy quantum number $n$ is directly included in such a designation as $n\,l$ and all states of the hydrogen atom are doublet states when the intrinsic angular momentum of the electron is taken into account, to yield a term symbol such as 1 $^2$S, 2 $^2$S, 2 $^2$P et cetera. The presence of $\kappa$ for spheroconical coordinates instead of equatorial quantum number $m$ for spherical polar coordinates has no implication for this nomenclature.

Cook and Fowler [6] sought to explore solutions of the temporally independent Schroedinger equation for the hydrogen atom in spheroconical coordinates in terms of Lamé functions of the first and second kind, but produced neither an explicit formula for a spheroconical amplitude function nor a plot thereof.

**V     CONCLUSION**

Amplitude functions in spheroconical coordinates have several attractive features: they are entirely real and hence lend themselves directly to a physical depiction in cartesian coordinate space, they are associated with integer values of quantum numbers $k$ and $l$ that define, in a sum with unity, the energy, and spectral states are readily associated with those quantum numbers. Like their real counterparts in spherical polar coordinates, they lack a particular directional character, having mostly an overall oblate spheroidal shape. Further assessment of their character and a comparison with the amplitude functions in other systems of coordinates appears in part V of this series of articles.

# THE HYDROGEN ATOM ACCORDING TO WAVE MECHANICS – V IMPLICATIONS OF MULTIPLE COORDINATE SYSTEMS

*J. F. Ogilvie**

Centre for Experimental and Constructive Mathematics, Department of Mathematics, Simon Fraser University, Burnaby, British Columbia V5A 1S6 Canada
Escuela de Química, Universidad de Costa Rica, Ciudad Universitaria Rodrigo Facio, San Pedro de Montes de Oca, San José, 11501-2060 Costa Rica

**Abstract**

Because Schroedinger's temporally dependent or independent partial-differential wave equation for the hydrogen atom is solvable in spatial coordinates in four distinct systems, the properties of those solutions, and even the quantum numbers in sets that characterize those wave or amplitude functions, are parochial to each such system. The relations among the wave functions of the hydrogen atom, wave mechanics and molecular structure are discussed. Quantum mechanics is argued to be largely irrelevant in chemical education.

**Resumen**

La ecuación de onda parcial-diferencial temporalmente dependiente o independiente de Schroedinger para el átomo de hidrógeno se puede solucionar en coordenadas espaciales en cuatro sistemas distintos. Sin embargo, las propiedades de dichas soluciones e incluso los números cuánticos en conjunto que caracterizan esas funciones de onda o amplitud son muy propias a cada sistema. Por lo anterior, se discuten las relaciones entre las funciones de onda del átomo de hidrógeno, la mecánica ondulatoria y la estructura molecular. Por otro lado, se argumenta que la mecánica cuántica es en gran medida irrelevante en la educación química.

**Key words:** hydrogen atom, wave mechanics, orbitals, atomic spectra

**Palabras clave:** átomo de hidrógeno, mecánica de onda, orbitales, espectro atómico

## I. INTRODUCTION

In London in 1815, an English chemist and physician named William Prout published an hypothesis, based on inaccurate measurements of molar masses of the known chemical elements, that the hydrogen atom was the only truly fundamental object and that atoms of other chemical elements comprised aggregates of hydrogen atoms of varied number [*1*]. In Manchester in 1915, Ernest Lord Rutherford, a physicist who anyhow became Nobel laureate in chemistry for his discovery of the transmutation of chemical elements, concluded from experiments in which he bombarded nitrogen atoms with α particles that perhaps the nuclei of all elements were made of

---

* Corresponding author: ogilvie@cecm.sfu.ca

J. F. OGILVIEhydrogen nuclei; in Cambridge in 1920, he named the atomic nucleus of hydrogen a proton. In Cambridge also in 1920, following his production of the first mass spectrograph in 1919, Francis Aston formulated the *whole-number rule* whereby the masses of atomic isotopes are practically integer multiples of the mass of the hydrogen atom. In Cambridge in 1932, Sir James Chadwick's discovery of the neutron completed a simple interpretation of an atomic nucleus of mass number $A$ as comprising $Z$ protons and $A - Z$ neutrons, supplanting Prout's hypothesis. The spectre of Prout's hypothesis, however, lingers in chemistry in that many calculations, and more numerous qualitative explanations, of atomic and molecular properties are based on an assumption that all electrons in any atom behave according to the central field of the model of atomic hydrogen.

In preceding articles in this sequence on the hydrogen atom according to Schroedinger's wave mechanics, we solved the pertinent partial-differential equation in coordinates of four systems -- spherical polar [2], paraboloidal [3], ellipsoidal [4] and spheroconical [5] -- to yield explicit algebraic formulae [6] for amplitude functions that generate the distribution of density of negative charge associated with one electron in the vicinity of an atomic nucleus of charge $+Z\,e$; atomic number $Z = 1$ for H itself. We distinguish between amplitude functions, which arise from the solution of the temporally independent Schroedinger equation in terms of only three spatial variables, in various systems, and wave functions, resulting from solution of the temporally dependent equation involving three spatial variables and time [7]. In this essay we compare the results for those four coordinate systems for the hydrogen atom and discuss the validity of a contemporary form of Prout's hypothesis.

A treatment of the hydrogen atom in coordinates in any system within quantum mechanics must reproduce the energies of the discrete states as being approximately inversely proportional to the square of a positive integer, generally denoted $n$; $n$ thus becomes an *energy quantum number*. The latter result is a conclusion purely from experiment, specifically the numerological deduction made initially by Balmer in 1885 and elaborated on a more physical basis by Rydberg in 1888 from wave lengths $\lambda$/m of spectral lines in the visible region attributed to atomic hydrogen. Balmer's formula is equivalent to

$$\lambda = B\, n^2/(n^2 - 2^2)$$

in which fitted parameter $B = 3.6456 \times 10^{-7}$ m became known subsequently as the Balmer constant; integer $n$ assumes values 3, 4, 5, 6 for only four lines $\alpha, \beta, \gamma, \delta$ in the so-called Balmer series, respectively. Rydberg deduced a more general formula for these spectral lines, measurable as circular frequency $\nu$ or wave length $\lambda$ in the optical spectrum, equivalent to

$$\Delta E = E_2 - E_1 = R\,h\,c \left( \frac{1}{n_1^2} - \frac{1}{n_2^2} \right) = h\,\nu = h\,c/\lambda$$

containing, with Planck constant $h$ and speed of light $c$, rydberg constant $R$ in wavenumber unit that is the most accurately known fundamental physical constant (as $R_\infty$); that experimentally observable quantity thus corresponds to an energy difference between states characterised with positive integers $n_1$ and $n_2$, with $n_1 < n_2$. These lines are associated with transitions between states of the hydrogen atom, according to the interpretation originally proffered by Nicholson and Ritz. The energies of these discrete states of the hydrogen atom are hence implied to be expressible as

$$E = -R\,h\,c/n^2 + C,$$

in which *C* is a constant that includes all other energy of the atomic system, such as mass energy, that is not significantly involved in a transition between the states that yield an observed spectral line, and that can hence be ignored for our present purpose. Without *C*, the energies are negative because work must be done to remove an electron from a region near a positively charged nucleus. We accordingly view *n* as *an integer quantity that is purely experimentally derived*, bereft of any intrinsic theoretical significance, but which any acceptable theoretical treatment must reproduce. This formula might be the first result of an analysis in quantum physics, and has no inherent connexion, except precursor, to quantum mechanics that it preceded by a few decades. We must, however, expect that any succeeding wave-mechanical derivation of a solution of Schroedinger's equations in coordinates of various systems for discrete states yield parameters, parochial to each treatment, of which an appropriate combination becomes equivalent to that positive integer, *n*.

## II. SOLUTION OF SCHROEDINGER'S EQUATIONS IN FOUR SYSTEMS OF COORDINATES

We summarise in table 1 the results for the four coordinate systems [*2 - 5*] that enable solutions of Schroedinger's temporally independent equation, specifying the coordinates and the respective quantum numbers, with the formula of the combination to express those quantum numbers to correlate with the energy quantum number. Note that the use of ξ and η in both ellipsoidal and spheroconical systems must not be taken to imply a relation of these coordinates between these systems.

**TABLE 1.** System of coordinates and associated quantities

| system | coordinates | quantum numbers | formula for *n* |
|---|---|---|---|
| spherical polar coordinates | *r*, θ, ϕ | *k*, *l*, *m* | *k* + *l* + 1 |
| paraboloidal coordinates | *u*, *v*, ϕ | $n_1$, $n_2$, *m* | $n_1 + n_2 + |m| + 1$ |
| ellipsoidal coordinates | ξ, η, ϕ | $n_\xi$, $n_\eta$, *m* | $n_\xi + n_\eta + |m| + 1$ |
| spheroconical coordinates | ξ, *r*, η | *k*, *l*, κ | *k* + *l* + 1 |

Some coordinates are common to two or three systems, such as equatorial angle ϕ for the former three systems and radial distance *r* for the first and fourth systems; equatorial or magnetic quantum number *m* and radial *k* and azimuthal *l* quantum numbers are correspondingly common to those particular systems. Energy quantum number *n*, which has an indisputable experimental basis as explained above, is likewise expressed as a varied combination of other quantum numbers depending on the system, as indicated in the table above. The shapes and nodal properties of surfaces, but not greatly their sizes for a given value of energy and hence energy quantum number *n*, of amplitude function ψ at a set constant value appropriately chosen analogously vary appreciably with the system of coordinates, although common trends of nodal surfaces are perceptible between the systems, as demonstrated in the figures of the preceding four parts of this series [*2 - 5*].

Of amplitude functions in the four systems of coordinates, which should one choose? As wave mechanics is one method within a collection of such algorithms for calculations on an atomic scale, the choice must depend on the purpose of a calculation on the hydrogen atom, or other atom with only one electron, in which the amplitude functions serve as working formulae. The overwhelmingly best known system comprises, of course, spherical polar coordinates, which are described in Schroedinger's paper simply as polar coordinates [*7*]; as the properties of the Laguerre and Legendre polynomials, also discussed in the third paper of Schroedinger [7], involved therein are highly developed, calculations are generally rapid.  This system is applicable to a hydrogen atom, or to any other atom with only one electron, that is isolated -- no other matter in the vicinity, no applied electric field apart from an electromagnetic wave in the form of light that might interact classically with the atom in absorption, emission or scattering. Practically all textbooks of chemistry allude to these amplitude functions, generally in mistaken contexts; some such textbooks, particularly in physical and inorganic chemistry, describe their properties with accurate formulae but more or less inaccurate figures depicting poorly defined surfaces and shapes. The authors of textbooks on quantum mechanics in physics typically content themselves with the mathematical details of this solution of the temporally independent Schroedinger equation in spherical polar coordinates, and present some exemplary formulae.  Following Schroedinger's own solution of his equation in paraboloidal coordinates [7], some textbooks of quantum mechanics in physics treat also this system, but no known textbook of chemistry even mentions that this system exists for the hydrogen atom.  Common to spherical polar, paraboloidal and spheroconical amplitude functions, Laguerre polynomials, applied for paraboloidal spatial variables both $u$ and $v$, are just as easy to manipulate, and calculations are generally rapid. Schroedinger applied [7] this paraboloidal system of coordinates to treat, with perturbation theory that he concurrently developed, the hydrogen atom in an homogeneous electric field; the purpose was to calculate the Stark effect, explicitly the shifting and splitting of spectral lines as a result of hydrogen atoms being subjected to a uniform electric field [3].  Other contexts of calculations in which these paraboloidal coordinates are particularly useful include the photoelectric effect, the Compton effect and a collision of an electron with a H atom; in each case, a particular direction in space is distinguished according to some external force [8].  In ellipsoidal coordinates, one focus of an ellipsoid is located at or near the atomic nucleus; the other focus, at distance $d$, is merely a dummy location; as the latter can become the location of a second atomic nucleus, the associated amplitude functions become formally applicable to a diatomic molecule, which has been the reason for the attention given to these coordinates [9]. These amplitude functions, derived directly, contain confluent Heun functions [4], which pose difficulties of calculation because they lack a simple polynomial expression.  Some indirect derivations of amplitude functions in the literature [9], through solutions of the differential equations in series, have hence been expressed in terms of polynomials; the shapes of these functions at a particular value of $\psi(\xi,\eta,\phi)$ depend appreciably on that distance $d$ [4].  For all three preceding coordinate systems, equatorial angle $\phi$ is one variable; its presence in a resulting derived amplitude function has invariably this form,

$$\Phi(\phi) = \frac{e^{(i\,m\,\phi)}}{\sqrt{2\,\pi}} = \frac{\cos(m\,\phi) + i\,\sin(m\,\phi)}{\sqrt{2\,\pi}}$$

in which the presence of $\mathbf{i} = \sqrt{-1}$ with equatorial quantum number $m$ dictates generally complex total amplitude functions; their intrinsic real, cosine, and imaginary, sine, parts hence preclude depiction of total surfaces in real space of three dimensions unless $m = 0$. As a further complication, confluent Heun functions in ellipsoidal coordinates $\xi$ and $\eta$ have also an intrinsically complex nature [4].  In contrast, each and every amplitude function in spheroconical coordinates as directly derived is prospectively entirely real [5, 6] -- thus no imaginary part, enabling a direct plot of each such surface.  Calculations with the general Heun functions in two spheroconical coordinates are easier than with confluent Heun functions; the third coordinate is just the separation $r$ between electron and nucleus, the same as in spherical polar coordinates [2]. These spheroconical

THE HYDROGEN ATOM – V IMPLICATIONS OF MULTIPLE COORDINATE SYSTEMS

coordinates have thus much to recommend them for a general discussion of the intrinsic wave-mechanical properties of the hydrogen atom, and should effectively supplant the spherical polar coordinates for this purpose.

The incontestable fact that the shape of a surface of an amplitude function in coordinates of the four specified systems depends on that system has profound implications for chemical or physical interpretations.  The most momentous implication is that not only is any such shape merely an artifact of one particular coordinate system, but even the quantum numbers, as presented in table 1, associated with any such amplitude function are equally artifacts of that system [*6*].  Such shapes are, of course, not entirely independent: an appropriate linear combination of amplitude functions in one system has an algebraic form that is subject to a transformation of coordinates to generate a particular amplitude function in another system corresponding to the same value of energy, and hence quantum number *n*. A shape of a surface of such a combination, at a selected value of ψ, is hence identical with a shape of a particular amplitude function directly derived in another system when plotted in common cartesian coordinates.  Which particular shape or system one might choose must thus be arbitrary; any conclusion in relation to specific properties of the hydrogen atom based arbitrarily on any such particular shape or the pertinent parochial quantum numbers is unwarranted and fallacious. The only constant quantity is the energy quantum number, *n*, which is independent of any system, consequent of its experimental origin as explained above.

About three quarters of all mass in the known universe is composed of hydrogen, in mostly atomic and plasma forms.  Being a system comprising two bodies, that atomic form, supposing point masses, is amenable to an exact mathematical treatment in classical or quantum mechanics, such as that in each of the four preceding parts [*2 – 5*] of essays in this series according to a particular system of coordinates.  The present importance of hydrogen in chemistry is related, however, not to its separate atomic nature, nor even to its incorporation in innumerable chemical compounds of diverse nature; to the contrary, the presumed importance lies in a gratuitous assumption and expectation that the calculated properties of atomic hydrogen, with $Z = 1$, might somehow be directly pertinent to both atoms of other elements, with $Z > 1$, and molecules or materials containing those elements -- virtually Prout's hypothesis. Employing such an assumption amounts to extrapolation from a point, a practice that anybody must agree in isolation to be indefensibly illogical, even insane [*10*].

At this point we recall the distinctions among quantum physics, quantum chemistry and quantum mechanics.  Quantum physics implies experiments or observations on an atomic scale and the principles that arise therefrom; a prototypical instance is the generation of a formula for the energies of the hydrogen atom in states of discrete energy, according to the work of Balmer and Rydberg, as explained above.  The first observation of quantum physics was the discovery by William Wollaston in Cambridge of black lines in the solar spectrum; of these lines, subsequently classified by Fraunhofer, those designated C, F, G', h correspond to lines α, β, γ, δ of the Balmer series, respectively.  Quantum mechanics is recognised [*11, 12*] to imply a collection of methods of calculation, or algorithms, applicable to a system on an atomic scale; among at least twelve such methods [*13, 14*] including relativistic wave mechanics of Dirac, non-relativistic wave mechanics [*6*] has been applied to generate the amplitude functions of the hydrogen atom in the four systems of coordinates presented in preceding parts of this series [*2 - 5*].  Quantum chemistry is generally understood to imply a programmed calculation of electronic structure of molecules or materials with atomic nuclei in more or less fixed relative positions; such calculations made with standard computer programs have been developed to an astonishing degree of sophistication, and have become an established accessory to the exercise of research in practical organic and inorganic chemistry, apart from innumerable separate calculations of atomic and molecular structure and properties of varied scientific worth. Although some practitioners might fancy a description of their work as being *ab initio* -- from first principles, apart from the typically calibrated basis sets, their restriction to electronic motion and electrostatic interactions defines an incontestably *semi-empirical* constituent, a reversion to classical mechanics in which the motion of the atomic nuclei is treated classically, if at all; such a restriction is unnecessary, but its avoidance imposes a cost that a traditional molecular structure is generally precluded as a result [*15*]. A separate treatment of electronic and nuclear motions is nearly invariably based on an approximation resulting from an analysis originally undertaken by Born and Oppenheimer [*16*], and has since been discussed continually and expansively in increasing sophistication. One aspect of that analysis of which a casual user of quantum-chemical programs might be unaware is that the concept of a curve or hypersurface of potential energy, supposed to govern the relative locations and motions of atomic nuclei, is valid only in the immediate vicinity of a point of a local minimum energy, corresponding to a particular molecular structure or conformation; any extrapolation of calculations away from that immediate vicinity must again be deprecated, with such results possibly at great variance with experimental data [*17*].

A crucial component of most computer programs for quantum chemistry is a set of functions, called basis functions, each likely centred on an atomic nucleus at a fixed relative location. In the early years of such calculations, functions of Slater type were used to diminish the effort of manual calculations: these Slater functions resembled amplitude functions of the hydrogen atom in spherical polar coordinates, but had no radial node; they conform to a correct cusp condition at the local origin corresponding to the location of the respective atomic nucleus. When computational resources expanded, the basis functions of form exactly those of the explicit hydrogen functions in spherical polar coordinates became tractable, but major computational efficiency was achieved on replacing each such hydrogen-like function with functions of gaussian type in a small set [*18*], even though the cusp condition at the atomic nucleus was forsaken. Amplitude functions for a molecule treated in such a calculation are prepared as linear combinations of these basis functions, each set on a separate atomic centre. The extent and success of these calculations are phenomenal: the derived molecular structures, with slightly adjusted relative nuclear locations to generate a local minimum of energy, have accuracies generally comparable with determinations from experiments of essentially classical nature -- electron or xray diffraction, for instance; molecular properties, such as electric-dipolar moments and polarizabilities are also generally reproduced in a reasonably satisfactory manner, but the best basis sets to generate a structure might not be the best for particular properties associated with that structure. An alternative approach, still involving protracted numerical calculation based on Schroedinger's equation, relies on the intermediacy of density functionals instead of basis functions that mimic amplitude functions of the hydrogen atom; in this case the spatially dependent densities of electronic charge replace those amplitude functions as quantities to be varied to obtain the best energy of the system. Although in some cases the density functionals are based on amplitude functions of form that of hydrogen, in other cases, for instance [*19*], no such amplitude function is involved. The results from such calculations with density functionals might be less accurate, with reference to experimental quantities, than those directly based on amplitude functions or their gaussian mimics, but, as a compromise with substantially decreased duration and hence cost of those calculations, their accuracy suffices for various purposes, especially with large molecules or aggregates of atomic centres. In contrast, the application of software to implement *molecular mechanics* [*20*] is nearly as effective to calculate a molecular structure and selected properties, at greatly decreased computational cost, and with no pretence of wave-mechanical provenance.

What is critically necessary that one understand about these calculations on systems containing multiple (i.e. more than one) electrons, either separate atoms or their aggregates in molecules or materials, is the distinction between the explicit amplitude functions of the hydrogen, or other one-electron, atom and the selected basis sets that might or might not rely on those functions. The amplitude functions of the hydrogen atom have become traditionally called *orbitals*, a term that Mulliken with characteristic obfuscation invented to signify a mathematical function as near a physical trajectory or orbit, in the context of a Bohr atom, as is possible in wave mechanics [*21*]. Apart from such an atomic orbital, there is a possibility of a molecular orbital, which corresponds to an exact solution of Schroedinger's equation for a system of one electron in the vicinity of two or more atomic nuclei that are accorded fixed relative locations. The shape of a surface of such a molecular orbital at a particular value depends definitively on a conformation or the relative spatial locations of those atomic nuclei; for instance, a surface of a molecular orbital for $H_2^+$ in its state of least energy must have a shape disparate from that of a surface of a respective molecular orbital for $H_3^{2+}$ in its state of least energy. These direct molecular orbitals have little practical interest; for this reason we neglect them. The typical method to generate a molecular orbital for use in systems of multiple electrons and multiple nuclei is to form a linear combination of atomic basis functions, as mentioned above; such a molecular orbital is really a molecular basis function. The crucial point is that one must not confound an *orbital*, which is an amplitude function derived as a *result* of a calculation with Schroedinger's equation for an atom with one electron, i.e., the *output* from such a calculation, with a *basis function* that is an assumed component to enable, and is *within*, a calculation for a system of multiple electrons, i.e. the *input* for the latter calculation; that basis function has no intrinsic meaning apart from that calculation. For these systems of multiple electrons, quantum numbers are no longer uniquely defined, which condition is characteristic of a classical system. For only an atomic system of one electron is the energy defined with a single quantum number, i.e. *n*.

The *aufbauprinzip* -- building-up principle -- that has been, since Bohr in 1921, applied to formulate a supposed electronic configuration of atoms with multiple electrons is another casualty of a recognition that, among the four sets that we have derived for separate systems of coordinates, a choice of quantum numbers for the hydrogen atom is arbitrary, apart from the fact that an atom with multiple electrons in any case suffers the loss of identifiable quantum numbers. As Millikan recognised [*22*] even before the present context arose of multiple sets of quantum numbers to describe the hydrogen atom depending on the coordinate system, the *aufbauprinzip* is an illusion: the periodic chart of the chemical elements is not a theoretical result, but rather the product of experiment not derivable according to any physical or chemical theory, notwithstanding the fact that sufficiently extensive quantum-chemical calculations can, through brute force, reproduce satisfactorily the properties of atoms that might be measurable or supposed to be predictable. Although the diagonal rule of Madelung, about 1926, makes a slight concession to the loss of central symmetry in the presence of multiple electrons in the vicinity of a single atomic nucleus, hence eliminating the degeneracy attributed to quantum number *l*, the principle is still essentially based on an extrapolation from the hydrogen atom. Bohr formulated this rule of thumb, another manifestation of Prout's hypothesis, before the development of quantum mechanics. Nearly a century afterward, there continues naïve and superficial debate about the ordering of some elements in columns of the periodic chart to avoid long rows. The state of an atom, or molecule, is defined purely by its energy and its angular momentum; only changes of energy, with possible associated changes of angular momentum, are observable in transitions between states of an atom,

J. F. OGILVIE

but, unlike the formula of Rydberg presented above for transitions of the hydrogen atom, absolute quantum numbers associated with energy, but not angular momentum, are inevitably undefined.

Electrons are fundamentally indistinguishable: there is no *s* electron, no *p* electron ... in an atom, no σ electron, no π electron, no bonding electron, no lone pair ... in a molecule; there are only electrons [10]. The culprit for the original flagrant violation of this undeniable physical principle was G. N. Lewis, then in Harvard University USA, apparently beginning shortly after the discovery of electrons as individual physical particles by J. J. Thomson in 1897, following a concept by R. Laming published in 1838 and the naming by G. J. Stoney in 1891, all in England. The subsequent promulgations of electron pairs by Kossel and by Lewis and the octet rule and various elaborations by I. Langmuir inspired L. C. Pauling, on the basis of an inadequate understanding of the then new quantum mechanics -- despite his study of mathematical physics during his doctoral research, to develop his ideas about the nature of the chemical bond [23]. As Pauling was a highly effective orator [24], his evangelistic fervour motivated many other authors whose understanding of the physical principles and of the mathematics of the wave-mechanical method was much less than his own; these were gullible scientists or teachers whom Valiunas described as "that sad benighted chemistry professoriate" [25]. Pauling's approach to the application of quantum-mechanical concepts in chemistry was deeply intuitive and qualitative, in contradistinction with the fundamentally mathematical nature of quantum mechanics; for practical purposes, his resonance and valence-bond theories are largely ignored at present, although organic chemists continue blindly to insert various related terms into their qualitative explanations of molecular structure and reactions. With regard to the hydrogen atom, although Pauling and Wilson [26] cited Schroedinger's third article in the series *Quantisation as a Problem of Proper Values* [7] in which the latter author solved the hydrogen atom in paraboloidal coordinates with the respective quantum numbers, Pauling failed to recognise the significance of the fact that a second coordinate system already existed in the fact of this solution, which implies an arbitrary choice between them of both amplitude functions and quantum numbers to describe that hydrogen atom. Likewise, although Pauling and Wilson mentioned matrix mechanics [26] in a book that appeared a few years after Teller confirmed the applicability of ellipsoidal coordinates [27] that Pauling likewise ignored, he failed to recognise that any amplitude function as a solution of Schroedinger's equation is an artifact of one particular method -- wave mechanics -- of many methods within quantum mechanics, not a physically observable quantity. Although Thomson generated a stream of electrons in an evacuated chamber, and with sufficiently sensitive methods might have identified the arrival of an individual, single electron at his detector, in chemical matter there is only a total density of negative electronic charge, variably distributed in the vicinity of positively charged atomic nuclei, not capable of being partitioned into single particles in particular regions or volumes of relative space. Any such attempted partition, according to some capriciously chosen criterion, is inevitably arbitrary. That density can be effectively calculated with quantum-chemical programs, or even measured -- although accuracy is a challenge -- in experiments involving diffraction of electrons and xrays.

### III. SIGNIFICANCE FOR CHEMICAL EDUCATION

In year 1990 there appeared an essay of title *The Nature of the Chemical Bond 1990* [28], which deliberately recalled the same words as in the title of Pauling's highly influential book; the latter first edition appeared a half century before [29], in turn based on articles in a series published during the preceding decade. The subtitle of that essay, *There are no such things as orbitals*, provides a more cogent indication of its tone; in that subtitle, *thing* is supposed to imply a tangible object rather than an abstract entity. According to Bohr, a wave function is an abstract object – simply an element of a theory used to make predictions about observations, with which interpretation one can fully concur. Although that essay attracted the positive attention of theoretical chemists and physicists, which led to its republication in a monograph [*10*] in an expanded form in the company of other theoretically or computationally oriented essays, it was effectively ignored by most chemists, or seemed to be treated as lacking relevance for the way that practising chemists and poorly qualified instructors -- "that sad benighted chemistry professoriate" [*25*] indoctrinated with uncomprehended and fallacious ideas for two or three generations -- proceed to discuss their experimental and computational results and to teach further generations the same nonsense.

That essay [*28*, *10*], which contained information and ideas about the explanation of molecular structure and chemical binding collected during two decades, was essentially a qualitative critique of contemporary abuse of various terms mistakenly derived from not only quantum mechanics but also its preceding *old quantum theory* of Bohr, W. Wilson, Sommerfeld and others; a fatal deficiency of this old quantum theory was a failure to find methods to calculate the intensity of a spectral line, which Schroedinger achieved *at a stroke* in his third paper introducing wave mechanics [*7*]. This melange of ideas about definite orbits and a primitive understanding of the hydrogen atom calculated in only spherical polar coordinates originated during a period when a quantitative understanding of the fundamental bases of chemical structure and reactions exceeded the then current ability to test their correctness or objectivity. Despite the enormous advances in experimental, theoretical and computational capabilities and activities in chemistry that might be deployed at present to combat the obscurity and inaccuracy of those ill formed ideas, the latter linger, to the detriment of chemical education and an improved understanding of the structure of chemical matter and its reactions. Within those computational activities, one must recognise and emphasize the importance of advanced mathematical software, such as *Maple*, that enables the direct algebraic solution of Schroedinger's equations in multiple systems of coordinates, as presented in preceding parts of this series [*2 - 5*]. Equally compelling is the power of contemporary computers applied to purely numerical quantum-chemical calculations, generally described as *quantum chemistry*, of the properties of both molecules, small and large, and materials -- crystalline phases, atomic layers and intermediate matter on a nanometre scale that engenders novel properties and capabilities for applications in diverse areas of science and technology. We reiterate that orbitals, as amplitude functions for the hydrogen atom, arise as a result of the former algebraic calculations but are superfluous as basis functions to facilitate the latter numerical calculations.

We mention above the fallacy perpetrated by G. N. Lewis in attempting to locate individual electrons at particular points in relative space, contrary to the fundamental indistinguishability of electrons and the indeterminacy of such individual location. One might recognise the next major fallacy as being due to N. Bohr; his *aufbauprinzip* essentially amounts to extrapolation from a point, regardless whether Madelung's modification is taken into account. The grossest harm to chemical education arose, however, from Pauling's failure to recognise, among other aspects, that amplitude functions in spherical polar coordinates represent an arbitrary choice of the two coordinate systems that Schroedinger himself applied in the solution of his equations [*7*]. Pauling's attention was directed to the structure of chemical matter on an atomic scale: the full title of his book [*29*] was *The Nature of the Chemical Bond and the Structure of Molecules and Crystals*. Before that book or its preceding papers of the same title appeared, Teller [*27*] had recognised the importance of ellipsoidal amplitude functions of the hydrogen atom that are applicable to a chemical bond, unlike both spherical polar and paraboloidal amplitude functions. Within Pauling's book [*29*] in its three editions appear innumerable descriptions of such structure in terms of orbitals, which are merely amplitude functions -- mathematical formulae -- appropriate in spherical polar, paraboloidal and spheroconical coordinates to only an atom with only one electron and that is isolated from other atoms. As such, these formulae and their nearly universal depictions in spherical polar coordinates constitute an arbitrary choice; the alternative, ellipsoidal functions, would have been a superior choice for Pauling but he ignored them. All these amplitude functions must be perceived as purely mathematical formulae and abstract quantities lacking finite spatial extent, parochial to wave mechanics and foreign to other methods of quantum mechanics, as Bohr recognised: according to wave mechanics, a single hydrogen atom in its ground state might formally occupy the entire universe, but without exclusion of other atomic centres. Even though most amplitude of any such function for small values of the pertinent quantum numbers occurs at distances comparable with internuclear separations in molecules or crystals, as our figures in preceding parts directly demonstrate in coordinates in any of the four systems [*2 - 5*], the point is that these figures pertain expressly to a hydrogen, or other one-electron, atom: their extrapolated application to atoms with multiple electrons is unwarranted and unjustifiable.

      A half century elapsed after the origination of wave mechanics before Woolley proclaimed a truth [*30*] that should have been obvious much earlier, namely that a calculation according to wave mechanics in which the electrons and atomic nuclei of a selected system are treated on a par practically forfeits the possibility of an interpretation of the results in terms of a somewhat rigid frame of atomic nuclei in a particular relative geometric arrangement surrounded by the associated electronic density: the latter is the essence of molecular structure. As atomic nuclei are distinguishable from electrons, and even from one another if the atomic and mass numbers differ, a structure of a diatomic molecule, expressed as an internuclear distance, is formally and practically definable, but, ironically, such a diatomic molecule lacks other than a trivial shape or structure. If one simply writes, for a particular assembly of $N$ nuclei and $n$ electrons in a polyatomic molecule, a full hamiltonian operator that includes only terms for the kinetic energy of both electrons and nuclei with the electrostatic potential energy of all their interactions -- which corresponds to the standard hamiltonian in the practice of quantum chemistry apart from an inclusion here of nuclear kinetic energy, one finds clearly that the result of the solution of the Schroedinger equation, necessarily by purely numerical means, yields only an energy, or rather prospectively a manifold of energies of all states discrete and continuous. For a particular collection of nuclei and electrons, those energies would encompass all possible conformational isomers, such as ethanol and dimethyl ether for $C_2H_6O$ [*11*], or cyclopropene, allene and propyne for $C_3H_4$ [*31*]. As such a classical molecular structure is incompatible with quantum mechanics, to justify or to rationalise such a structure with arbitrarily selected amplitude functions based on wave mechanics for a hydrogen atom is profoundly illogical, despite the fact that semi-empirical calculations, based roughly on Schroedinger's equation used selectively without nuclear kinetic energy, can reproduce or even predict such structure in favourable cases. In the most accurate such calculations including perturbation theory to large order to take extensive account of electron correlation, any relation between the details of the basis set and the eventual structure is lost in the quest for an ever more negative minimum energy of the system. In typical contemporary discussions of molecular structure in chemical education, an invocation of orbitals of one kind or another divorced from such calculations is the mechanism of a circular argument, whether implicit or explicit: a known molecular structure is considered to imply chosen orbitals or their combinations as hybrids on particular atoms, and then that orbital configuration is applied to explain the structure. The process is blatantly fraudulent [*32*], and incomprehensible to students unless and until they themselves become indoctrinated, so as to repeat, in the manner of a parrot, or to echo -- replete with distortions, such explanations. To abandon such arguments incorporating

orbitals and to teach, honestly and frankly, molecular structure as a consequence of experimental measurements of a classical nature is unquestionably a more effective heuristic strategy.

Although some introduction to *quantum mechanics*, generally only wave mechanics, has become universally an integral component of the teaching of physical chemistry, somewhat displacing chemical kinetics that constitutes the treatment of transformations of chemical matter that are the *raison d'être* of chemical science, within an undergraduate curriculum there might be insufficient time to transmit an intimate knowledge of the theoretical basis and construction of quantum-chemical programs and their effective use that must precede their competent application in other than a superficial manner. In any case, such applications are of largely marginal interest in the general practice of chemistry. One contemporary pretext for the teaching of quantum mechanics is as a basis of molecular structure, but such molecular structure is incompatible with rigorous quantum mechanics [*30*]! Clearly not only orbitals but also quantum mechanics, as a collection of mathematical methods [*13*], can be opportunely discarded from general chemical education until the post-graduate level; if topics of quantum physics, such as the details of atomic spectra and the photoelectric effect of molecules or solid materials, be deemed germane for significant objectives, there is no objection to their discussion, free from the baggage of orbitals. For the purposes of analytical chemistry, electronic transitions of atoms play an important role in practical quantitative analysis, but the traditional treatment of atomic spectra in analytical or physical chemistry is inadequate for other than a superficial description of the underlying phenomena. Photoelectron spectra in typical practice must be recognised to be concerned with transitions of a molecule from a, generally neutral, ground electronic state to various electronic states of a molecular cation, not to a loss of an electron from a fictitious atomic or molecular orbital.

Despite the astonishing progress of both experimental techniques to characterise the structure of molecules and other chemical matter and of computational schemes to reproduce that structure and its associated properties, the chemical bond [*33*, *34*], whatever that might be, remains just as much an enigma as when the first chemists and physicists sensed the presence of geometrical order at an atomic level during the mid-nineteenth century. What is a chemical bond? One might respond with the same answer attributed to Thomas Aquinas who was asked to define time: "I knew before you asked me". There are strong bonds and weak bonds, short bonds and long bonds; bonds might form or break, oscillate or rotate. A chemical bond might exist in the minds of chemists who perceive a qualitative description of diverse chemical matter, but neither experiment nor theory nor calculation unequivocally elucidates its palpable existence. What we can measure and calculate are the mean distances between centres of electronic charge associated with atomic nuclei and the density of electronic charge in the vicinity of those atomic nuclei; any attribution of a chemical bond between two such nuclei is, from a quantitative point of view, inevitably a figment of one's chemical imagination.

Atomic and molecular spectra are integral and invaluable tools of the practice of chemistry; their introduction and treatment are essential components of chemical education, but their discussion can rely on classical description and explanation, in combination with quantum laws or the laws of discreteness [*28*]; in practice, that classical description occurs anyhow, despite the pretence to embellish with terms of ostensibly quantum-mechanical aggrandizement. Many textbooks of physical chemistry introduce quantum-mechanical -- nearly invariably merely wave-mechanical -- concepts before discussing the spectra of simple molecules. Other textbooks of physical or inorganic chemistry, increasingly generated, usurp the primary role of macroscopic chemical thermodynamics by beginning with microscopic quantum mechanics. Although the latter practice might seem logical, apart from the schism between quantum mechanics and molecular structure, its systematic development to encompass, for instance, liquid crystals, or even van't Hoff's equation for osmotic pressure that has likewise practical applications, would take forever. With regard to quantum mechanics and spectra, the historical order was the reverse of the relative placement in current textbooks. The first observation in quantum physics was likely the discovery of dark lines within the emitted continuum of the solar spectrum, by Wollaston in Cambridge in 1802; these lines were subsequently classified by Fraunhofer, as mentioned above. The regularities in atomic spectra, deduced by Balmer, and in molecular spectra, by Deslandres nearly concurrently, then became the first quantitative analyses in quantum physics. The recording of structure in the bands of infrared spectra of gaseous diatomic molecules after a few years created a further impetus for the understanding of the *quantum laws of matter and radiation* [*10*, *28*]. As Bjerrum's first quantum theory of molecules, related to these infrared spectra, that appeared in 1912, so preceding Bohr's quantum theory of the hydrogen atom, also preceded Rutherford's recognition of the nuclear atom, it was bound to be unsuccessful [*35*]. By 1920 the distinction among rotational, vibrational and electronic motions in simple molecules in relation to their spectra was appreciated. Such rotational and vibrational motions are incontestably a classical interpretation -- nobody has ever directly observed a molecule undergoing a vibrational motion, on a time scale ~ $10^{-13}$ s, or even a rotational motion, on a time scale ~ $10^{-10}$ s. What one can observe through xray diffraction is that the electronic density around an atomic nucleus in a crystalline sample might become more compact as the temperature of the crystal is decreased toward 0 K. The electronic motions associated with spectral transitions at photon frequencies ~ $10^{15}$ Hz are more difficult to picture in classical terms than the vibrational -- internal -- motion of a molecule, or rotational -- external -- motion about an axis within the molecule; a simplistic description as involving a density of electronic charge near some particular nuclei that is altered between the states involved in an electronic transition, whether or not accompanied with altered internuclear distances in the case of molecules, might serve for that purpose. According to quantum mechanics, there are no such rotational and vibrational motions, just as there is no molecular structure, and for the same reason; as mentioned above, there are only energies of states of which some energy differences between discrete states might be associated, classically, with rotational or vibrational transitions. The relative order of rotational, vibrational and electronic transitions with generally increasing frequency or energy of photons is no guide to the nature of such a transition; for instance, a transition between two electronic states of nitrogen oxide, NO, distinguished by their angular momenta expressed in their term symbols, $^2\Pi_{3/2} \leftarrow {}^2\Pi_{1/2}$, occurs in the midst of transitions associated with rotational motion of $H_2O$. An appeal to quantum mechanics to explain rotational or vibrational motions is clearly yet another logical fallacy. In particular, a canonical linear harmonic oscillator, possessing a quadratic dependence on displacement, is a farcical basis for a model of a diatomic molecule; apart from its evenly spaced and uncountable discrete energies with thus no finite dissociation energy, and apart also from transitions only between states of adjacent energies, its rotational parameters increase systematically with vibrational energy, contrary to the general systematic decrease of these parameters for any real diatomic molecular species. That canonical oscillator serves as a useful exercise in physics to introduce the diversity of quantum-mechanical methods [*11*], but has little relevance to chemistry. Even its invocation to explain the continuous spectral distribution from a *black body* is superseded [*36*]; according to its continuous nature, that distribution is, in any case, inconsistent with a necessity of an interpretation involving discrete quantities. Associations of roughly evenly spaced lines in the far infrared region with rotational motions and (more) roughly evenly spaced bands in the mid- and near-infrared regions with vibrational motions are readily argued on the basis of the isotopic effects, between $^1H^{35}Cl$ and $^2H^{35}Cl$ for instance, and, by analogy, for spectral features of other molecules in those regions [*37*]. On the same basis, the lack of appreciable effect of nuclear mass on spectral lines at the onset of a spectral system in the visible and ultraviolet regions warrants an association with a transition between electronic states, which might be accompanied with vibrational and rotational energies altered from those of the ground state. With a separation of nuclear and electronic motions, quantum-mechanical methods are, however, useful to generate the relative energies of states of an asymmetric rotor, for instance.

For atomic or molecular spectra based directly on properties described formally as intrinsic angular momenta of electrons and nuclei, such as nuclear magnetic resonance or electron paramagnetic resonance, wave mechanics is essentially useless, because Schroedinger's equations involve spatial coordinates that are inapplicable to these *spin* phenomena. For instance, in discussing these molecular spectra on the basis of chemical shifts and coupling parameters, some textbooks of physical chemistry present a matrix, with its component matrix elements to be made diagonal to yield the energies of states, or a determinant of that matrix, without admitting the relation to the original matrix mechanics. The eventual description of magnetic-resonance spectra in those books proceeds to become qualitative and pragmatic; this approach is typical in organic chemistry, which suffices for the effective use of NMR spectra that play an enormous role in the conduct of organic and inorganic chemistry. As there is no classical basis of these magnetic-resonance spectra, unlike spectra associated with rotational and vibrational motions, a pragmatic approach is unavoidable. Incidentally, Dirac considered matrix mechanics to be more fundamental than wave mechanics [*38*], in part because Schroedinger's approach applied in quantum electrodynamics led to intractable infinities whereas Heisenberg's approach was practicable. Ironically, particular textbooks on quantum mechanics and theoretical chemistry such as that by Eyring, Walter and Kimball [*39*], and more recently those by J. Simons [*40*, *41*], for instance, make no concession to the fact that matrix mechanics was ever developed -- even though it was the instigation for wave mechanics. This myopic view of quantum mechanics for chemical purposes is deprecable.

IV.    CONCLUSION

In summary, on the basis of the preceding arguments, one can cogently argue to abandon not only the use of orbitals of a hydrogen, or other one-electron atom, except within that specific context -- to eliminate a vestige of Prout's hypothesis, but also the teaching of quantum mechanics in chemistry, before the post-graduate level in chemistry for perceived specialist purposes. Without that extrapolation from a point, even the teaching of the solution of the hydrogen atom, as presented in the four preceding parts [*2 - 5*], seems to be worthless other than as a mathematical exercise in physics, unessential for chemistry. A reader might assess the authority of this author who makes such an apparently radical proposal: the author has demonstrated and lectured in chemistry in several branches, and in mathematics and physics, for several decades; our qualifications include books on molecular spectrometry [*37*], models for structural chemistry [*42*] and mathematics for chemistry [*43*]. A knowledge and practice of wave mechanics are demonstrated in the preceding parts of this series [*2 - 6*], and a broader practical application of quantum mechanics in three methods elsewhere [*11*], although our knowledge is far from complete [*12*]. Quantum mechanics, we reiterate, constitutes undeniably a collection of mathematical methods or algorithms [*13*, *14*], applicable to calculations pertaining to phenomena on an atomic scale. If students are not expected to make significant use of these methods, apart from esoteric exercises in their development, what is the heuristic value of consuming valuable time and resources in their presentation, to the detriment of other and genuinely chemical topics?

That Planck's flawed derivation of a formula for radiation from a black body, which is a continuous spectral distribution, initiated the era of quantum mechanics is a fallacy as mentioned above [*35*]. Einstein's derivation of the photoelectric effect depended critically on the quantum laws of matter and radiation [*10*]; his treatment appears simple, but its value at the time of its generation was that it was seminal in establishing such discrete or *quantum* properties, which are an essential basis for the understanding of molecular spectra and molecular structure. Heisenberg's principle of indeterminacy that limits the precision with which complementary variables, such as the position and momentum of a particle on an atomic scale, can be measured simultaneously is applicable to an experimental description of measurements on that scale; for instance, although one can generate flashes of light of duration on an attosecond scale, i.e. ~$10^{-17}$ s, the consequent uncertainty or spread of energy precludes the observation of purported atomic vibrations. Aware of these conditions, an instructor of physical chemistry can astutely design courses that genuinely prepare a student to appreciate the structures, properties and transformations of molecules and chemical matter, without the distracting and resisted mathematical digressions that reflect a lack of comprehension of the global scope of chemistry. Authors of textbooks for chemistry in all its branches should revise their content accordingly. The pernicious cycle of instructors, or educational administrators, selecting textbooks to prescribe for their students on the basis of their own superficial understanding, or even ignorance, and then authors pandering to the crudity of those selectors, must be severed.

Writing before the emergence of the present revelations about the wave mechanics of the hydrogen atom, Pritchard advocated a revision of the theory of chemical binding, or "the teaching of valence theory" [*44*]. Such a proposal might presuppose that a description based on electrons being distinguishable or their distribution depending on amplitude functions in an arbitrarily selected system of coordinates is a legitimate objective, whereas the preceding discussion tends to demolish such a description. Because the structure of molecules and chemical materials is a quintessential concern of chemistry, as a basis of a description of chemical reactions, and as that structure is classical in nature, seeking a quantum-mechanical explanation of that incompatible structure is illogical and bound to fail. One might hope for, and work toward, an innovative development of a theory or models to yield an interpretation of the structure of molecules and materials that lacks obvious artifacts, whilst recognizing and applying the practical value of mathematical tools and the software of quantum chemistry, and molecular mechanics, in the praxis of chemistry. Through analytical chemistry that defends the quality of life, and organic chemistry that enables great advances in medicine, and inorganic chemistry with material science that creates an ever improved and expanding range of materials, not to mention the associated chemical industry, chemistry remains the central science. Let us obliterate the pseudo-science based on orbitals and irrelevant quantum mechanics so that chemistry and chemists can legitimately contribute to the solutions of global problems.

**Acknowledgement**

Much common or general historical information contained in this essay has been confirmed from readily accessible sources through internet and the world-wide web, mostly wikipedia entries. I am grateful to several colleagues for valuable discussion and information.